\documentclass[11pt]{article}
\pdfoutput=1 

\usepackage{jheppub} 

\usepackage[T1]{fontenc} 

\usepackage{euscript,amsmath,amsthm,amsfonts,amssymb}
\usepackage{subcaption}

\newcommand{\ket}[1]{|{#1}\rangle}

\newcommand{\bra}[1]{\langle{#1}|}

\newcommand{\ev}[1]{\langle{#1}\rangle}

\newcommand{\pd}[2]{\frac{\partial #1}{\partial #2}}
\newcommand{\R}{\mathbf{R}}

\newcommand{\HH}{\mathcal{H}}
\newcommand{\OO}{\mathcal{O}}
\newcommand{\GN}{G_{\rm N}}

\DeclareMathOperator{\Tr}{Tr}

\DeclareMathOperator{\Real}{Re}

\DeclareMathOperator{\area}{area}

\title{\boldmath Lectures on entanglement entropy in field theory and holography
}

\author{Matthew Headrick}
\affiliation{Martin Fisher School of Physics, Brandeis University, Waltham MA, USA}

\abstract{
These notes, based on lectures given at various schools over the last few years, aim to provide an introduction to entanglement entropies in quantum field theories, including holographic ones. We explore basic properties and simple examples of entanglement entropies, mostly in two dimensions, with an emphasis on physical rather than formal aspects of the subject. In the holographic case, the focus is on how the Ryu-Takayanagi formula geometrically realizes general features of field-theory entanglement, while revealing special properties of holographic theories. In order to make the notes somewhat self-contained for readers whose background is in high-energy theory, a brief introduction to the relevant aspects of quantum information theory is included.
}

\preprint{BRX-TH-6653}

\begin{document} 
\maketitle
\flushbottom

\section{Introduction}
\label{sec:intro}

The phenomenon of \emph{entanglement} is fundamental to quantum mechanics, and one of the features that distinguishes it sharply from classical mechanics. Although sometimes viewed as an exotic phenomenon, entanglement is in fact generic and ubiquitous in quantum systems. Given a way to divide a quantum system into two parts, and a state, it is natural to ask, ``How entangled are they?'' This turns out to be a very fruitful question in almost any area of physics where quantum mechanics plays a role, including the study of quantum many-body systems, quantum field theories, and even quantum gravity theories. \emph{Entropy}, the bedrock concept of statistical mechanics and thermodynamics, is deeply related to entanglement. In particular, the entropy of each subsystem quantifies how entangled they are (in a pure state), giving rise to the so-called \emph{entanglement entropy} (EE). Entanglement thereby connects several areas of theoretical physics, including quantum information theory, condensed-matter theory, high-energy theory, and statistical mechanics.

The aim of these notes is to give a basic introduction to some of the most important and interesting ideas in the study of entanglement in quantum field theories, including both general quantum field theories (in section \ref{sec:EEfields}) and holographic ones, that is, ones with a dual description in terms of classical gravity (in section \ref{sec:EEfieldsgravity}). The key word in the previous sentence is ``some''; no attempt is made to be comprehensive or to bring the reader up to date in this wide-ranging and rapidly-developing field. The target audience is students with a background in high-energy theory. Since such students may not have studied classical or quantum information theory, we begin in sections \ref{sec:1} and \ref{sec:entropy and entanglement} with a brief review of the most relevant aspects of those subjects. On the other hand, the reader is assumed to have some familiarity with quantum field theory, general relativity, and holography. To make them easy to find, \textbf{key words} are put in bold where they are defined.

These notes are based on lecture series given at various schools over the years, as well as a number of overview talks. The lecture series include the following (links are included for the reader interested in watching the lectures):
\begin{itemize}
\item TASI '17: Physics at the Fundamental Frontier (CU Boulder): \\ \href{https://physicslearning.colorado.edu/tasi/tasi_2017/tasi_2017.html}{https://physicslearning.colorado.edu/tasi/tasi\_2017/tasi\_2017.html}. These are the only ones to cover the introduction to information theory from sections \ref{sec:1} and \ref{sec:entropy and entanglement}. An earlier and less complete version of these notes was published as part of the proceedings of TASI '17.
\item 2018 Theory Winter School (MagLab, Tallahassee): \href{https://nationalmaglab.org/news-events/events/2018-theory-winter-school}{https://nationalmaglab.org/news-events/events/2018-theory-winter-school}. Short and aimed at condensed-matter theorists; includes a lightning introduction to holography.
\item PiTP 2018: From Qubits to Spacetime (IAS):\\ \href{https://static.ias.edu/pitp/2018/node/1796.html}{https://static.ias.edu/pitp/2018/node/1796.html}. Covers some advanced material, including the covariant holographic entanglement entropy formula (not discussed in this notes).
\item Quantum Information and String Theory 2019: It from Qubit School/Workshop (Kyoto University): \href{https://www2.yukawa.kyoto-u.ac.jp/~qist2019/4th5th.php}{https://www2.yukawa.kyoto-u.ac.jp/-qist2019/4th5th.php}. Includes an introduction to holography, as well as an overview of bit threads (not discussed in these notes).
\end{itemize}
Of course, these notes contain significantly more material, both in breadth and depth, than can be covered even in a week-long lecture series (at least by this lecturer).

There are many places to learn more about this subject.  A discussion of entanglement and von Neumann entropy can be found in any of the many textbooks on quantum information theory, such as the one by Nielsen-Chuang \cite{NielsenChuang}, as well as many articles such as the classic review by Wehrl \cite{Wehrl} and the recent introduction by Witten \cite{Witten:2018zva}. Useful review articles and lecture notes on EE in field theory and holography include those by Casini-Huerta \cite{Casini:2009sr}, Calabrese-Cardy \cite{Calabrese:2009qy}, Nishioka-Ryu-Takayanagi \cite{Nishioka:2009un}, Van Raamsdonk \cite{VanRaamsdonk:2016exw}, Witten \cite{Witten:2018lha}, and Nishioka \cite{Nishioka:2018khk}. There is also the book devoted to holographic EE by Rangamani-Takayanagi \cite{Rangamani:2016dms}. Finally, chapter 27 of the book by Nastase \cite{Nastase:2015wjb} covers holographic EE.

\subsection{Historical sketch}

Finally, a few comments are in order about the history of the subject. Like so many advances in quantum field theory, the study of EE has been pushed forward by two superficially unrelated goals: on one hand the desire to understand quantum gravity and black holes, and on the other the desire to understand condensed-matter systems. Following Bekenstein and Hawking's work on black-hole entropy in the 1970s, it was understood that in the vacuum of a quantum field theory, even in flat spacetime, a region of space is effectively in a mixed state; specifically, in the Rindler case, this state is the Gibbs state with respect to the boost generator \cite{PhysRevD.14.870,Bisognano:1976za}. It was noted repeatedly by different authors during the 1980s and 1990s (including 't Hooft \cite{tHooft:1984kcu}, Bombelli-Koul-Lee-Sorkin \cite{Bombelli:1986rw}, Susskind \cite{Susskind:1993ws}, and Srednicki \cite{Srednicki:1993im}) that the leading term in the entropy of a region is proportional to its surface area (and UV divergent); hence the Bekenstein-Hawking entropy could be understood at least partly as arising from entanglement of quantum fields across the horizon. This led to a significant amount of work during the 1990s on computing EEs in field theory (and in string theory, which in principle cuts off the UV divergences of field theory). Wilczek and collaborators \cite{Callan:1994py,Holzhey:1994we}, in particular, recognized that these are interesting quantities to consider in quantum field theories even independently of their possible relevance for black holes. They developed useful Euclidean techniques to compute EEs, including the replica trick, and found the famous $\frac c3\ln\frac L\epsilon$ behavior in 2d conformal field theories.

During the 2000s, partly due to the above developments and partly for autonomous reasons (for example, having to do with the question of when and how many-body systems can be efficiently simulated), EEs became a standard tool for condensed-matter theorists for characterizing phases of many-body systems. Landmarks included Calabrese-Cardy's work on EEs in 2d CFTs, starting in \cite{Calabrese:2004eu} and then studying its time dependence in \cite{Calabrese:2005in}; Hasting and collaborators' work showing that gapped systems satisfy an area law \cite{2006PhRvB..73h5115H,Wolf:2007tdq}; Kitaev-Preskill \cite{Kitaev:2005dm} and Levin-Wen's \cite{2006PhRvL..96k0405L} definition of the topological EE; and Li-Haldane's use of the entanglement spectrum to characterize fractional quantum Hall states \cite{2008PhRvL.101a0504L}.\footnote{The characterization of entanglement and the study of various entropies have also always been part and parcel of quantum information theory. Obviously we cannot begin to do justice to the history of this subject, but we note that, in addition to having close ties to condensed-matter theory, quantum information theory has also long benefitted from contributions by high-energy theorists, including Bell, Wheeler, Feynman, Preskill, Farhi, and Goldstone.}

During the same decade, however, only a few isolated groups within high-energy theory were studying EEs. The most important of these were Casini-Huerta, starting with \cite{Casini:2003ix}, and especially Ryu-Takayanagi (RT). By conjecturing a simple formula for EEs in holographic theories \cite{Ryu:2006bv,Ryu:2006ef}, this collaboration between a condensed-matter theorist (Ryu) and a string theorist (Takayanagi) brought the subject full circle back to gravity. The RT formula was a groundbreaking discovery, arguably the most important in quantum gravity since the discovery of holography itself. However, despite some notable early work (including by Emparan \cite{Emparan:2006ni}, Fursaev \cite{Fursaev:2006ih}, Headrick-Takayanagi \cite{Headrick:2007km}, Hubeny-Rangamani-Takayanagi \cite{Hubeny:2007xt}, Klebanov-Kutasov-Murugan \cite{Klebanov:2007ws}, Solodukhin \cite{Solodukhin:2008dh}, Buividovich-Polikarpov \cite{Buividovich:2008gq}, and Swingle \cite{Swingle:2009bg}), it was not until the early 2010s that the validity and utility of RT became fully established and the subject really entered the mainstream of high-energy theory research. The interest generated by RT, along with ongoing work by Casini-Huerta and collaborators and continued strong mixing with the condensed-matter community, brought much progress on EEs not only in holography but in field theories more generally. The developments in this decade have come fast and furious, and are far too numerous to list here. So it is now time to leave off the history and turn to the physics.

\section{Entropy}
\label{sec:1}

\subsection{Shannon entropy}

We start with a little bit of classical information theory, because many of the concepts that carry over to the quantum case can be understood more simply here.

We consider a classical system with a discrete state space labelled by $a$. Although classical systems usually have continuous state (or phase) spaces, we will stick to the discrete case both to avoid certain mathematical complications and because that will provide the relevant math when we turn to the quantum case.

Our knowledge (or lack thereof) of the state of the system is described by a probability distribution $\vec p$, where
\begin{equation}
p_a\ge0\,,\qquad\sum_ap_a = 1\,.
\end{equation}
(For example, the system could be a message about which we have partial information.) The expectation value of an observable $\OO_a$ (or other function of $a$; an observable is a function of $a$ that does not depend on $\vec p$) on the state space is
\begin{equation}
\ev{\OO_a}_{\vec p} = \sum_a\OO_ap_a\,.
\end{equation}
The \textbf{Shannon entropy} \cite{MR0026286} is
\begin{equation}
S(\vec p) := \ev{-\ln p_a}_{\vec p} = -\sum_ap_a\ln p_a
\end{equation}
(where $0\ln0$ is defined to be 0). The entropy detects indefiniteness of the state in the following sense:
\begin{eqnarray}\label{diagignorance}
&&S(\vec p) = 0 \,\Leftrightarrow\,p_a=\delta_{aa_0}\text{ for some $a_0$ }\,\Leftrightarrow\,\text{ all observables have definite values}\,, \nonumber \\
&&\text{otherwise }S(\vec p)>0\,.
\end{eqnarray}
(By ``definite value'' we mean vanishing variance, i.e.\ $\ev{\OO_a^2}=\ev{\OO_a}^2$.)

More than just detecting indefiniteness, the Shannon entropy is a quantitative measure of the \emph{amount} of our ignorance, in other words of how much information we've gained if we learn what actual state the system is in. This assertion is backed up by two basic facts.

First, the Shannon entropy is \emph{extensive}. This means that if $A,B$ are \emph{independent}, so that the joint distribution $\vec p_{AB}$ is 
\begin{equation}
\vec p_{AB} = \vec p_A\otimes\vec p_B\,,\quad
\text{i.e.}\quad(p_{AB})_{ab}=(p_A)_a(p_B)_b\,, 
\end{equation}
then the entropies add:
\begin{equation}
S(\vec p_{AB}) = S(\vec p_A)+S(\vec p_B)\,.
\end{equation}
In particular, the total entropy of $N$ independent copies of $A$ with identical distributions is $NS(\vec p_A)$.

Second, the \emph{Shannon noiseless coding theorem} states that the state of our system can be specified using a binary code requiring on average $S(\vec p)/\ln 2$ bits. For example, for a 3-state system with $p_1=1/2$, $p_2=p_3=1/4$, then $S(\vec p)=(3/2)\ln2$, the following coding uses 3/2 bits on average:
\begin{equation}
1\mapsto0\,,\quad
2\mapsto10\,,\quad
3\mapsto11\,.
\end{equation}
(More precisely, the theorem states that the state of $N$ independent copies of the system, all with the same distribution, can be coded with $NS(\vec p)/\ln 2$ bits in the limit $N\to\infty$.) Thus, all forms of information are interconvertible, as long as the entropies match. The impressive thing about this theorem is that, once we've computed $S(\vec p)$, we know many bits we will need \emph{without having to actually construct a coding} (and, if we do construct a coding, we know how far it is from being optimal). The ability to straightforwardly calculate a quantity of information-theoretic interest is unfortunately the exception rather than the rule: many interesting quantities in information theory are simply \emph{defined} as the optimal rate at which some task can be performed, and there is usually no practical way to calculate them. As physicists, we are mostly interested in quantities that, like the Shannon entropy, are \emph{calculable}.

\subsection{Joint distributions}

Above we saw that the entropy is additive for independent systems. Now we will consider a general joint probability distribution $\vec p_{AB}$, and ask what we can learn from the Shannon entropy.

The \textbf{marginal distribution} $\vec p_A$ is defined by
\begin{equation}\label{marginal}
(p_A)_a = \sum_b(p_{AB})_{ab}
\end{equation}
and likewise for $\vec p_B$. (Note that, in the special case of independent systems, $\vec p_{AB}=\vec p_A\otimes\vec p_B$, the notation is consistent, in the sense that $\vec p_A$ is indeed the marginal distribution as defined by \eqref{marginal}.) $\vec p_A$ is the distribution that gives the right expectation value for any observable that depends only on $a$, i.e.\ $\OO_{ab}=\OO_a$:
\begin{equation}
\ev{\OO_{ab}}_{\vec p_{AB}} = \ev{\OO_a}_{\vec p_A}\,.
\end{equation}

For a state $b$ of $B$ with $(p_B)_b\neq0$, the conditional probability on $A$, denoted $\vec p_{A|b}$, is
\begin{equation}
(p_{A|b})_a = \frac{(p_{AB})_{ab}}{(p_B)_b}\,.
\end{equation}
A few lines of algebra reveal that its entropy, averaged over $b$, can be written in a very simple form:
\begin{equation}\label{condent1}
\ev{S(\vec p_{A|b})}_{\vec p_B} = S(\vec p_{AB})-S(\vec p_B)\,.
\end{equation}
From now on, we will abbreviate $S(\vec p_{AB})$ as $S(AB)$, etc. The quantity on the right-hand side of \eqref{condent1} is called the \textbf{conditional entropy}, and denoted $H(A|B)$:
\begin{equation}
H(A|B):=S(AB)-S(B)\,.
\end{equation}
This is how much, on average, you \emph{still} don't know about the state of $A$ even after learning the state of $B$. Note that, as the expectation value of a non-negative function, it is non-negative:
\begin{equation}
H(A|B)\ge 0\,.
\end{equation}
This implies that, if $S(AB)=0$ then $S(B)=0$, in other words if the full system is in a definite state then so are all of its parts. We emphasize this rather obvious point since it fails in the quantum setting.

The amount of information about $A$ you've gained (on average) from learning the state of $B$ is
\begin{equation}
S(A)-H(A|B) = S(A)+S(B)-S(AB)=:I(A:B)\,.
\end{equation}
$I(A:B)$ is called the \textbf{mutual information}. Note that it is symmetric between $A$ and $B$. By extensivity of the entropy, the mutual information vanishes between independent systems:
\begin{equation}
\vec p_{AB}=\vec p_A\otimes\vec p_B\quad\Rightarrow\quad S(AB)=S(A)+S(B)\quad\Rightarrow\quad I(A:B)=0\,.
\end{equation}
(Clearly, if the systems are independent, then learning $b$ tells you nothing about $a$.) It turns out that in the presence of correlations the mutual information is necessarily positive:\footnote{For this and other claims made without proof in this and the following section, the proofs can be found in \cite{Wehrl}, \cite{NielsenChuang}, or most other quantum information textbooks.}
\begin{equation}
\vec p_{AB}\neq\vec p_A\otimes\vec p_{B}\quad\Rightarrow\quad
S(AB)<S(A)+S(B)\quad\Rightarrow\quad
I(A:B)>0\,.
\end{equation}
For example, for a perfectly correlated pair of bits, with $p_{00}=p_{11}=1/2$,
\begin{eqnarray}
I(A:B)=\ln2\,.
\end{eqnarray}
The inequality
\begin{equation}
S(AB)\le S(A)+S(B)
\end{equation}
is called \textbf{subadditivity} of Shannon entropy.

Correlations can also be detected by correlation functions between observables on $A$ and on $B$:
\begin{equation}
\ev{\OO_a\OO'_b}_{\vec p_{AB}}-\ev{\OO_a}_{\vec p_A}\ev{\OO'_b}_{\vec p_B}\,.
\end{equation}
The correlation functions vanish for \emph{all} observables $\OO_a$, $\OO'_b$ if and only if the systems are independent. In practice, one is unlikely to compute the correlation function for \emph{all} possible observables. The correlations may be \emph{hidden}, in the sense that they may be invisible to an accessible set of observables. The advantage of the mutual information, if it can be computed, is that it detects correlations of any form.

\begin{figure}[tbp]
\centering
\includegraphics[width=0.7\textwidth]{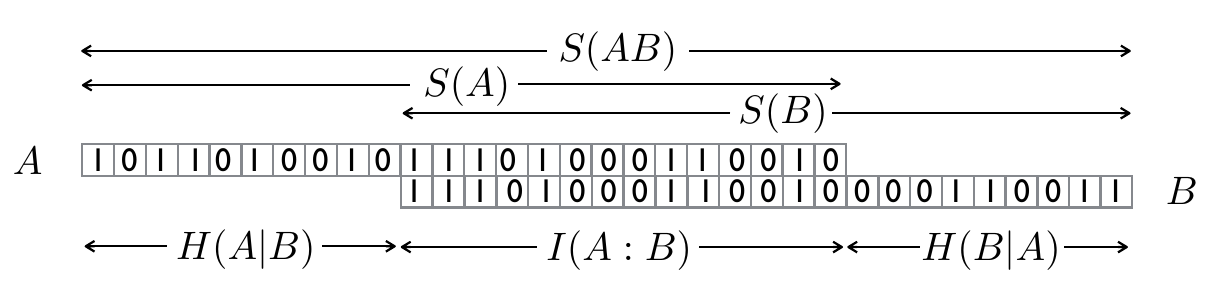}
\caption{\label{fig:MIclassical}
Schematic illustration of the mutual information and conditional entropy in a classical bipartite system: Two correlated systems $A$ and $B$ can be encoded into $S(A)$ and $S(B)$ bits respectively, such that $I(A:B)$ bits of each are perfectly correlated, $H(A|B)$ bits of $A$ are uncorrelated with those of $B$, and $H(B|A)$ bits of $B$ are uncorrelated with those of $A$. (Factors of $\ln2$ have been dropped for clarity.)
}
\end{figure}

There is a simple picture, shown in fig.\ \ref{fig:MIclassical}, which helps to understand the meaning of the conditional entropy and mutual information. $AB$ is coded into $S(AB)/\ln 2$ bits, divided into three groups containing $H(A|B)/\ln2$, $I(A:B)/\ln2$, and $H(B|A)/\ln2$ bits respectively. The first two groups together are a coding of $A$, while the last two are a coding of $B$. Thus $H(A|B)/\ln2$ bits belong purely to $A$, $H(B|A)/\ln2$ bits belong purely to $B$, and $I(A:B)/\ln2$ bits are perfectly correlated between the two systems. (More precisely, this is a coding of $N$ independent identical copies of $AB$, with the number of bits in each group multiplied by $N$.) Thus, $I(A:B)$ does not just detect the \emph{existence} of correlation, but quantifies the \emph{amount} of correlation.

\subsection{Von Neumann entropy}\label{sec:vne}

We now move on to quantum mechanics. One of the many strange features of quantum mechanics is that it does not allow us to separate the notion of ``the actual state of the system'' from the notion of ``our knowledge of the state of the system''. Together, they are encoded in what is simply called a \textbf{state}, or \textbf{density matrix}\footnote{``State'' is more common in the quantum information literature; ``density matrix'' is more common in the physics literature, where ``state'' often implicitly means pure state.}, which is an operator $\rho$ with the following properties:
\begin{equation}\label{rhodef}
\rho^\dag = \rho\,,\qquad
\rho\ge0\,,\qquad
\Tr\rho = 1\,.
\end{equation}
The expectation value of an observable (or other operator) $\OO$ is
\begin{equation}
\ev{\OO}_\rho = \Tr(\OO\rho)\,.
\end{equation}
The most ``definite'' states are the projectors, of the form
\begin{equation}\label{pure def}
\psi=\ket{\psi}\bra{\psi}\,.
\end{equation}
Such a state is called \textbf{pure}; the rest are \textbf{mixed}. In \eqref{pure def}, we have followed the quantum-information convention of using the variable name for a pure state as the label for the corresponding ket and bra. 
Whereas, in classical mechanics, in a definite state all observables have definite values, in quantum mechanics by the uncertainty principle even in a pure state not all observables have definite values.

More generally, based on \eqref{rhodef}, $\rho$ can be diagonalized,
\begin{equation}
\rho = \sum_ap_a\ket{a}\bra{a}\,,
\end{equation}
and its eigenvalues $p_a$ form a probability distribution. The Shannon entropy of this distribution,
\begin{equation}\label{vN def}
S(\rho) := S(\vec p) = -\Tr\rho\ln\rho = \ev{-\ln\rho}_\rho\,,
\end{equation}
is the \textbf{von Neumann entropy} of $\rho$. From \eqref{diagignorance}, we have $S(\rho)\ge0$, and $S(\rho)=0$ if and only if $\rho$ is pure. Thus the von Neumann entropy is a diagnostic of mixedness. An important property, obvious from its definition, is invariance under unitary transformations:
\begin{equation}
S(U\rho U^{-1}) = S(\rho)\,.
\end{equation}

When we consider joint systems, much but---as we will see---not all of the classical discussion carries through. The Hilbert space for the joint $AB$ system is the tensor product of the respective $A$, $B$ Hilbert spaces,
\begin{equation}\label{AB factorization}
\HH_{AB}=\HH_A\otimes\HH_B\,.
\end{equation}
Given a state $\rho_{AB}$, the effective state on $A$ which reproduces the expectation value of an arbitrary observable of the form $\OO_A\otimes I_B$ is given by the partial trace on $B$:
\begin{equation}
\rho_A:=\Tr_B\rho_{AB}
\end{equation}
(where $\Tr_B$ is shorthand for $\Tr_{\HH_B}$). This is called the \textbf{marginal state} or \textbf{reduced density matrix} (the former is more common in the quantum information theory literature, and the latter in the physics literature). As in the classical case, if $\rho_{AB}$ is a product state, $\rho_{AB}=\rho_A\otimes\rho_B$, then all correlation functions of operators on $A$ and $B$ vanish:
\begin{equation}
\ev{\OO_A\otimes\OO_B}_{\rho_{AB}} - \ev{\OO_A}_{\rho_A}\ev{\OO_B}_{\rho_B} = 0\,;
\end{equation}
in this case we say that $A$ and $B$ are uncorrelated.

Like the Shannon entropy, the von Neumann entropy is extensive and subadditive:
\begin{eqnarray}
\rho_{AB}=\rho_A\otimes\rho_B\quad&\Leftrightarrow&\quad S(AB)=S(A)+S(B)\\
\rho_{AB}\neq\rho_A\otimes\rho_b\quad&\Leftrightarrow&\quad S(AB)<S(A)+S(B)
\end{eqnarray}
Hence the mutual information, defined again by
\begin{equation}
I(A:B):=S(A)+S(B)-S(AB)\,,
\end{equation}
detects correlation. It can also be viewed as a quantitative measure of the amount of correlation. This is more subtle than in the classical case, since, as we will see, in general there is no simple coding of the state like the one described in fig.\ \ref{fig:MIclassical}. However, there are two facts that support the interpretation of the mutual information as the amount of correlation. The first is that it bounds correlators:
\begin{equation}
\frac12\left(\frac{\ev{\OO_A\otimes\OO_B} - \ev{\OO_A}\ev{\OO_B}}{\|\OO_A\|\|\OO_B\|}\right)^2\le I(A:B)\,,
\end{equation}
where $\|\cdot\|$ is the operator norm \cite{HiaiOT81,PhysRevLett.100.070502}. (Unfortunately, this bound is not so useful in the field-theory context, where the observables of interest are rarely bounded operators.) The second is that it is non-decreasing under adjoining other systems to $A$ or $B$:
\begin{equation}
I(A:BC)\ge I(A:B)\,.
\end{equation}
In terms of the entropy, this inequality becomes
\begin{equation}\label{SSA1}
S(AB)+S(BC)\ge S(B)+S(ABC)\,,
\end{equation}
called \textbf{strong subadditivity} \cite{LiebRuskai}. Strong subadditivity is a cornerstone of quantum information theory.

The entropies of subsystems also obey two other useful inequalities:
\begin{equation}\label{arakilieb}
S(AB)\ge|S(A)-S(B)|\,,
\end{equation}
called the \textbf{Araki-Lieb} inequality,
and
\begin{equation}\label{SSA2}
S(AB)+S(BC)\ge S(A)+S(C)\,,
\end{equation}
called either the \emph{second form of strong subadditivity} or \emph{weak monotonicity}. A useful special case of Araki-Lieb occurs when $AB$ is pure; then it implies
\begin{equation}\label{pure}
S(A)=S(B)\,.
\end{equation}

For any $\rho_A$, we can define a formal ``Hamiltonian'' $H_A$, called the \textbf{modular Hamiltonian}, with respect to which it is a Gibbs state:
\begin{equation}\label{modHamdef}
\rho_A = \frac1Ze^{-H_A}\,.
\end{equation}
Here the ``temperature'' is by convention set to 1. $H_A$ is defined up to an additive constant. (If $\rho_A$ has vanishing eigenvalues, then $H_A$ has infinite eigenvalues.) From \eqref{vN def}, we have
\begin{equation}\label{modularHamiltonian}
S(A) = \ev{H_A}+\ln Z\,.
\end{equation}
(Note that the expectation value can be evaluated either on $A$ or on the full system. In the latter case, strictly speaking one should write $H_A\otimes I_B$; however, it is conventional to leave factors of the identity operator implicit in cases like this.) The fact that any subsystem can, in any state, be thought of as being in a canonical ensemble will allow us to employ our intuition from statistical physics. As we will see, there are also a few interesting cases where $H_A$ can be written in closed form and $S(A)$ can be evaluated using \eqref{modularHamiltonian}. The spectrum of $H_A$ is sometimes referred to as the \textbf{entanglement spectrum}, especially in the condensed-matter literature.

\subsection{R\'enyi entropies}
\label{sec:Renyis}

The \textbf{R\'enyi entropies} (sometimes called \emph{$\alpha$-entropies}) $S_\alpha$ are a very useful one-parameter generalization of the Shannon entropy, defined for $\alpha\in[0,\infty]$. Since we will apply them in the quantum context, we will define them directly in terms of $\rho$:
\begin{eqnarray}
S_\alpha &:=& \frac1{1-\alpha}\ln\Tr\rho^\alpha=\frac1{1-\alpha}\ln\left(\sum_ap_a^\alpha\right) \qquad(\alpha\neq0,1,\infty) \\
S_0&:=&\lim_{\alpha\to0}S_\alpha = \ln(\text{rank}\rho) \\
S_1&:=&\lim_{\alpha\to1}S_\alpha = S \\
S_\infty&:=&\lim_{\alpha\to\infty}S_\alpha = -\ln\|\rho\| = -\ln(\max_ap_a)\,.
\end{eqnarray}

$S_\alpha$, for any fixed $\alpha$, shares several important properties with $S$. First, it is unitarily invariant. Second, it is non-negative and vanishes if and only if $\rho$ is pure. Therefore it detects mixedness.\footnote{$S_\alpha$ is also a good quantitative measures of mixedness in the sense that it increases under mixing, in other words replacing $\rho$ by $\sum_ip_iU_i^\dag\rho U_i$, where $\{p_i\}$ is a probability distribution and the $U_i$'s are unitaries. See \cite{Wehrl} for details.} Third, it is extensive. Therefore the R\'enyi mutual information,
\begin{equation}
I_\alpha(A:B):=S_\alpha(A)+S_\alpha(B)-S_\alpha(AB)\,,
\end{equation}
detects correlations, in the sense that if it is non-zero then $A$ and $B$ are necessarily correlated. However, the R\'enyi entropy is not subadditive (except for $\alpha=0,1$), so the converse does not hold; $I_\alpha(A:B)$ can be zero, or even negative, in the presence of correlations. Related to this, $S_\alpha$ does not obey strong subadditivity, so $I_\alpha(A:B)$ can decrease under adjoining other systems to $A$ or $B$. For these reasons (and another one given below), $I_\alpha(A:B)$ is a poor quantitative measure of the amount of correlation.

$S_\alpha$ also has some useful properties as a function of $\alpha$ (for fixed $\rho$), for example
\begin{equation}
\frac{dS_\alpha}{d\alpha} \le0\,,\qquad
\frac{d^2S_\alpha}{d\alpha^2} \ge0\,.
\end{equation}

Our interest in the R\'enyi entropies stems mainly from the fact that some of them are much more easily computed than the von Neumann entropy. A direct computation of the von Neumann entropy from its definition requires an explicit representation of the operator $\ln\rho$, which essentially requires diagonalizing $\rho$. On the other hand, computing $\Tr\rho^2$, $\Tr\rho^3$, \ldots, and therefore $S_2$, $S_3$, . . ., is relatively straightforward given the matrix elements of $\rho$ in any basis. What can we learn from these R\'enyis? First, as discussed above, we learn whether $\rho$ is pure, and, in the case of joint systems, we may learn whether they are correlated. Second, it can be shown that $\Tr\rho^\alpha$ is an analytic function of $\alpha$, which moreover can in principle be determined from its values at $\alpha=2,3,\ldots$.\footnote{Since the eigenvalues $p_a$ of $\rho$ are a probability distribution, $\Tr\rho^\alpha=\sum_ap_a^\alpha$ is a uniformly convergent sum of analytic functions, and is therefore itself an analytic function, in the region $\Real\alpha\ge1$. Furthermore, it is bounded (in absolute value) by 1, and therefore satisfies the assumptions of Carlson's theorem, which makes it uniquely determined by its values at the integers $\alpha>1$. Note that here we are discussing a fixed state $\rho$; taking a \emph{limit} of states (such as a thermodynamic limit) may introduce non-analyticities, as we will discuss futher below.} If we can figure out what that function is, and take its limit as $\alpha\to1$, we learn the von Neumann entropy. In many cases, especially in field theories, this is the most practical route to the von Neumann entropy \cite{Holzhey:1994we}. Knowing $S_\alpha$ for all $\alpha$ also in principle determines, via an inverse Laplace transform, the full spectrum of $\rho$.

To get a feeling for the R\'enyis, let us consider two examples. First, if $\rho$ is proportional to a rank-$n$ projector $P$,
\begin{equation}
\rho = \frac1nP\,,
\end{equation}
then the R\'enyis are independent of $\alpha$:
\begin{equation}
S_\alpha = \ln n\,.
\end{equation}
This would apply, for example, to the microcanonical ensemble
\begin{equation}
\rho = \frac1{n(E)}\sum_{E\le E_a\le E+\Delta E}\ket{a}\bra{a}\,,
\end{equation}
(where the $\ket{a}$ are energy eigenstates), in which case
\begin{equation}\label{mcrenyi}
S_\alpha = S(E) := \ln n(E)\,,
\end{equation}
where $S(E)$ is the microcanonical entropy.

In the canonical ensemble, or Gibbs state,
\begin{equation}
\rho=\frac1{Z_\beta}e^{-\beta H}\,.
\end{equation}
we find instead
\begin{equation}\label{crenyi}
S_\alpha = \frac{\alpha\beta}{1-\alpha}\left(F_\beta-F_{\alpha\beta}\right)\qquad (\alpha\neq0,1,\infty)\,,
\end{equation}
where $Z_\beta$ and $F_\beta$ are the partition function and free energy respectively at temperature $1/\beta$.

These two examples illustrate an important fact about R\'enyi entropies, which explains why they are not very familiar to most physicists: they are not thermodynamic quantities. By this we mean the following. In a thermodynamic limit, with $N$ degrees of freedom, extensive quantities (such as the free energy and von Neumann and R\'enyi entropies) are, to leading order in $1/N$, proportional to $N$. Furthermore, for certain quantities, this leading term depends only on the macroscopic state; it is independent of the particular ensemble---microcanonical, canonical, grand canonical, etc.---used to describe that state. These leading parts are the thermodynamic quantities, the ones we learn about in high-school physics. The subleading (order $N^0$ and higher in $1/N$)\ terms, meanwhile, are sensitive to statistical fluctuations and therefore depend on the choice of ensemble. From \eqref{mcrenyi} and \eqref{crenyi}, we immediately see that the leading term in $S_\alpha(\rho)$ for $\alpha\neq1$ is of order $N$ but depends on the ensemble, and is therefore \emph{not} a thermodynamic quantity.

We can say this another way: Thermodynamics is related to statistical mechanics essentially by a saddle-point approximation. The operations of calculating a R\'enyi entropy and making a saddle-point approximation do not commute, since taking the $\alpha$ power of $\rho$ will typically shift the saddle-point. This can be seen in \eqref{crenyi}, where $S_\alpha$ depends on the free energy at the temperature $1/(\alpha\beta)$ rather than the system's actual temperature $1/\beta$. The shift in the saddle-point has several implications: (1) for fixed $\alpha$, $S_\alpha$ will have phase transitions at different values of the temperature and other parameters than the actual system; (2) for fixed parameters, $S_\alpha$ may have phase transitions (i.e.\ be non-analytic) as a function of $\alpha$; (3) for joint systems that are correlated only via their fluctuations (for example, a cylinder with two chambers separated by a piston), such that $I(A:B)$ is of order $N^0$, $I_\alpha(A:B)$ is typically of order $N$---a drastic overestimate of the amount of correlation.\footnote{See \cite{Headrick:2013zda} for further discussion.}

All of which is to say that, when working in a thermodynamic limit, the R\'enyis should be treated with extreme caution.

\section{Entropy and entanglement}
\label{sec:entropy and entanglement}

\subsection{Entanglement in pure states}\label{sec:entanglement}

So far, we have emphasized the parallels between the concepts of entropy in the classical and quantum settings. Now we will discuss some of their crucial differences. We will start by consider the \emph{pure} states of a bipartite system $AB$, where these differences are dramatically illustrated. The analogous states in the classical setting are, as we saw, quite boring: a definite state of a bipartite system is necessarily of the form $\vec p_{AB}=\vec p_A\otimes\vec p_B$, with no correlations, and with each marginal state $\vec p_A$, $\vec p_B$ itself definite. These statements are not true in quantum mechanics.

Consider a \textbf{Bell pair}, a two-qubit system in the following pure state:
\begin{equation}
\ket{\psi}=\frac1{\sqrt{2}}\left(\ket{0}_A\otimes\ket{0}_B+\ket{1}_A\otimes\ket{1}_B\right).
\end{equation}
This state clearly cannot be factorized, i.e.\ written in the form
\begin{equation}\label{factorized}
\ket{\psi}=\ket{\psi'}_A\otimes\ket{\psi''}_{B}
\end{equation}
for any choice of $\ket{\psi'}_A$, $\ket{\psi''}_B$. Furthermore, its marginals are easily computed,\footnote{Recall that $\psi:=\ket{\psi}\bra{\psi}$ (see \eqref{pure def}). We write $\ket{0}_A\bra{0}$ for the projector onto $\ket{0}_A$, etc. We strongly advise the reader to check \eqref{marginals} by writing out the terms in $\psi$ and taking the partial traces.}
\begin{align}\label{marginals}
\psi_A &:= \Tr_B\psi
=\frac12\left(\ket{0}_A\bra{0}+\ket{1}_A\bra{1}\right),\\
\psi_B &:=  \Tr_A\psi=\frac12\left(\ket{0}_B\bra{0}+\ket{1}_B\bra{1}\right),\nonumber
\end{align}
and they are mixed!

A pure state that cannot be factorized is called \textbf{entangled}. As mentioned at the very start of these lectures, entanglement is one of the key features that distinguishes quantum from classical mechanics. It is at the root of many seemingly exotic quantum phenomena, such as violation of the Bell inequalities, and it plays a crucial role in essentially all potential quantum technologies, including quantum cryptography and quantum computation.

The example of the Bell pair suggests a connection between entanglement and mixedness of the two subsystems. For a factorized state, of the form \eqref{factorized}, the subsystems are clearly pure:
\begin{equation}
\psi_A = \ket{\psi'}_A\bra{\psi'}\,,\qquad
\psi_B = \ket{\psi''}_B\bra{\psi''}\,.
\end{equation}
The converse is easily established. Using the singular value decomposition, an arbitrary pure state on $AB$ can be written in the form
\begin{equation}\label{Schmidt}
\ket{\psi} = \sum_{a}\sqrt{p_a}\ket{a}_A\otimes\ket{a}_B\,,
\end{equation}
where $\{\ket{a}_A\}$, $\{\ket{a}_B\}$ are orthonormal (not necessarily complete) sets of vectors in $\HH_A$ and $\HH_B$ respectively, and $\{p_a\}$ is a probability distribution. \eqref{Schmidt} is called the \textbf{Schmidt decomposition}. (It is unique up to a unitary acting jointly on any subsets $\ket{a}_A$, $\ket{a}_B$ with degenerate $p_a$s.) The marginals are then
\begin{equation}\label{ABmarginals}
\psi_A = \sum_ap_a\ket{a}_A\bra{a}\,,\qquad
\psi_B = \sum_ap_a\ket{a}_B\bra{a}\,.
\end{equation}
Thus $\psi_A$ is pure if and only $\psi_B$ is pure, and this is true if and only if $\ket{\psi}$ is factorized.

We can therefore use the R\'enyi entropy $S_\alpha(A)$ (which equals $S_\alpha(B)$ by \eqref{ABmarginals}) for any $\alpha$ to detect entanglement between $A$ and $B$. For this reason, physicists often refer to the entropies of subsystems as \textbf{entanglement entropies} (EEs)---even when the full system is \emph{not} (as we have been discussing here) in a pure state, and therefore $S(A)$ does \emph{not} necessarily reflect entanglement! For this and other reasons, this terminology is not ideal, and it is generally frowned upon by qunatum information theorists. Nonetheless, since it is common in the physics literature, as a synonym for ``subsystem entropy'', we will follow it in these notes.

The von Neumann entropy $S(A)$ not only detects entanglement, but is also a quantitative measure of the \emph{amount} of entanglement, in the following sense. A unitary acting only on $A$ or $B$ cannot change $S(A)$ or $S(B)$, and therefore cannot turn a factorized state into an entangled one. More generally, no combination of \emph{local operations} on $A$ and $B$ and \emph{classical communication} between $A$ and $B$ (abbreviated \textbf{LOCC}) can create a pure entangled state from a factorized one. (We will consider the mixed case below.) Therefore, if $A$ and $B$ are separated, creating entanglement requires actually sending some physical system between them (or from a third party), so entanglement between separated systems should be viewed as a scarce resource. Now, LOCC can transform entangled states into other entangled states, for example a general entangled state $\ket{\psi}$ into a set of Bell pairs. More precisely, $N$ identical copies of $\ket{\psi}$ can be transformed into $NS(A)/\ln2$ Bell pairs and vice versa, in the limit $N\to\infty$. This is very similar in spirit to the Shannon noiseless coding theorem. It shows that different forms of entanglement are interconvertible under LOCC, as long as the \emph{amount} of entanglement matches, as quantified by the EE. As with the Shannon theorem, the utility of this statement derives in large part from the fact that the entropy is an independently defined, and often computable, quantity.

The existence of entangled states is fairly obvious mathematically from the definition of a tensor product of Hilbert spaces. Nonetheless, it is worth stepping back to ask how it is possible physically for a state to be pure with respect to the whole system, yet mixed with respect to a subsystem. As we emphasized above, a classical state that is definite with respect to the whole system has definite values for \emph{all} observables, including all the observables on the subsystem; therefore, it must be definite with respect to the subsystem. The point where this logic breaks down in quantum mechanics is that, by the uncertainty principle, even in a pure state only some observables have definite values. In particular, the observables necessary to establish that the state is pure may not reside entirely in the subalgebra of $A$-observables. If this is true, then to an $A$-observer, the state is effectively mixed.

\begin{figure}[tbp]
\centering
\includegraphics[width=0.35\textwidth]{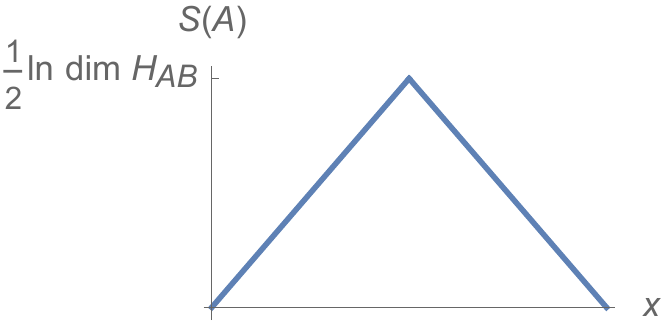}
\caption{\label{fig:Page}
Page curve: The EE $S(A)$ in a random pure state of a large Hilbert space $\HH_{AB}$ as a function of the fractional size $x$ of the $A$ subsystem, defined as $\ln\dim\HH_A/\ln\dim\HH_{AB}$.
}
\end{figure}

As emphasized above, most pure states on $AB$ are entangled. In fact, for \emph{large} systems, most states are essentially maximally entangled. It is clear from the definition that the von Neumann entropy cannot exceed the log of the dimension of the Hilbert space:
\begin{equation}\label{SAbound1}
S(A) \le \ln\dim\HH_A\,.
\end{equation}
Furthermore, if $AB$ is in a pure state, then $S(A)=S(B)$, so we also have
\begin{equation}\label{SAbound2}
S(A) \le \ln\dim\HH_B\,.
\end{equation}
It turns out that, in the limit of \emph{large} Hilbert spaces, a \emph{typical} (or random) pure state in $\HH_A\otimes\HH_B$ is as large as possible subject to the contraints \eqref{SAbound1}, \eqref{SAbound2}, in other words it approximately saturates either one or the other bound \cite{Page:1993wv}:
\begin{equation}\label{page}
S(A) = \min\{\ln\dim\HH_A,\ln\dim\HH_B\} + \mathcal{O}(1)\,.
\end{equation}
For example, if $AB$ is made up of many spins, with a fraction $x$ of them belonging to $A$ and the rest to $B$, then we obtain the curve shown in fig.\ \ref{fig:Page} for $S(A)$. This is called the \textbf{Page curve}. It originally arose as a model for the entropy of a black hole which is initally in a pure state. As it evaporates the black hole is entangled with the emitted Hawking photons. The black hole's entropy then follows the Page curve under the assumption that the dynamics governing the radiation process is essentially random, except for determining the amount of radiation---and therefore the sizes of the black hole and radiation's respective Hilbert spaces---as a function of time. For us, the Page curve will serve as a useful benchmark when we consider entropies in field theories, as these also have large Hilbert spaces.

\subsection{Purification and thermofield double}
\label{sec:purification}

Now suppose we have a single system $A$, and a mixed state $\rho_A$ on it. Whatever the physical origin of $\rho_A$, we can always mathematically construct a larger system $AB$ and a pure state $\ket{\psi}$ on it, such that $\psi_A=\rho_A$. This is called \textbf{purification}, and it is accomplished by reverse-engineering the Schmidt decomposition. Write $\rho_A$ in diagonal form,
\begin{equation}
\rho_A = \sum_a p_a\ket{a}_A\bra{a}
\end{equation}
(where $p_a\neq0$). Let $\HH_B$ be a Hilbert space whose dimension is at least the rank of $\rho_A$ (for example, it can be a copy of $\HH_A$), and let $\{\ket{a}_B\}$ be an orthonormal set of vectors. Then
\begin{equation}
\ket{\psi} := \sum_a\sqrt{p_a}\ket{a}_A\otimes\ket{a}_B
\end{equation}
is the desired purification.

Purification shows that mixedness of a state, whatever its physical origin, is indistinguishable from that arising from entanglement. Therefore in principle ``entanglement'' and ``entropy'' are interchangeable concepts (arguably making the phrase ``entanglement entropy'' redundant). Thinking of entropy as resulting from entanglement marks a radical change from the classical conception of entropy as fundamentally a measure of ignorance. In classical mechanics, you can never ``unmix'' a system by adjoining another one.

Purification is also a handy technical device. For example, by purifying the $AB$ system, the subadditivity inequality $S(AB)\le S(A)+S(B)$ is seen to imply the Araki-Lieb inequality \eqref{arakilieb} and vice versa; similarly, the two forms of strong subadditivity \eqref{SSA1}, \eqref{SSA2} are equivalent.

As an example of purification, consider the Gibbs state
\begin{equation}\label{Gibbs}
\rho_A = \frac1Ze^{-\beta H_A} = \frac1Z\sum_ae^{-\beta E_a}\ket{a}_A\bra{a}\,,
\end{equation}
where $\ket{a}_A$ are the energy eigenstates. Then we can choose $\HH_B$ to be a copy of $\HH_A$; the purification is
\begin{equation}\label{TFD}
\ket{\psi} = \frac1{\sqrt Z}\sum_ae^{-\beta E_a/2}\ket{a}_A\otimes\ket{a}_B\,.
\end{equation}
This is called the \textbf{thermofield double} state.

The thermofield double state can be represented in terms of a Euclidean path integral as a semicircle of length $\beta/2$. It is worth going over this in detail, as it provides good practice for the path-integral manipulations we will do in field theories. We start by noting that the operator $e^{-\beta H_A}$ can be represented by a Euclidean path integral on an interval of length $\beta$; concretely, its matrix element between position eigenstates $\ket{x_0}_A$ and $\ket{x_1}_A$ is given by
\begin{equation}\label{Gibbs PI}
_A\ev{x_0|e^{-\beta H_A}|x_1}_A =
\raisebox{-2\baselineskip}{
\includegraphics[width=0.16\textwidth]{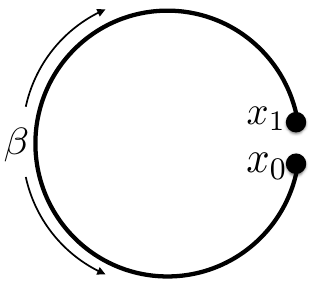}
}
\,,
\end{equation}
where we've chosen to draw the interval as an open circle, and the path integral is evaluated with boundary conditions $x=x_0$ and $x=x_1$ respectively on the two endpoints. The partition function is the trace of this operator:
\begin{equation}
Z = \Tr e^{-\beta H_A} = \int dx_0\,{}_A\ev{x_0|e^{-\beta H_A}|x_0}_A = \int dx_0
\raisebox{-2\baselineskip}{
\includegraphics[width=0.15\textwidth]{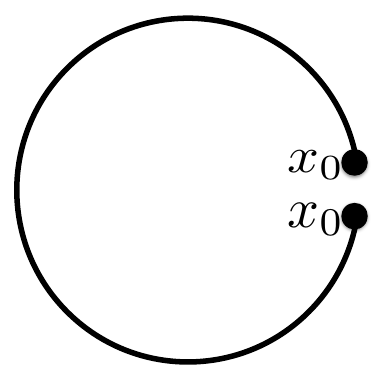}
} = 
\raisebox{-2\baselineskip}{
\includegraphics[width=0.15\textwidth]{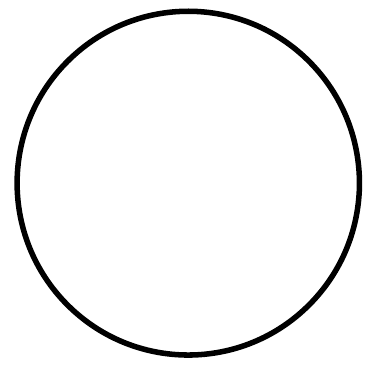}
}\,;
\end{equation}
integrating over the boundary conditions ``sews together'' the endpoints, leaving a circle of circumference $\beta$. The thermofield double state \eqref{TFD} is then given in terms of a path integral as a semicircle of length $\beta/2$, meaning that its components in the position basis $\ket{x_0}_A\otimes\ket{x_2}_B$ are given as follows:
\begin{equation}
\left({}_A\bra{x_0}\otimes{}_B\bra{x_2}\right)\ket{\psi} = \frac1{\sqrt Z}\times
\raisebox{-1.5\baselineskip}{
\includegraphics[width=0.14\textwidth]{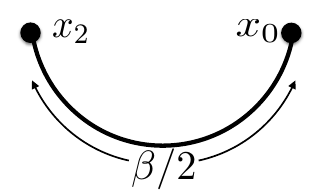}
}\,.
\end{equation}
To check that this state $\ket{\psi}$ reproduces the Gibbs state on $A$ when traced over $B$, we take the outer product with the Hermitian conjugate $\bra{\psi}$, represented by an upper semicircle, and sew together the two $B$ ends:
\begin{align}
_A\ev{x_0|\psi_A|x_1}_A &= {}_A\bra{x_0}\left(\Tr_B\ket{\psi}\bra{\psi}\right)
\ket{x_1}_A \nonumber \\
&=\int dx_2\left({}_A\bra{x_0}\otimes{}_B\bra{x_2}\right)\ket{\psi} \bra{\psi}\left(\ket{x_1}_A\otimes\ket{x_2}_B\right) \nonumber \\
&= \frac1Z\int dx_2
\raisebox{-1.8\baselineskip}{
\includegraphics[width=0.15\textwidth]{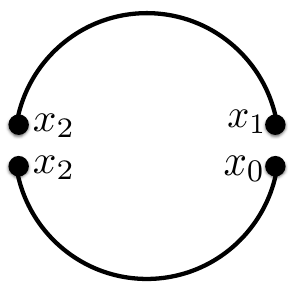}
} \\
&=\frac1Z\times
\raisebox{-1.8\baselineskip}{
\includegraphics[width=0.15\textwidth]{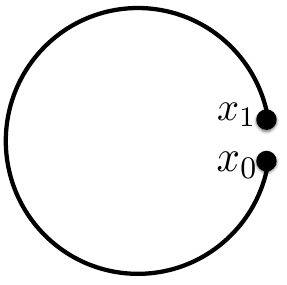}
}\nonumber  \\
&={}_A\bra{x_0}\rho_A\ket{x_1}_A\,.\nonumber 
\end{align}

\subsection{Entanglement in mixed states}
\label{sec:entangled}

The discussion in subsection \ref{sec:entanglement} assumed that the joint system $AB$ is in a pure state and explained that $S(A)$ is a measure of the amount of entanglement between $A$ and $B$. It is clear that this interpretation is lost once we allow $AB$ to be in a mixed state. For example, $S(A)$ may be non-zero even in a product state $\rho=\rho_A\otimes\rho_B$. (Consider, for example, the Gibbs state of two non-interacting systems.) This is why (as mentioned in subsection \ref{sec:entanglement}) the phrase ``entanglement entropy'' is misleading.

Can we nonetheless define what it means for $A$ and $B$ to be entangled, and quantify the amount of entanglement, when $AB$ is in a mixed state? First, we define a \textbf{separable}, or \textbf{classically correlated}, state as a convex mixture of product states:
\begin{equation}\label{separable}
\rho = \sum_ip_i\rho_A^i\otimes\rho_B^i\,,
\end{equation}
where $\{p_i\}$ is a probability and $\rho_A^i$, $\rho_B^i$ are arbitrary states. This is precisely the class of states that can be obtained by LOCC starting from a product state. A state that is not separable is then considered \emph{entangled}. It is easy to see that this definition reduces to the previous one in the case that $\rho$ is pure.

As noted in subsection \ref{sec:vne}, the mutual information $I(A:B)$ quantifies the amount of correlation. This includes both classical correlation and entanglement. For example, a maximally classically correlated pair of qubits
\begin{equation}\label{correlated}
\rho = \frac12\left(\ket{00}\bra{00}+\ket{11}\bra{11}\right)
\end{equation}
has $I(A:B)=\ln2$, while a Bell pair has $I(A:B)=2\ln 2$ (thereby confirming the intuition that entanglement is a stronger form of correlation than classical correlation). However, it does not distinguish between the two forms of correlation; for example, two classically correlated pairs of qubits have the same mutual information as a single Bell pair.

How then can we tell whether a given state is separable or entangled? This turns out to be a very hard (in fact, NP-hard) problem. Here we will discuss two tests for entanglement that are useful in the field-theory context. First, we saw that, for classical probability distributions,
\begin{equation}\label{poscondent}
H(A|B)\ge0\,.
\end{equation}
In fact, \eqref{poscondent} holds generally for separable states. However, it does not always hold in the presence of entanglement; for example, for an entangled pure state,
\begin{equation}
H(A|B) = -S(B)<0\,.
\end{equation}
Therefore, negativity of the conditional entropy can be used to test for the presence of entanglement. However, the test has ``false negatives''; it's easy to construct entangled states with positive conditional entropy. 

Another calculable test for entanglement involves the partial transpose \cite{PhysRevLett.77.1413,HORODECKI19961}. This is based on the fact that the transpose of a density matrix is also a density matrix. (The transpose depends on the basis, but in any basis it is a density matrix.) Suppose we define $\tilde\rho$ to be the operator obtained by transposing $\rho$ only on the $B$ indices:
\begin{equation}\label{ptdef}
(\tilde\rho)_{ab,a'b'}:=(\rho)_{ab',a'b}\,.
\end{equation}
If $\rho$ is separable, \eqref{separable}, we have
\begin{equation}
\tilde\rho = \sum_ip_i\rho_A^i\otimes\rho_B^{iT}\,.
\end{equation}
For all $i$, $\rho_B^{iT}$ is still a density matrix, so $\tilde\rho$ is a density matrix as well. On the other hand, if $\rho$ is entangled, then nothing guarantees that $\tilde\rho$ is a density matrix: while $\tilde\rho$ is Hermitian and has unit trace, it may not be a positive operator. Therefore the presence of negative eigenvalues of $\tilde\rho$ establishes that $\rho$ is entangled. (While $\tilde\rho$ depends on the choice of basis used in \eqref{ptdef}, its spectrum does not.) This is called the \textbf{positive partial transpose} criterion. In particular, if $\tilde\rho$ has any negative eigenvalues, then $\Tr|\tilde\rho|>\Tr\tilde\rho=1$; hence the \textbf{logarithmic negativity} (or \textbf{entanglement negativity}) $\ln\Tr|\tilde\rho|$ detects entanglement \cite{PhysRevA.65.032314}. However, like the conditional entropy, it is not reliable detector, in the sense that an entangled state can have vanishing logarithmic negativity. These are the only two mixed-state entanglement detectors that have been calculated in field theories.

Quantifying the \emph{amount} of entanglement in a given mixed state $\rho$ is similarly hard, and there exists a large body of theory around this problem. Unlike in the pure-state case, there is no single, canonical measure of entanglement; rather, many different measures of entanglement have been defined, which are useful in different operational and theoretical senses. These include:
\begin{itemize}
\item the \emph{entanglement of formation}, the number of Bell pairs required to create $\rho$ by LOCC; 
\item the \emph{distillable entanglement}, the number of Bell pairs that can be created out of $\rho$ by LOCC; 
\item the \emph{squashed entanglement}, the minimum of $I(A:B|C)_\sigma$ over extensions $\HH_{ABC}$ of $\HH_{AB}$ and states $\sigma$ on $\HH_{ABC}$ such that $\Tr_C\sigma=\rho$, where 
\begin{equation}
I(A:B|C):=S(AC)+S(BC)-S(C)-S(ABC)
\end{equation}
is the \emph{conditional mutual information}; 
\item the \emph{relative entropy of entanglement}, the minimum of $S(\rho||\sigma)$ over separable states $\sigma$ on $\HH_{AB}$, where
\begin{equation}
S(\rho||\sigma):=\Tr(\rho\ln\rho-\rho\ln\sigma)
\end{equation}
is the \emph{relative entropy}; 
\item and the \emph{logarithmic negativity}, defined above.
\end{itemize}
See \cite{Plenio:2007zz} for an overview and references. Again, except for the logarithmic negativity, they are hardly ever computable in practice, so we will not dwell on them.

\subsection{Entanglement, decoherence, and monogamy}

Although it is sometimes depicted as an exotic phenomenon, entanglement is the rule rather than the exception. Mathematically, given a bipartite system, factorized states are a measure-zero subset of the pure states, and separable states are measure-zero subset of the mixed states. Physically, interactions between two systems nearly always lead to entanglement.

One of the most important physical implications of this is that interactions with the environment cause \textbf{decoherence} of a system's state. Consider, for example, a spin-1/2 system $A$ interacting with some environmental degree of freedom $B$. Label the initial state of $B$ $\ket{0}_B$, and suppose the interaction causes $B$ to change to $\ket{1}_B$ if $A$'s spin is down:
\begin{equation}
\ket{\uparrow}_A\otimes\ket{0}_B \to\ket{\uparrow}_A\otimes\ket{0}_B \,,\qquad
\ket{\downarrow}_A\otimes\ket{0}_B \to\ket{\downarrow}_A\otimes\ket{1}_B \,.
\end{equation}
At first it would appear that the state of $A$ is not affected. Suppose, however, that $A$ starts in an arbitrary pure state:
\begin{equation}
\alpha\ket{\uparrow}_A+\beta\ket{\downarrow}_A\,.
\end{equation}
After interaction with $B$, this becomes
\begin{equation}
\alpha\ket{\uparrow}_A\otimes\ket{0}_B+\beta\ket{\downarrow}_A\otimes\ket{1}_B\,.
\end{equation}
If the environmental degree of freedom is inaccessible, then $\rho_A$ has undergone the following transformation (in the $\ket{\uparrow}_A,\ket{\downarrow}_A$ basis):
\begin{equation}
\begin{pmatrix}|\alpha|^2 & \alpha^*\beta \\ \alpha\beta^* & |\beta|^2 \end{pmatrix} \to
\begin{pmatrix}|\alpha|^2 & 0 \\ 0 & |\beta|^2 \end{pmatrix}\,.
\end{equation}
Assuming neither $\alpha$ nor $\beta$ is zero, the state has become mixed! In general, entanglement with inaccessible environmental degrees of freedom causes the off-diagonal matrix elements of $\rho_A$ (in a basis determined by the interactions) to decrease in magnitude, which increases $S(A)$.\footnote{It is important to note that entanglement alone does not produce decoherence; rather, the environmental degrees of freedom need to be both inaccessible and physically independent. Slave degrees of freedom, such as the fast modes in a system with a separation of energy or time scales (such as the electrons in the Born-Oppenheimer approximation, or short-wavelength field-theory modes that have been integrated out in a Wilsonian scheme), do not decohere the slow modes (the nuclei, or the modes kept), although the fast and slow modes are entangled. For example, the quarks may be inaccessible to a low-energy nuclear physicist, but they do not decohere the state of the nucleons. For this reason, it is not generally useful to think of the slow modes as being in a mixed state, even though they are entangled with the fast modes.}

Environmental decoherence is the basic reason it's harder to build a quantum than a classical computer: not only must the qubits be protected from being disturbed by the environment (as the bits in a classical computer must be), but they must also be protected from \emph{influencing} the environment, even microscopic environmental degrees of freedom (stray photons, etc.). The latter is much harder, particularly if one nevertheless wants to be able to manipulate and measure those qubits.

It has also been argued that environmental decoherence explains why the world appears to be classical (i.e.\ to lack superpositions) to macroscopic observers such as ourselves, and that decoherence due to interactions between a system and measurement apparatus is behind the apparent collapse of the wave function upon measurement \cite{Zurek}.

Paradoxically, while entanglement causes decoherence, decoherence destroys entanglement. Consider, for example, the following pure state on three qubits, called the \textbf{GHZ state}:
\begin{equation}\label{GHZ}
\ket{\text{GHZ}} = \frac1{\sqrt2}\left(\ket{000}+\ket{111}\right).
\end{equation}
Are $A$ and $B$ entangled? At first sight, it may look like they are. But, as we emphasized above, this is really a question about $\rho_{AB}$, which is given by the separable state \eqref{correlated}. So, while $A$ is entangled with $BC$, it is actually \emph{not} entangled with $B$, due to decoherence by $C$. Thus, although ubiquitous, entanglement is also rather fragile if subject to environmental decoherence.

More generally, entanglement within a system excludes entanglement with other systems: if $A$ and $B$ are in an entangled pure state, then they cannot be entangled---or even classically correlated---with $C$. A more general statement can be made if we use a negative conditional entropy as a diagnostic of entanglement. The second form of SSA can be written as follows:
\begin{equation}
H(A|B)+H(C|B)\ge0\,,
\end{equation}
showing that at most one of the pairs---$AB$ or $BC$---can have a negative conditional entropy. This exclusivity property is called \textbf{monogamy of entanglement}. It is quite different from the classical case, where nothing stops $A$ from being simultaneously correlated with $B$ and with $BC$ (in fact, if $A$ is correlated with $B$ then it \emph{must} be correlated with $BC$). Monogamy of entanglement plays an important role in many aspects of quantum information theory. For example, in quantum cryptography one establishes entanglement between $A$ and $B$; monogamy then guarantees that any eavesdropping---which necessarily requires creating some correlation between the eavesdropper and $AB$---destroys the entanglement and is therefore detectable. Monogamy also appears in many ideas concerning the black-hole information paradox, such as the so-called firewall argument \cite{Mathur:2009hf,Almheiri:2012rt}.

\subsection{Subsystems, factorization, and time}
\label{sec:factorization and time}

We noted above that interactions between two systems nearly always lead them to become entangled. Specifically, for a generic Hamiltonian $H_{AB}$ on a joint system coupling $A$ and $B$, if at time $t_0$ the system is in a pure, unentangled state, then at another time $t_1$ they will---absent some fine-tuning---be entangled, just because almost all pure states in the joint Hilbert space $\HH_{AB}$ are entangled. In the Schr\"odinger picture, it is clear that the state evolves from an unentangled to an entangled one, and more generally the EE $S(A)$ evolves with time. In the Heisenberg picture (which we will be using in the next section, when we discuss field theories), however, this is confusing: the state doesn't evolve, so how can it become entangled? The answer is that the \emph{factorization} of the Hilbert space into $\HH_A\otimes\HH_B$ evolves! More precisely, instead of \eqref{AB factorization}, we really should have written
\begin{equation}
\HH_{AB}\cong\HH_A\otimes\HH_B\,,
\end{equation}
where $\cong$ means isomorphic. Normally we wouldn't bother with such a fine point, but in the Heisenberg picture the isomorphism is \emph{time-dependent}. Specifically, as one can see from the usual relation between the Schr\"odinger and Heisenberg pictures, the unitary map $U:\HH_A\otimes\HH_B\to\HH_{AB}$ evolves according to
\begin{equation}
U^\dag\frac{dU}{dt} = iH_{AB}\,.
\end{equation}
Whether a given state $\ket{\psi}_{AB}$ in $\HH_{AB}$ is isomorphic to a tensor product $\ket{\psi'}_A\otimes\ket{\psi''}_B\in\HH_A\otimes\HH_B$ depends on the isomorphism and therefore on the time. Thus, in the Heisenberg picture, when factorizing a Hilbert space into $\HH_A\otimes\HH_B$ and computing $\rho_A$, $S(A)$, etc., we need to specify not only the subsystems $A$ and $B$ but also the \emph{time} at which we are looking at them.

\section{Entropy, entanglement, and fields}
\label{sec:EEfields}

In this section, our aim will be to understand a few essential facts about entanglement entropies (EEs) in \emph{relativistic quantum field theories}. In the first two subsections, we consider field theories in any dimension, but when once we consider concrete examples starting in subsection \ref{sec:Rindler}, we will mostly restrict ourselves to $(1+1)$-dimensional field theores, with a very brief discussion of higher dimensions in subsection \ref{sec:higherdim}. This will suffice for the limited selection of topics we'll cover, although obviously many interesting new phenomena arise in higher dimensions. The focus here will mostly be on the physics contained in EEs, rather than on methods for calculating them, which is a broad and interesting but somewhat technically involved topic. Thus in many cases we will simply cite a result or sketch a derivation without fully explaining how it is obtained.

\subsection{What is a subsystem?}
\label{sec:subsystem}

\begin{figure}[tbp]
\centering
\includegraphics[width=0.45\textwidth]{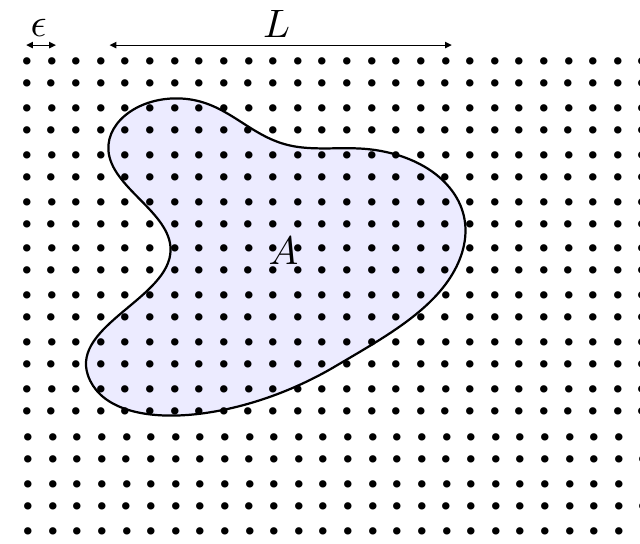}
\caption{\label{fig:lattice}
A region $A$ of size $L$ in a lattice system with lattice spacing $\epsilon$.
}
\end{figure}

What makes a field theory a field theory is the existence of \emph{space}, that is, the observables depend not only on time but also on a position in some (fixed) spatial manifold. In order to gain intuition, let us start by considering a discretized space, i.e.\ a spatial lattice such as a spin chain; many interesting field theories can be obtained by taking a continuum limit of such a lattice system. At each site $x$ there is a local Hilbert space $\HH_x$, and the full Hilbert space is their tensor product over all the sites:
\begin{equation}
\HH = \bigotimes_x\HH_x\,.
\end{equation}
Therefore associated to any decomposition of the lattice into a region $A$ and its complement $A^c$ (see fig.\ \ref{fig:lattice}), there is a corresponding factorization of the Hilbert space:
\begin{equation}\label{latticefact}
\HH = \HH_A\otimes\HH_{A^c}\,,\qquad
\HH_A := \bigotimes_{x\in A}\HH_x\,,\qquad
\HH_{A^c} := \bigotimes_{x\in A^c}\HH_x\,.
\end{equation}
While this is not the only way to factorize the Hilbert space, given the spatial structure it is certainly the most obvious and natural one.

Based on such the factorization \eqref{latticefact}, and given any particular state $\rho$, we can define the EE $S(A)$ and related quantities such as R\'enyi entropies, mutual information, and so on, as in the previous sections. Let us attempt to make a crude estimate of the entropy $S(A)$. Since we are not too interested in the physics at the scale of the lattice spacing $\epsilon$, we take $A$ to be a region of size $L\gg\epsilon$ (and suppose the full system is much larger still). $S(A)$ could be as large as the log of the dimension of $\HH_A$, which is the number $N_A$ of lattice sites in $A$ times the log of the dimension of the local Hilbert space $\HH_i$; in fact, according to the Page formula \eqref{page}, if we simply pick a random state in the full Hilbert space, it will come close to saturating this bound:
\begin{equation}\label{volumelaw}
S(A) \propto N_A \sim \frac{L^{D-1}}{\epsilon^{D-1}}\,,
\end{equation}
where $D-1$ is the number of spatial dimensions. This is called a \textbf{volume-law} growth of entanglement.

However, we are usually not interested in \emph{random} states, but rather those that arise physically, such as the ground states and low-lying excited states of physically interesting Hamiltonians. Such Hamiltonians respect the spatial structure of the lattice, i.e.\ they preserve some notion of locality, for example by containing interactions only between nearest neighbors. This locality will manifest itself in the entanglement structure of the ground state (and low-lying excited states). We might guess that most of the entanglement is short-ranged, roughly speaking between nearby lattice sites. In that case, the entropy will be proportional to the number of bonds between neighboring sites cut by the boundary between $A$ and $A^c$, and will therefore grow only like the area of this boundary:
\begin{equation}\label{arealaw}
S(A) \sim \frac{L^{D-2}}{\epsilon^{D-2}}\,.
\end{equation}
The remarkable thing about such an \textbf{area-law} growth is how small it is compared to the volume-law growth of a random state. Thus, in some sense, we expect physical states to contain very \emph{little} entanglement. It turns out that the estimate \eqref{arealaw} is correct for many lattice systems (and can even be proven rigorously in some cases). As we will see, there are exceptions---for example, gapless systems in $D=2$---in which the ground state has enough long-range entanglement to add a logarithmically growing factor to \eqref{arealaw}; however, the point remains that such states have very \emph{little} entanglement compared to the naive expectation \eqref{volumelaw}.\footnote{An interesting case with a volume-law growth of entropy, which we will study below, is a thermal state. However, the entropy density depends on the temperature. If we are viewing the lattice system as a cut-off version of a field theory, then we would take the temperature far below the cutoff, so again $S(A)$ is far less than that of a random state \eqref{volumelaw}.}

If we consider a quantum field theory to be a continuum limit $\epsilon\to0$ of such a lattice system, then it seems natural to consider our subset $A$ again to be a spatial region, and to assume that again we have a factorization of the Hilbert space,
\begin{equation}\label{fact}
\HH = \HH_A\otimes\HH_{A^c}\,.
\end{equation}
More careful consideration shows that \eqref{fact} does not \emph{quite} hold, due to subtleties at the boundary $\partial A$ of $A$, which is called the \textbf{entangling surface}. There is some interesting physics associated with these subtleties, but nothing that affects what we will say in the rest of these lectures. Suffice it to say that, with a bit of care, one can nonetheless define the reduced density matrix $\rho_A$ and EE $S(A)$ even when \eqref{fact} fails. The essential principle allowing us to consider $A$ and $A^c$ to have independent degrees of freedom (again, up to subtleties at the shared entangling surface), and which therefore makes \eqref{fact} morally correct, is that operators at the same time and different points commute, a principle that holds in any (local) quantum field theory.

Another point that we can anticipate from the lattice discussion is that, based on the estimate \eqref{arealaw}, we should expect EEs in field theories to be \emph{ultraviolet divergent} (at least in $D>2$, although they often turn out to be divergent in $D=2$ as well).

\subsection{Subsystems in relativistic field theories}
\label{sec:subsystem relativistic}

The notion of a subsystem as a spatial region can be further refined in a \emph{relativistic} quantum field theory. In such a theory, we have not just a spatial manifold but a \emph{spacetime} manifold $M$ equipped with a causal structure. Several notions implied by the causal structure then come into play. The \textbf{causal domain} $D(S)$ of set $S\subseteq M$ is the set of points $p\in M$ such that every inextendible causal curve through $p$ intersects $S$. A set $S$ is \textbf{acausal} if no two distinct points in $S$ are causally related (i.e.\ on the same causal curve). A \textbf{Cauchy slice} $\Sigma$ is an acausal set whose domain is the entire manifold, $D(\Sigma) = M$; indeed, such a set is necessarily a slice.\footnote{We use \emph{slice} to denote a spacelike codimension-one submanifold, and \emph{surface} to denote a spacelike submanifold which is codimension two in spacetime or codimension one in space.} Finally, if there exists a Cauchy slice then $M$ is \textbf{globally hyperbolic}; this is required for the consistent definition of a field theory on the spacetime. An example of a Cauchy slice for Minkowski space is a flat spacelike hyperplane (a constant-time slice in a Cartesian coordinate system). However, there are many others; in fact, any acausal hypersurface that extends (in every spatial direction) to the point $i^0$ at spatial infinity is a Cauchy slice. In general, globally hyperbolic manifolds admit many different Cauchy slices.

The Heisenberg equations of motion which evolve observables in time must respect the causal structure of $M$. These can be used to evolve any observable onto a given Cauchy slice $\Sigma$. The observables on $\Sigma$ must therefore be complete, in the sense that any observable can, using the equations of motion, be written in terms of those on $\Sigma$. Furthermore, an important axiom of relativistic quantum field theory states that spacelike-separated observables commute. Therefore, $\Sigma$ plays the role that the spatial manifold played in the previous subsection, when we were finding a natural way to define a ``subsystem''. We thus expect a region $A$ of $\Sigma$ to be associated with a factorization of the Hilbert space of the form \eqref{fact} (again, up to subtleties related to the entangling surface).

\begin{figure}[tbp]
\centering
\includegraphics[width=0.55\textwidth]{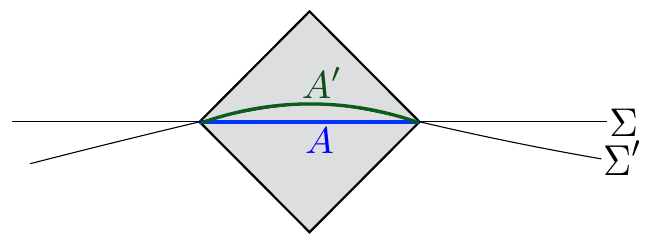}
\caption{\label{fig:causaldomain}
A region $A$ of a Cauchy slice $\Sigma$ and a region $A'$ of another Cauchy slice $\Sigma'$ with the same causal domain (shown in gray), $D(A')=D(A)$.
}
\end{figure}

In fact, we can go further: Given a region $A$ of $\Sigma$, any observable in the causal domain $D(A)$ can be written (again, using the equations of motion) in terms of those in $A$. Therefore the expectation value of any observable in $D(A)$ is determined by those in $A$ and therefore in turn by the reduced density matrix $\rho_A$. Hence we should really associate $\HH_A$ and $\rho_A$, not with $A$ itself, but with $D(A)$. To put it another way, given a region $A'$ of a different Cauchy slice $\Sigma'$ such that $D(A')=D(A)$ (see fig.\ \ref{fig:causaldomain}), we have $\HH_{A'}=\HH_A$, $\rho_{A'} = \rho_A$, and therefore $S(A')=S(A)$.\footnote{In the Schr\"odinger picture, states on difference Cauchy slices are related by unitary maps, and if $D(A')=D(A)$ then the map acts separately on $A$ and $A^c:=\Sigma\setminus A$, so $\rho_{A'}$ and $\rho_A$ are unitarily related. The conclusion that $S(A')=S(A)$ therefore still holds.}

On the other hand, two regions $A,A'$ that do not share the same causal domain, $D(A')\neq D(A)$, should not be thought of as the same ``subsystem'': $\HH_{A
}\neq\HH_{A}$. This is true even if, in some coordinate system, they occupy the same part of space at different times. This accords with the fact emphasized in subsection \ref{sec:factorization and time} that (in the Heisenberg picture) the factorization of the Hilbert space is itself time-dependent.

The two features we emphasized above---the flexibility in the choice of Cauchy slice and the fact that $\rho_A$ depends only on $D(A)$---have no analogues in non-relativistic theories. As we will see, they imply powerful constrains on EEs and related quantities.

\subsection{General theory: half-line}
\label{sec:Rindler}

\begin{figure}[tbp]
\centering
\includegraphics[width=0.35\textwidth]{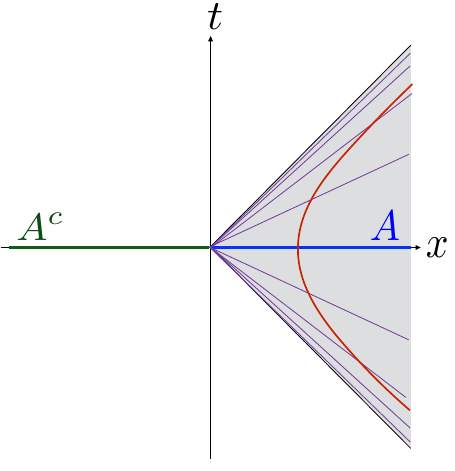}
\caption{\label{fig:Rindler}
The half-line $A$ \eqref{half-line}, its complement $A^c$, and its causal domain $D(A)$, the Rindler wedge \eqref{Rindler} (shaded), in $(1+1)$-dimensional Minkowski space. Also shown are some lines of constant $\chi$ in the coordinates $(\chi,r)$ of \eqref{Rindler metric} (purple), each of which is a Cauchy slice for the Rindler wedge, as well as a trajectory of constant $r$ (red).
}
\end{figure}

For the rest of this section, we will restrict ourselves to field theories in $1+1$ dimensions. We start with a field theory in two-dimensional Minkowski space. A simple example of a Cauchy slice is the line $t=0$ (in standard coordinates), and a simple example of a region is the half-line:
\begin{equation}\label{half-line}
A = \{t=0,x\ge0\}\,.
\end{equation}
(See fig.\ \ref{fig:Rindler}.) Its causal domain is called the \textbf{Rindler wedge}, or---considered as a globally hyperbolic spacetime in its own right---$(1+1)$-dimensional \textbf{Rindler space}:
\begin{equation}\label{Rindler}
D(A) = \{|t|\le x\}\,.
\end{equation}
This causal domain preserves the boost subgroup of the Poincar\'e group. The entangling surface is just the origin. 
For the state we choose the vacuum, which is also boost-invariant:
\begin{equation}\label{vacuum}
\rho = \ket{0}\bra{0}\,.
\end{equation}
We will see that this state is highly entangled. Therefore, for an observer who spends her whole life inside the Rindler wedge, the world appears to be in a mixed state.

\subsubsection{Reduced density matrix}
\label{sec:reduced}

The boost symmetry will allow us to solve this example, in the sense of writing down a closed-form expression for $\rho_A$ in terms of the stress tensor. This will allow us to gain a lot of intuition. Also, the derivation illustrates a very useful technique using Euclidean path integrals.

We formally write the full set of fields as $\phi$, and work in a ``position'' basis $\ket{\phi_0}$, where $\phi_0(x)$ is a field configuration on a fixed time slice. The matrix elements of $\rho$ are given as the $\beta\to\infty$ limit of the matrix elements of the Gibbs state $e^{-\beta H}/Z$. The latter are calculated as in \eqref{Gibbs PI} by a Euclidean path integral on a strip of height $\beta$ (in the Euclidean time direction), which we draw as a cylinder cut along the $\tau=0$ axis, with boundary conditions dictated by the bra and ket:
\begin{equation}\label{finite T}
\ev{\phi_0|e^{-\beta H}|\phi_1} =
\raisebox{-2.3\baselineskip}{
\includegraphics[width=0.4\textwidth]{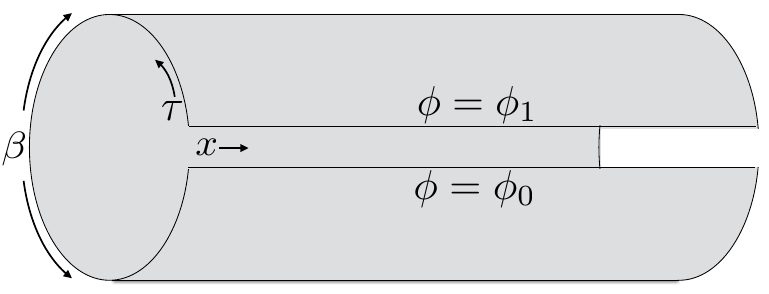}
}\,.
\end{equation}
Now take the limit $\beta\to\infty$. What is left is the $(\tau,x)$ plane cut open along $\tau=0$:
\begin{equation}\label{rho matrix}
\ev{\phi_0|0}\ev{0|\phi_1}=
\ev{\phi_0|\rho|\phi_1} =\frac1Z\times
\raisebox{-3.25\baselineskip}{
\includegraphics[width=0.35\textwidth]{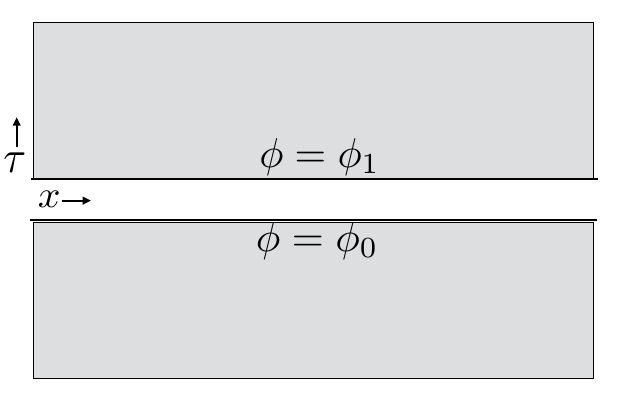}
}\,.
\end{equation}
As expected, this factorizes, with the path integral on the lower half-plane giving $\ev{\phi_0|0}$:
\begin{equation}\label{vacuum wf}
\ev{\phi_0|0} = \frac1{\sqrt Z}\times\raisebox{-1.5\baselineskip}{
\includegraphics[width=0.35\textwidth]{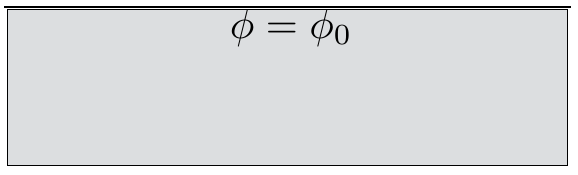}
}\,.
\end{equation}

To calculate $\rho_A$, we trace $\rho$ over $\HH_{A^c}$. This amounts to writing
\begin{equation}
\ket{\phi_0}=\ket{\phi_0^{A^c}}_{A^c}\otimes\ket{\phi_0^A}_A\,,
\end{equation}
where $\phi_0^{A^c}$ and $\phi_0^A$ represent the function $\phi_0(x)$ restricted to $x<0$ and $x\ge0$ respectively, and summing the matrix element \eqref{rho matrix} over $\phi_0^{A^c}$ (henceforth to avoid cluttering the notation we leave the subscripts on the bras and kets implicit, for example writing $\ket{\phi_0^A}$ for $\ket{\phi_0^A}_A$):
\begin{equation}
\ev{\phi_0^A|\rho_A|\phi_1^A} =\ev{\phi_0^A|\left(\Tr_{A^c}\rho\right)|\phi_1^A} = \int D\phi_0^{A^c}
\left(\bra{\phi_0^{A^c}}\otimes\bra{\phi_0^A}\right)
\rho\left(\ket{\phi_0^{A^c}}\otimes\ket{\phi_0^A}\right).
\end{equation}
The sum over the field value on $A^c$ ``glues'' the top and bottom sheets together along $A^c$, leaving a path integral on the plane cut only along $A$, i.e.\ the half-line $\{\tau=0,x\ge 0\}$, with boundary conditions given by the field values appearing in the bra and ket:
\begin{equation}\label{cut plane A}
\ev{\phi_0^A|\rho_A|\phi_1^A} =\frac1Z\times
\raisebox{-3.25\baselineskip}{
\includegraphics[width=0.35\textwidth]{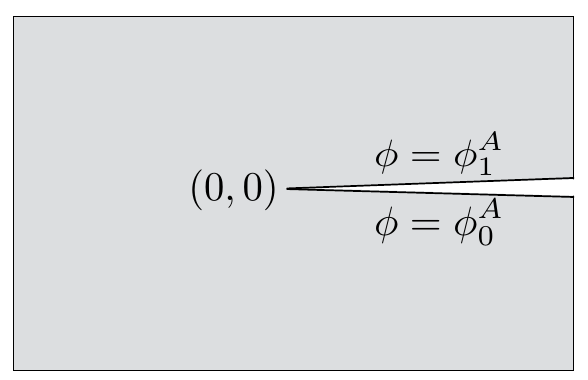}
}\,.
\end{equation}

Looking at the right-hand side of \eqref{cut plane A}, it makes sense to change from Cartesian coordinates $(\tau,x)$ to polar coordinates $(\theta,r)$, since then the cut plane can be described as $r\in[0,\infty)$, $\theta\in[0,2\pi]$. Normally, in polar coordinates, for a given $r$ the two points $(\theta=0,r)$ and $(\theta=2\pi,r)$ would be identified, but the cut along $\tau=0$, $x\ge0$ makes them distinct. In Euclidean space, we have some freedom in choosing which coordinate we consider the ``Euclidean time'' coordinate. Let us change our point of view and consider $\theta$, rather than $\tau$, to be the Euclidean time coordinate. From this point of view, the cut plane is simply a time interval of length $2\pi$ over the half-line $r\ge0$. The metric is
\begin{equation}
ds^2_{\rm E} = r^2d\theta^2+dr^2\,.
\end{equation}
If we Wick rotate with respect to $\theta$, writing
\begin{equation}
\theta = i\chi\,,
\end{equation}
the map to standard Minkowski coordinates becomes
\begin{equation}
x = r\cos\theta=r\cosh\chi\,,\qquad
t = -i\tau = r\sinh\chi\,.
\end{equation}
With $\chi$ running from $-\infty$ to $\infty$, the $(r,\chi)$ coordinates do not cover all of Minkowski space, but only the Rindler wedge. This is to be expected, since $\rho_A$ is the state of the field theory restricted to the Rindler wedge. The metric in these coordinates is
\begin{equation}\label{Rindler metric}
ds^2_{\rm L} = -r^2d\chi^2+dr^2\,.
\end{equation}

If we compare \eqref{cut plane A} to \eqref{Gibbs PI}, we see that $\rho_A$ is a Gibbs state, with $\beta=2\pi$ and the role of the Hamiltonian being played by the generator of $\chi$ translations. The latter is, in the Lorentzian picture, the generator $K$ of boosts:
\begin{equation}\label{A thermal}
\rho_A =\frac1Z e^{-2\pi K}\,.
\end{equation}
Thus the modular Hamiltonian is
\begin{equation}
H_A = 2\pi K\,.
\end{equation}
In general, the conserved quantity associated to a Killing vector $k^\mu$ is $\int_\Sigma\sqrt{h}k^\mu n^\nu T_{\mu\nu}$, where $\Sigma$ is a Cauchy slice, $h$ is the induced metric, and $n$ is the unit normal. The Killing vector for boosts is
\begin{equation}
\pd{}{\chi} = x\pd{}{t}+t\pd{}{x}\,,
\end{equation}
so, choosing $\chi=0$ as our Cauchy slice for the Rindler wedge, the boost generator is
\begin{equation}\label{boost}
K = \int_0^\infty dr\,\frac1rT_{\chi\chi}(r,\chi=0) = \int_0^\infty dx \,x\,T_{tt}(t=0,x)\,.
\end{equation}

The simple result \eqref{A thermal} can be derived by various methods aside from the path integral one we used here (which is due to Callan-Wilczek \cite{Callan:1994py}). For example, in a free theory it can be derived from the mode expansion of the fields \cite{PhysRevD.14.870}. It can even be proven rigorously within the framework of algebraic quantum field theory, a result known as the \textbf{Bisognano-Wichmann theorem} \cite{Bisognano:1976za}.

According to \eqref{TFD}, the thermofield double state is constructed by a path integral in which the Euclidean time runs over an interval $\beta/2$. Here this means that $\theta$ runs over an angle $\pi$, giving a half-plane. \eqref{vacuum wf} shows that this state is nothing but the vacuum. We thus recover the vacuum as the thermofield double of the Rindler state $\rho_A$, with $A^c$ being the purifying system.

\subsubsection{Entropy: estimate}
\label{sec:estimate}

Eq.\ \eqref{A thermal} contains a lot of physics. Recall that we are just talking about the vacuum of a field theory. Eq.\ \eqref{A thermal} tells us that, to an observer confined to the Rindler wedge, that state appears to be a thermal state with respect to the boost generator (its only continuous symmetry) at a temperature $T_\chi=1/(2\pi)$. An example of such an observer is one who follows a worldline of constant $r$ (see fig.\ \ref{fig:Rindler}). In the original, inertial coordinates $(t,x)$, this trajectory is
\begin{equation}
x(t) = \sqrt{r^2+t^2}\,,
\end{equation}
which is accelerating with constant proper acceleration $a=1/r$. To get the physical temperature experienced by such an observer, we have to multiply by the local redshift factor:
\begin{equation}\label{Unruh}
T_{\rm phys}(r) = (-g_{\chi\chi})^{-1/2}T_\chi = \frac1{2\pi r}=\frac a{2\pi}\,.
\end{equation}
This can also be understood from the fact that the physical inverse temperature $\beta_{\rm phys}$ is the proper length of the circle of constant $r$ in the Euclidean plane, which is $2\pi r$. The fact that an observer in flat space accelerating at a proper rate $a$ experiences a temperature $a/(2\pi)$ is called the \textbf{Unruh effect} \cite{PhysRevD.14.870}. We see that, close to the entangling surface (small $r$), the fields are very \emph{hot}. Roughly speaking, the observer sees the modes that have been decohered by tracing over $A^c$, and the closer she is to the entangling surface, the more UV modes are seen to be decohered. Note that, although the physical temperature is spatially varying, the system is in perfect thermal equilibrium; this is possible due to the spatially varying gravitational potential.

The fact that $\rho_A$ is a thermal state with respect to the boost generator means we can calculate $S(A)$ as a thermodynamic entropy. It is worth emphasizing that this EE is a real physical entropy, as seen by an observer confined to the Rindler wedge, not just a mathematical abstraction.

We will use \eqref{Unruh} to estimate $S(A)$ by adding up local thermal entropies. This is just an estimate, since the thermal entropy density $s(T)$ is defined in flat space at \emph{constant} temperature, whereas here the temperature is spatially varying. Nonetheless we can get useful intuition from this estimate. For a field of mass $m$, the entropy density $s(T)$ essentially vanishes at temperatures below $m$, as the field is frozen out. Thus, much past a distance $r=\xi$, where $\xi:=1/m$ is the \textbf{correlation} (or \textbf{Compton}) \textbf{length}, the field does not contribute significantly to the entropy. In other words, the field is entangled only within a neighborhood $r\lesssim\xi$ of the entangling surface. This very intuitive fact holds in any dimension and for any entangling surface.

On the other hand, for temperatures well above $m$, by dimensional analysis $s(T)\propto T$. Thus for $r\ll\xi$, the entropy density is diverging like $1/r$. The integral of $1/r$ diverges, so we need to impose a UV cutoff at $r=\epsilon$. We thus estimate $S(A)$ as follows:
\begin{equation}\label{S(A) estimate1}
S(A) \approx \int_\epsilon^\infty dr\,s(T_{\rm phys}(r)) \propto\int_\epsilon^\xi dr\,\frac 1r=\ln\frac\xi\epsilon\,.
\end{equation}
We can be more precise by assuming that the theory has a UV fixed point which is a CFT. The entropy density of a CFT with central charge $c$ is
\begin{equation}
s(T) = \frac{2\pi c}6T\,,
\end{equation}
so we get
\begin{equation}\label{S(A) estimate}
S(A) \approx \int_\epsilon^\infty dr\,s(T_{\rm phys}(r)) \approx\int_\epsilon^\xi dr\,\frac c{6r}=\frac c6\ln\frac\xi\epsilon\,.
\end{equation}
In addition, there are $\epsilon$-independent terms, which depend on the details of the theory as well as on the precise form of the UV regulator. We learn from \eqref{S(A) estimate} that the entropy has a logarithmic UV divergence proportional to the central charge of the fixed point, and that in a massive theory it is IR-finite.

\subsubsection{Entropy: Replica trick}

A quantitative calculation of the entropy of the half-line in a given theory can be carried out either numerically (using a lattice discretization, a mode expansion, or some combination) or analytically. The analytic calculation usually proceeds by computing the R\'enyi entropies $S_\alpha(A)$ for integer $\alpha>1$, fitting the result to an analytic function of $\alpha$, and taking the limit $\alpha\to1$ (see subsection \ref{sec:Renyis}).

To calculate $S_\alpha(A)$, one needs $\Tr\rho^\alpha$, and this can be obtained in various ways, including by a Euclidean path integral, as follows. Consider the expression \eqref{cut plane A} for the matrix element of $\rho_A$ in the field basis. The matrix element of $\rho_A^2$ can be computed in the same way:
\begin{align}\label{S2}
\ev{\phi_0^A|\rho_A^2|\phi_2^A} &=
\int D\phi_1^A\,\ev{\phi_0^A|\rho_A^2|\phi_1^A}\ev{\phi_1^A|\rho_A^2|\phi_2^A} \\
&=\frac1{Z^2}\times
\raisebox{-3.25\baselineskip}{
\includegraphics[width=0.45\textwidth]{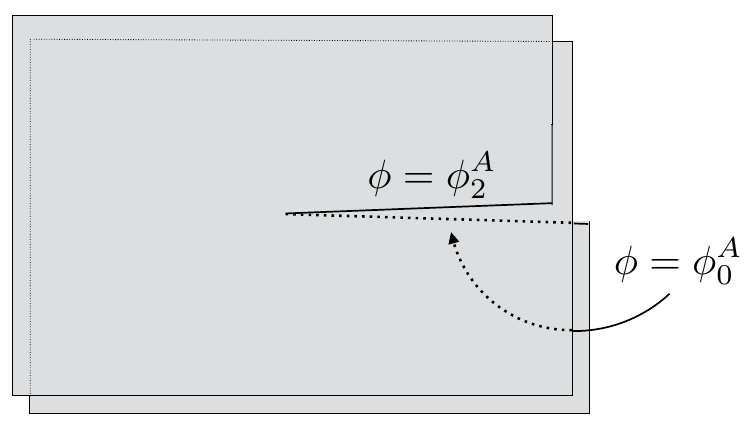}
}\,.\nonumber
\end{align}
This is two copies of the cut plane representing $\rho_A$, with the upper edge of one plane glued to the lower edge of the other. To compute $\Tr\rho_A^2$, we now set $\phi_2^A=\phi_0^A$ and integrate over $\phi_0^A$, which amounts to gluing the two remaining edges of the surface to each other. The result is a two-sheeted surface with a branch cut running along $A$ connecting the two sheets. In complex coordinates $z=x+i\tau$, this is the Riemann surface for the function $z^{1/2}$. Similarly, $\Tr\rho_A^\alpha$ is given by the path integral on the $\alpha$-sheeted surface with the sheets cyclically glued along $A$ (the Riemann surface for $z^{1/\alpha}$), divided by $Z^\alpha$. This way of writing $\Tr\rho^\alpha$ in terms of an $\alpha$-sheeted surface is called the \textbf{replica trick}.

This path integral can be computed using a mode expansion, heat-kernel, or other method (we will give an example in the next subsection). It can even be computed numerically using a Euclidean lattice. The important point is that the surface has a conical singularity at the origin with excess angle $2\pi(\alpha-1)$. This singularity leads to a divergence in the path integral which must be cut off by introducing an ultraviolet regulator. The singularity is thus the origin, within this method of calculating the entropy, of the divergence \eqref{S(A) estimate1}.

\subsection{CFT: interval}
\label{sec:one interval}

We can get even further by assuming more symmetries for our system. We will therefore next consider a conformal field theory.

If we take the limit $m\to0$ or $\xi\to\infty$ in \eqref{S(A) estimate}, we get, in addition to the UV divergence, an IR divergence. To cut off this divergence, we consider a finite interval instead of the half line. The entangling surface now consists of the two endpoints. By translational symmetry, the entropy can depend only on their relative position, the length $L$ of the interval. One might think that by conformal symmetry the entropy (which is dimensionless) cannot even depend on $L$. However, as we already saw the entropy is UV divergent and therefore depends on the UV cutoff $\epsilon$. This breaks the conformal symmetry and allows the entropy to depend on the ratio $L/\epsilon$. A quick and dirty derivation of this dependence can be made along the same lines as the calculation \eqref{S(A) estimate}. There we found that the UV-divergent part of the entropy near each endpoint is $-(c/6)\ln\epsilon$. Taking into account the fact that the interval has two endpoints, the total divergent part is $-(c/3)\ln\epsilon$. Since, as explained above, by conformal invariance, the entropy can only depend on $L/\epsilon$, the $\epsilon$ dependence determines the $L$ dependence, we must have
\begin{equation}\label{HLW}
S(A) = \frac c3\ln\frac L\epsilon+\text{constant}\,.
\end{equation}

The constant term in \eqref{HLW} shifts under changes of the regulator $\epsilon$, so, unlike the coefficient of the logarithm, it is not a universal quantity. However, the fact that it is not universal does not imply that it is physically meaningless, as its value for a \emph{particular} regulator is meaningful, and can be extracted by subtracting other quantities with the same divergence. For example, the entropy in an excited state has the same divergent part as in the vacuum, and therefore we can meaningfully compare their finite parts (as long as we are careful to use the same regulator, i.e.\ the same definition of $\epsilon$). Similarly, as we will discuss in subsection \ref{sec:two intervals} below, the divergences cancel in the calculation of the mutual information between separated regions, leaving a meaningful finite residue.

\subsubsection{Replica trick}

The result \eqref{HLW} can be confirmed by an honest calculation using path integral and CFT techniques \cite{Holzhey:1994we,Calabrese:2004eu}. We follow the same path as in the previous subsection. The matrix elements of $\rho_A$ are given by the path integral on the Euclidean plane cut along the interval $A$ on the $\tau=0$ axis:
\begin{equation}\label{cut plane A CFT}
\ev{\phi_0^A|\rho_A|\phi_1^A} =
\frac1Z\times
\raisebox{-3.25\baselineskip}{
\includegraphics[width=0.35\textwidth]{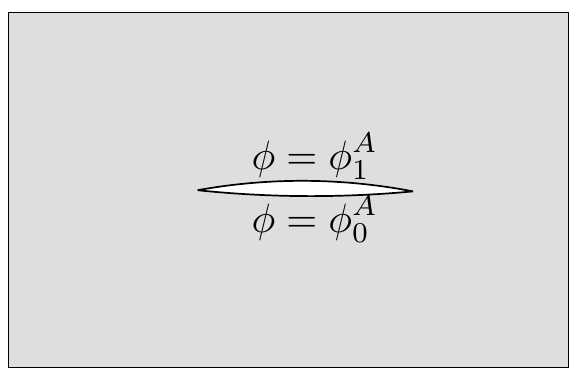}
}\,.
\end{equation}

\begin{figure}[tbp]
\centering
\includegraphics[width=0.4\textwidth]{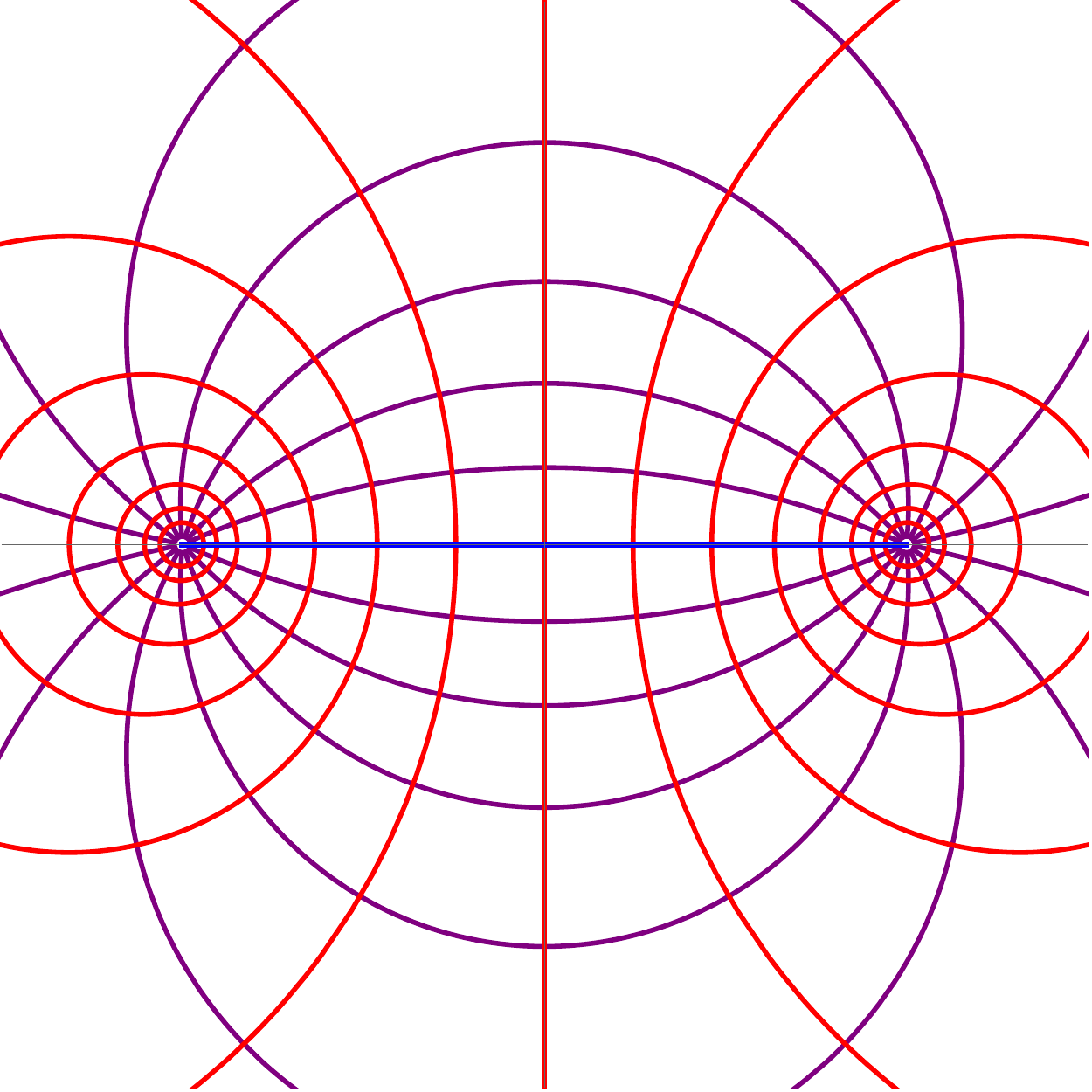}
\hspace{0.75in}
\includegraphics[width=0.4\textwidth]{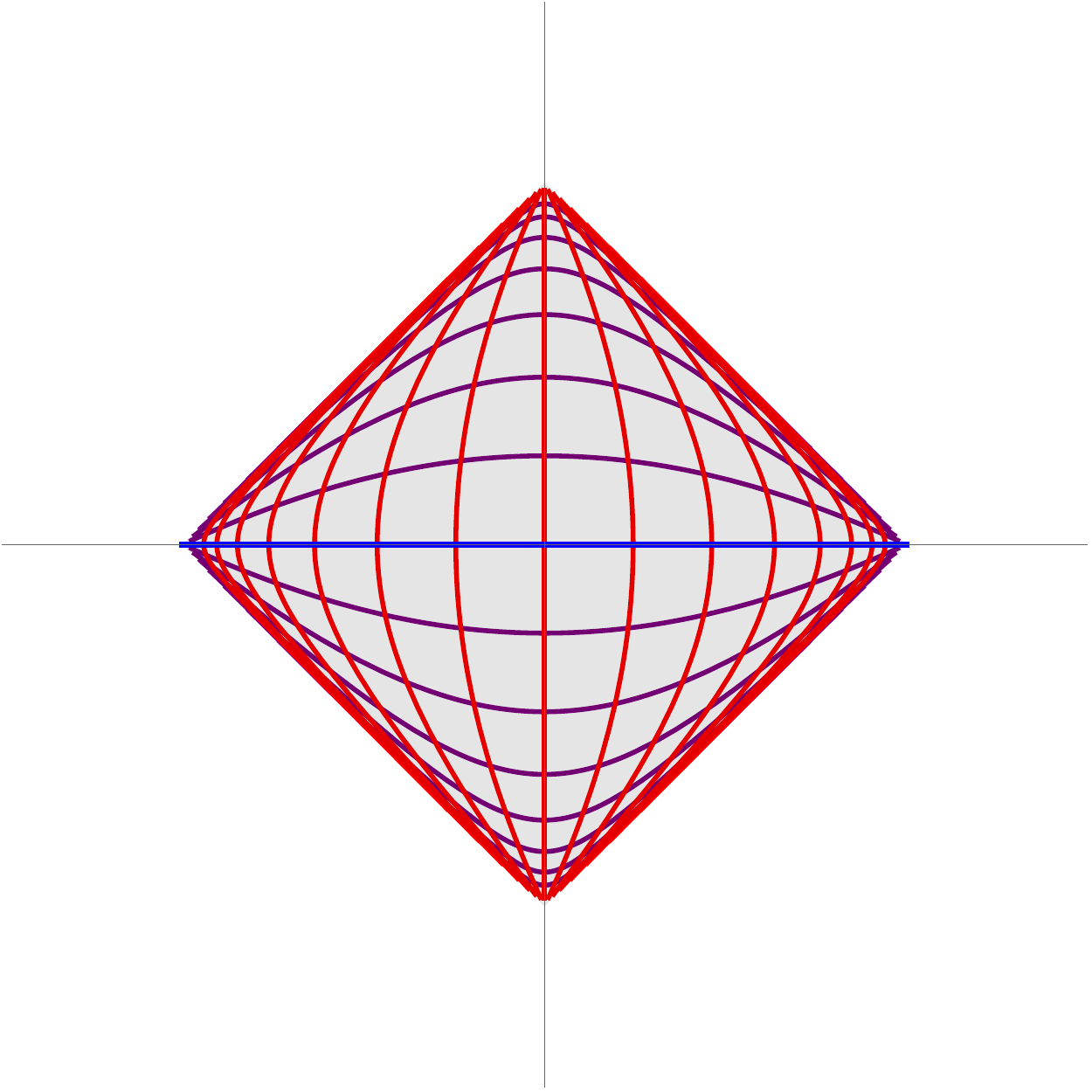}
\caption{\label{fig:bipolar}
Left: Bipolar coordinates defined by \eqref{bipolar} for the Euclidean plane; curves of constant $\theta$ are in purple and curves of constant $\rho$ are in red. Right: Bipolar coordinates defined by \eqref{bipolarL} for Minkowski space; curves of constant $\chi$ are in purple and curves of constant $\rho$ are in red. The causal diamond $D(A)$ covered by these coordinates is in gray.
}
\end{figure}

In just a moment we will use \eqref{cut plane A CFT} to compute the entropy via the replica trick, but first we pause to find the modular Hamiltonian. Just as we switched to polar coordinates in the half-line case, we now switch to bipolar coordinates $(\rho,\theta)$, where $\rho$ runs from $-\infty$ to $\infty$ and $\theta$ from 0 to $2\pi$ (see fig.\ \ref{fig:bipolar}):
\begin{equation}\label{bipolar}
x = \frac L2\frac{\sinh\rho}{\cosh\rho+\cos\theta}\,,\qquad
\tau = \frac L2\frac{\sin\theta}{\cosh\rho+\cos\theta}\,.
\end{equation}
The endpoints of the interval are at $x=\pm L/2$ or $\rho=\pm\infty$. The top of the cut is at $\theta=0$ and the bottom at $\theta=2\pi$. The metric in these coordinates is
\begin{equation}
ds^2_{\rm E} = \left(\frac L2\right)\frac{d\rho^2+d\theta^2}{(\cosh\rho+\cos\theta)^2}\,.
\end{equation}
Translations of the $\theta$ coordinate are not isometries but they are \emph{conformal} isometries, as the metric changes by a Weyl transformation. Therefore these ``conformal rotations'' are symmetries of the CFT. Writing their generator $K$, the reduced density matrix again takes the form
\begin{equation}
\rho_A = \frac1Ze^{-2\pi K}\,,
\end{equation}
and the modular Hamiltonian is again $H_A=2\pi K$. Making the Wick rotation
\begin{equation}
\theta = i\chi\,,
\end{equation}
the map to standard Minkowski coordinates becomes
\begin{equation}\label{bipolarL}
x = \frac L2\frac{\sinh\rho}{\cosh\rho+\cosh\chi}\,,\qquad
t = \frac L2\frac{\sinh\chi}{\cosh\rho+\cosh\chi}\,,
\end{equation}
which covers only the causal diamond $D(A)$. Translations of $\chi$ are again conformal isometries, which act like boosts near the endpoints of the interval. The associated conformal Killing vector is
\begin{equation}
\pd{}{\chi} = \frac1L\left[\left(\left(\frac L2\right)^2-x^2-t^2\right)\pd{}{t}+xt\pd{}{x}\right],
\end{equation}
so the generator can be written in terms of the energy density on $A$:
\begin{equation}
K = \frac1L\int_{-L/2}^{L/2}dx\left(\left(\frac L2\right)^2-x^2\right)T_{tt}\,.
\end{equation}

In the two examples we've seen so far, the half-line and the interval, the modular Hamiltonian is a weighted integral of the stress tensor. This is a consequence of the large amount of symmetry possessed by these examples, and it does not hold more generally. In fact, thes two examples we have studied are almost the \emph{only} cases where the modular Hamiltonian can be expressed as an integral of a local operator. Typically it is some non-local operator, and is not the generator of any geometric symmetry.

We now return to the computation of the entropy. Using \eqref{cut plane A CFT}, we can compute $\Tr\rho_A^\alpha$ for $\alpha=2,3,\ldots$ and thereby the R\'enyi entropies very similarly to the Rindler case discussed in the previous subsection. Again, the matrix elements of $\rho_A^\alpha$ are given by taking $\alpha$ copies of the cut plane and gluing the top of the slit on each sheet to the bottom on the next sheet, and the trace is computed by gluing the bottom of the slit on the first sheet to the top on the last. This gives a Riemann surface, which is the $\alpha$-fold branched cover of the plane with branch cut $A$. To compute the partition function on the resulting surface, we note that, after compactifying the plane, it is topologically a sphere (for any $\alpha$), and can therefore be related by a Weyl transformation to the unit sphere. The partition function on the unit sphere is independent of $L$, but the Weyl transformation depends on $L$, and the $L$-dependence is thus given entirely by the Weyl anomaly. The Weyl factor diverges near the conical singularity at each endpoint; to cut off the divergence in the resulting Weyl anomaly we must remove a disk of radius $\epsilon$ around each endpoint. The Weyl anomaly is proportional to the central charge $c$ of the CFT. We omit the calculation of the anomaly; the final result for the R\'enyi entropy is
\begin{equation}\label{Renyis}
S_\alpha(A) = \frac c6\left(1+\frac1\alpha\right)\ln\frac L\epsilon+\text{finite}\,.
\end{equation}
It is an easy step to find an analytic function of $\alpha$ that matches \eqref{Renyis} at $\alpha=2,3,\ldots$, since the function as written is already analytic. The von Neumann entropy is then just given by \eqref{HLW}, as anticipated.

\subsubsection{Thermal state and circle}
\label{sec:CFT thermal circle}

We will now give two straightforward generalizations of \eqref{HLW}, which can be obtained by the same method \cite{Calabrese:2004eu}. The first is to put the full system at finite temperature $T$. To do this, we simply \emph{don't} take the limit $\beta\to\infty$ in going from \eqref{finite T} to \eqref{rho matrix}. As a result, $\rho_A$ is computed by the path integral on a cylinder of circumference $1/T$ in the Euclidean time direction with a slit along $A$. Via the replica trick, we obtain the following entropy:
\begin{equation}\label{thermal line}
S(A) = \frac c3\ln\frac{\sinh(\pi TL)}{\pi T\epsilon}\,.
\end{equation}
For $L$ small compared to the thermal correlation length $1/T$, the result reduces to the zero-temperature one \eqref{HLW}, but for $L$ large compared to $1/T$ the growth with $L$ switches from being logarithmic to linear, with slope $2\pi cT/6$, precisely the thermal entropy density at temperature $T$. Roughly speaking, the entropy is receiving two contributions: a divergent ``area-law'' contribution due to entanglement across the entangling surface, and an extensive ``volume-law'' thermal contribution.

The second generalization is to put the CFT on a circle of length $2\pi R$. Here again we work in the vacuum. (If we put the CFT on a circle \emph{and} at finite temperature, then the Euclidean surface which computes the matrix elements of $\rho_A$ is a torus. The non-trivial topology makes the calculation much more difficult, and can only be carried out in certain theories or limits.) The result is very simple:
\begin{equation}\label{circle EE}
S(A) = \frac c3\ln\left(\frac{2R}\epsilon\sin\frac L{2R}\right)\,.
\end{equation}
Again, in the limit $R\to\infty$ we recover the Minkowski-space result \eqref{HLW}. Two features are noteworthy here: First, the complement $A^c$ is also an interval, of length $L^c=2\pi R-L$, and we can easily check that $S(A^c)=S(A)$; this is expected from \eqref{pure}, since the full system is in a pure state. Second, the entropy \emph{decreases} with $L$ for $L>\pi R$. If we consider an interval $B$ adjacent to $A$, then $AB$ is also an interval. The conditional entropy $H(B|A):=S(AB)-S(A)$ is then \emph{negative} (and finite), a hallmark of entanglement as discussed in subsection \ref{sec:entangled} above.

\begin{figure}[tbp]
\centering
\includegraphics[width=0.6\textwidth]{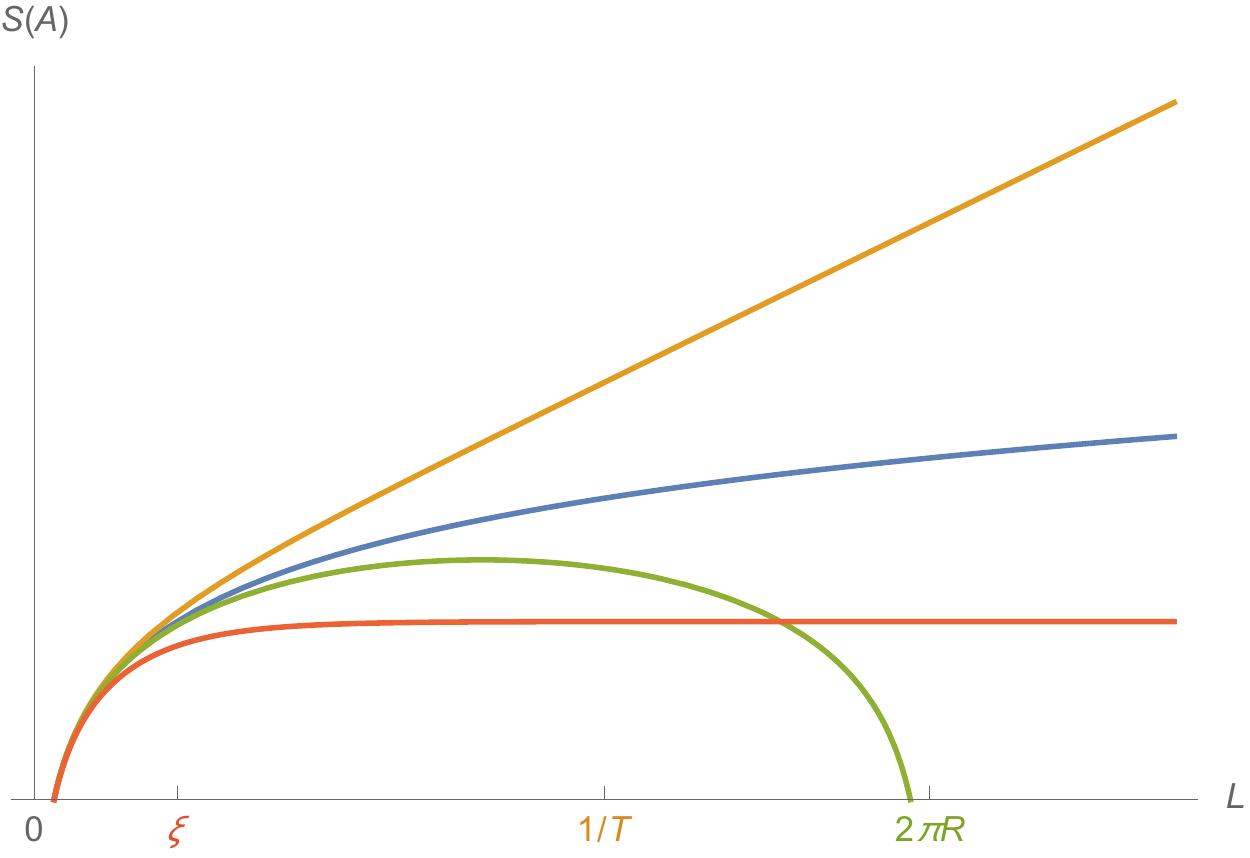}
\caption{\label{fig:EEs1d}
$S(A)$ vs.\ length $L$ for a single interval in various cases. Blue: vacuum of CFT on a line, \eqref{HLW}. Orange: thermal state at temperature $T$ of CFT on a line, \eqref{thermal line}; the slope approaches a positive constant for $L\gg1/T$. Green: vacuum of CFT on a circle of radius $R$, \eqref{circle EE}. Red: vacuum of theory with correlation length $\xi$ on a line; the value asymptotes to $(c/3)\ln(\xi/\epsilon)$ for $L\gg\xi$.
}
\end{figure}

The entropies for the three cases discussed here are plotted in fig.\ \ref{fig:EEs1d}.

\subsection{General theory: interval}
\label{sec:one interval general}

In a general $1+1$ dimensional QFT, we still have the formal expression \eqref{cut plane A CFT} for the reduced density matrix, but we no longer have the conformal symmetry that allows us to write down the modular Hamiltonian explicitly and compute the R\'enyi and von Neumann entropies as in the CFT case. It is possible to make further progress in specific cases, such as free theories; see \cite{Casini:2009sr}. Here we will limit ourselves to some qualitative observations.

Consider a theory with a UV fixed point and correlation length $\xi$. We can guess that for intervals much shorter than the correlation length, the physics is controlled by the UV fixed point and so the entropy is given approximately by \eqref{HLW}:
\begin{equation}\label{gapped short}
S(A)\approx\frac{c}3\ln\frac L\epsilon\qquad (L\ll\xi)\,.
\end{equation}
On the other hand, for an interval much longer than $\xi$, each endpoint looks locally like Rindler space, with the fields farther than a distance $\xi$ from each endpoint frozen out, so we expect the entropy to saturate at a value twice that of Rindler space, \eqref{S(A) estimate},
\begin{equation}\label{gapped long}
S(A)\approx\frac c3\ln\frac\xi\epsilon\qquad (L\gg\xi)\,.
\end{equation}
So the entropy as a function of $L$ should like qualitatively like the red curve of fig.\ \ref{fig:EEs1d}. While $S(A)$ has not been calculated exactly in any massive theory, semi-analytic calculations for free bosons and fermions confirm this picture \cite{Calabrese:2004eu,Casini:2005rm}.

Notice that all the curves in fig.\ \ref{fig:EEs1d} are concave. This is not a coincidence. In fact, it is a consequence of strong subadditivity \eqref{SSA1}. Choose  $A,B,C$ to be adjacent intervals, write the inequality in the form
\begin{equation}
\left(S(ABC)-S(BC)\right)-\left(S(AB)-S(B)\right)\le0\,,
\end{equation}
divide by $L_AL_C$ so it becomes
\begin{equation}
\frac{\frac{S(ABC)-S(BC)}{L_A}-\frac{S(AB)-S(B)}{L_A}}{L_C}\le0\,,
\end{equation}
and finally take the limit $L_A,L_C\to0$. We get
\begin{equation}\label{Sconcave}
\frac{d^2S(B)}{dL^2_B}\le0\,.
\end{equation}
(Note that in this argument we are never evaluating $S(A)$ or $S(C)$, so it's okay that we are taking their lengths to 0.)

\subsubsection{C-theorem}

\begin{figure}[tbp]
\centering
\includegraphics[width=0.4\textwidth]{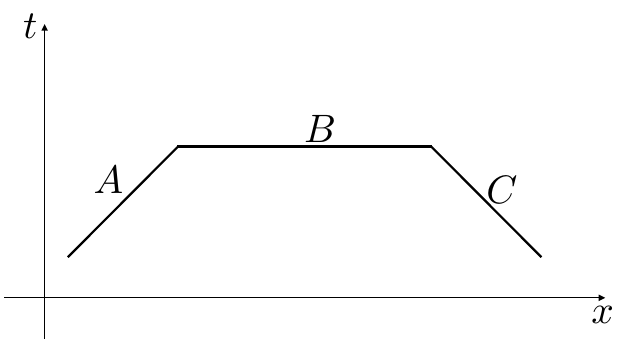}
\caption{\label{fig:trapezoid}
Null trapezoid configuration of intervals in Minkowski space used in the entropic proof of the C-theorem \eqref{C theorem}.}
\end{figure}

The concavity result \eqref{Sconcave} is a useful constraint on the entropy for a general translation\-ally-invariant state. (In fact, the idea that the entropy should never grow faster than linearly with system size was the original motivation for the conjecture of strong subadditivity \cite{doi:10.1063/1.1664685}.) However, by combining strong subadditivity with Lorentz symmetry, we can make a much more powerful statement \cite{Casini:2004bw}. For this purpose we need to work in the vacuum on Minkowski space, since then both the spacetime and state are invariant under the full Poincar\'e group. This implies that $S(A)$ is Poincar\'e invariant, and in particular if $A$ is a single interval on any time slice, then $S(A)$ can only be a function of the proper distance between its endpoints. We now consider the ``null trapezoid'' configuration shown in fig.\ \ref{fig:trapezoid}, in which $A,B,C$ are adjacent intervals and $A$ and $C$ are null.\footnote{Strictly speaking, since it includes null segments, this configuration does not lie on a Cauchy slice as we defined it in subsection \ref{sec:subsystem}. However, we can consider it as the limit of a sequence of Cauchy slices with $A$ and $C$ tending to become null; on each one the inequality must be obeyed, so in the limit it must also be obeyed. Note that in the following argument we are never evaluating $S(A)$ or $S(C)$.} A short exercise in Lorentzian geometry shows that the lengths of $B$, $AB$, $BC$, $ABC$ obey
\begin{equation}
L_{AB}L_{BC}=L_BL_{ABC}\,,
\end{equation}
or in other words
\begin{equation}
\ln L_{AB} + \ln L_{BC} = \ln L_B+\ln L_{ABC}\,.
\end{equation}
By the same logic as for \eqref{Sconcave} but working in terms of the logarithmic variables, strong subadditivity implies that $S(B)$ is a concave function of $\ln L$ (again $L:=L_B$):
\begin{equation}
\frac{d^2}{(d\ln L)^2}S(B)\le0\,.
\end{equation}
This means that the function
\begin{equation}\label{Cdef}
C(L):=3\frac{dS(B)}{d\ln L}\,,
\end{equation}
called the \textbf{renormalized EE}, is a non-increasing function of $L$. If the theory has a UV fixed points, then as argued above \eqref{gapped short}, for small $L$ the entropy should be that of the fixed point, \eqref{HLW} with $c=c_{\rm UV}$. This implies
\begin{equation}
\lim_{L\to0}C(L) = c_{\rm UV}\,.
\end{equation}
On the other hand, if the theory has an IR fixed point, then for large $L$ the entropy should again be given by \eqref{HLW}, with $c=c_{\rm IR}$ and a different constant than in the UV. Essentially, for large $L$, the whole RG flow can just be considered as a particular regulator for the IR CFT. We thus have
\begin{equation}
\lim_{L\to\infty}C(L) = c_{\rm IR}\,.
\end{equation}
The monotonicity of $C(L)$ then implies the \textbf{C-theorem}:
\begin{equation}\label{C theorem}
 c_{\rm IR}\le c_{\rm UV}\,.
\end{equation}
This proof of the C-theorem has the same ingredients as other proofs, such as the original one by Zamolodchikov \cite{Zamolodchikov:1986gt}, namely unitarity, locality, and relativity. But the way those ingredients are used seems very different. In particular, the function $C$ that interpolates monotonically between the UV and IR central charges is different from the one appearing in Zamolodchikov's proof.

The function $C(L)$ is very useful for diagnosing criticality and identifying the IR CFT in spin chains. In a numerical simulation of a spin chain, it is relatively straightforward to compute the entropy of an interval, and thereby compute $C(L)$. If the spin chain is critical, then it should approach a constant for long intervals, and that constant equals the central charge of the CFT, information which is difficult to obtain by other means.

Finally, we should mention a subtlety in the above C-theorem proof, which is of some interest in its own right. The proof required the existence of a UV regulator that  makes entropies finite, preserves Lorentz symmetry, and follows the rules of quantum mechanics---e.g.\ having a Hilbert space with a positive-definite Hilbert norm---in particular so that strong subadditivity is obeyed. However, the sad fact is that no such regulator is known. For example, a lattice breaks Lorentz symmetry, while a Pauli-Villars field introduces negative-norm states. In fact, there are arguments that no such regulator exists; or, more precisely, that the only such regulator is to couple the fields to quantum gravity, with $\epsilon$ being the Planck length \cite{Jacobson:2015hqa}. It may be possible to fix up the proof to avoid the use of a regulator, for example by using the techniques of \cite{Casini:2015woa} or working within the context of algebraic quantum field theory; however, as far as we are aware this remains an open problem.

\subsection{CFT: two intervals}
\label{sec:two intervals}

\begin{figure}[tbp]
\centering
\includegraphics[width=0.35\textwidth]{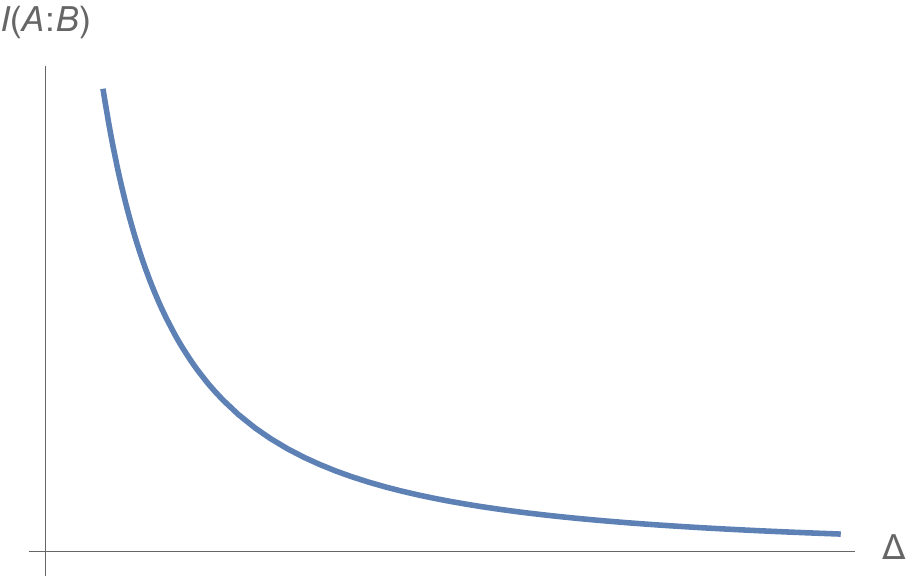}
\caption{\label{fig:MI}
Mutual information $I(A:B)$, given in \eqref{Dirac MI}, for two intervals separated by $\Delta$ for a free massless Dirac fermion.
}
\end{figure}

As a final example, we consider two separated intervals $A,B$ on a line in the vacuum of a CFT. Their union is the simplest example of a disconnected region. As discussed in subsection \ref{sec:vne}, the mutual information
\begin{equation}
I(A:B):=S(A)+S(B)-S(AB)
\end{equation}
quantifies the total amount of correlation between $A$ and $B$. Unfortunately, while we know $S(A)$ and $S(B)$, it is very hard to compute $S(AB)$, and indeed it has been computed analytically only for \emph{one} theory, namely the free massless Dirac fermion \cite{Casini:2004bw,Casini:2005rm}. The result in that case is very simple:
\begin{equation}\label{Dirac MI}
I(A:B) = \frac12\ln\frac{(L_A+\Delta)(L_B+\Delta)}{\Delta(L_A+L_B+\Delta)}\,,
\end{equation}
where $\Delta$ is the separation between $A$ and $B$. It is plotted as a function of $\Delta$ in fig.\ \ref{fig:MI}. In other CFTs, such as the compact boson, while it is possible to compute the R\'enyi entropies $S_\alpha(AB)$ for $\alpha=2,3,\ldots$ exactly, it is not known how to extrapolate the resulting values to $\alpha=1$ to obtain the von Neumann entropy \cite{Calabrese:2009ez,Calabrese:2010he,DeNobili:2015dla,Ruggiero:2018hyl}.

Although the massless Dirac fermion is the only case for which we know the mutual information exactly, many of its qualitative features can be shown to hold for the mutual information of separated intervals in the vacuum of any CFT, and some of them even more generally than that:
\begin{enumerate}
\item It is finite (independent of $\epsilon$). This is because the divergences are local on the endpoints of the intervals, and cancel in the mutual information (as long as the intervals are separated). This holds for separated regions in any state in any field theory in any dimension.
\item It is positive. By subadditivity, it must be non-negative, and if it vanished then we would have $\rho_{AB}=\rho_A\otimes\rho_B$, which would imply that all connected correlators between operators in $A$ and in $B$ would vanish, and we know this is not true in a field theory. Therefore, like property 1, this property holds in any state in any field theory.
\item It is conformally invariant. This is because it is finite and therefore independent of $\epsilon$, so it can only depend on $L_A$, $L_B$, and $\Delta$ in a conformally-invariant manner. Specifically, it must be a function of the cross-ratio of the four endpoints of the intervals. (The argument of the logarithm in \eqref{Dirac MI} is precisely the cross-ratio.)
\item It is a non-decreasing function of $L_B$. This follows from strong subadditivity, which says that, if we let $C$ be an interval adjacent to $B$ and define $B'=BC$, then  
\begin{equation}
I(A:B')=I(A:BC)\ge I(A:B)\,.
\end{equation}
Similarly, it is a non-decreasing function of $L_A$.
\item It is a non-increasing function of $\Delta$. This follows from the last two facts. It also makes sense intuitively: we would expect the amount of correlation to decrease as the separation increases.
\item It goes to $\infty$ as $\Delta\to0$.
\item It goes to 0 as $\Delta\to\infty$, and does so at the rate $\Delta^{-4d}$, where $d$ is the scaling dimension of the lightest non-trivial operator \cite{Calabrese:2010he}. This fact follows from applying the replica trick to the computation of $S(AB)$ and using an OPE expansion to compute the relevant partition function at large $\Delta$. In the case of the Dirac fermion, $d=1/2$.
\end{enumerate}

\subsection{Higher dimensions}
\label{sec:higherdim}

Our treatment of EEs in higher-dimensional field theories will be very brief, touching on only a few of the most important features. Except where indicated otherwise, we consider an arbitrary relativistic field theory.

We start with the half-space in $D$-dimensional Minkowski spacetime:
\begin{equation}\label{half-space}
A = \{t=0,x^1\ge0\}\,.
\end{equation}
Its causal domain is again the Rindler wedge $|t|\le x^1$. The derivation of the vacuum modular Hamiltonian in subsection \ref{sec:reduced} goes through essentially unchanged, simply with the replacement $x\to x^1$ and the addition of extra spatial coordinates $x^2,\ldots,x^{D-1}$ that go along for the ride. We thus arrive at the modular Hamiltonian
\begin{equation}\label{modularhigher}
K = \int dx^2\cdots dx^{D-1}\int_0^\infty dx^1\,x^1T_{tt}\,.
\end{equation}
Following the same strategy as in subsection \ref{sec:estimate}, we can use \eqref{modularhigher} to estimate the dependence of $S(A)$ on the cutoff $\epsilon$. For this we need to impose an IR regulator on the directions parallel to the entangling surface, for example by making them periodic. We leave the calculation as an exercise to the reader; you should find
\begin{equation}\label{area law}
S(A)\propto\frac\sigma{\epsilon^{D-2}}\,,
\end{equation}
where $\sigma$ is the area of the entangling surface. Unlike the logarithmic leading term in $D=2$, the coefficient of this term depends on the precise form of the UV regulator, and so is not physically meaningful.

For any spatial region with a smooth boundary, very close to the entangling surface it looks like a plane, so we would expect the leading divergence of the EE to be given by \eqref{area law}. This area-law divergence agrees with our expectation \eqref{arealaw} from lattice models.\footnote{Note the different use of the term \emph{area-law} here as compared to in subsection \ref{sec:subsystem} above. There, it referred to the growth of $S(A)$ for large size $L$ at fixed cutoff $\epsilon$; here it refers to the dependence of the divergent term in $S(A)$ on the area of the entangling surface. For example, in a CFT in $D=2$, $S(A)$ has logarithmic dependence on $L$, hence violates the first area law, yet it obeys the second one since the divergent term is $\frac16c\sigma\ln\epsilon$, with $\sigma$ the ``area'' (i.e.\ number of points) in the entangling surface.} There are also subleading divergent terms that depend on the particular field theory, as well as the geometry of $A$. The divergent terms arise from local UV physics near the entangling surface. Therefore, they must be independent of the state (for valid field-theory states, which have energy densities well below the cutoff scale) and take the form of integrals of local geometric invariants.\footnote{More precisely, this will hold for a sufficiently ``geometric'' UV regulator, i.e.\ one that does not add additional structure beyond the length scale $\epsilon$. For example, the anisotropy of a lattice will lead to a very complicated dependence on $\epsilon$.} Because they are integrals of local geometric quantities, the divergences cancel when computing for example the mutual information of separated regions, leaving a finite and cutoff-independent quantity, just as we saw in subsection \ref{sec:two intervals} in $D=2$. Similarly, they cancel when computing the difference in the EE between different states for the same region.

Let us see how this works in three- and four-dimensional CFTs. In $D=3$, the only divergent term is the area term:
\begin{equation}\label{3dCFT}
S(A) = \alpha\frac\sigma\epsilon +\text{constant}\,,
\end{equation}
where $\sigma$ is the ``area'' (in this case, length) of the entangling surface and $\alpha$ is a dimensionless coefficient. Again, since $\alpha$ changes under changing $\epsilon$, it is not meaningful. Why is there no $\ln\epsilon$ term in \eqref{3dCFT}? The answer is that the only local geometric invariant with the right dimensions, namely inverse length, is the extrinsic curvature $K$ of the entangling surface; however, under $A\to A^c$, $K\to-K$, whereas the EE must be invariant (at least in a pure state---but the divergent terms are state-independent). By a similar argument, for a CFT in any dimension the divergent terms skip alternating powers of $\epsilon$, going like $\epsilon^{2-D}$, $\epsilon^{4-D}$, etc. A log divergence is permitted in even $D$ but not odd $D$.

As we discussed in subsection \ref{sec:one interval}, in a 2d CFT, due to the presence of the term $-(c/3)\ln\epsilon$ in the entropy for an interval, the constant ($\epsilon$-independent) term in can be shifted by multiplying $\epsilon$ by a constant, and is therefore not meaningful on its own; only differences for different states or configuration are meaningful. However, in odd dimensions, since there is no $\ln\epsilon$ term, the constant term does \emph{not} change under changes of $\epsilon$, and is therefore meaningful on its own, or \textbf{universal}. Furthermore, while the constant may depend on the state and on the shape of $A$ (and the ambient geometry), by dimensional analysis it may not depend on the overall size of $A$, i.e.\ it must not change under a global rescaling of $A$ (and the ambient geometry, if the latter is not Minkowski spacetime). The case with the most symmetry is a round ball in Minkowski spacetime in the vacuum. By manipulating the Euclidean path integral and using Weyl invariance (specifically, the absence of a Weyl anomaly in odd dimensions), it can be shown that this constant equals the log of  the $D$-sphere partition function (more precisely, the $\epsilon$-independent term thereof) \cite{Casini:2011kv}.

For example, in $D=3$, for a disk in Minkowski spacetime in the vacuum, the constant is negative, and is called $-F$:
\begin{equation}
S(A) = \alpha\frac\sigma\epsilon -F\,.
\end{equation}
If the theory is not only conformal but topological, then $F$ equals the so-called \textbf{topological EE} (often denoted $\gamma$), which is related in a simple way to the spectrum of anyons \cite{Kitaev:2005dm,2006PhRvL..96k0405L}. One reason $F$ is interesting is that it is an RG monotone, in other words it obeys a C-theorem-like relation \cite{Casini:2012ei}:
\begin{equation}
F_{\rm UV}\ge F_{\rm IR}\,.
\end{equation}
The proof is parallel to that of the entropic C-theorem in 2d. One first defines the renormalized EE $F(R)$, similar to the function $C(L)$ defined in \eqref{Cdef}:
\begin{equation}
F(R):=-\frac d{dR}\left(RS(A)\right),
\end{equation}
where $S(A)$ is the entropy of a disk of radius $R$. In a conformal theory, we have $F(R)=F$, so in an RG flow, $F(R\to0)=F_{\rm UV}$ and $F(R\to\infty)=F_{\rm IR}$. An argument that (like the one in the C-theorem proof) relies crucially on Lorentz invariance and strong subadditivity (but is much more elaborate) then establishes that $F(R)$ is monotonically decreasing. Unlike in $D=2$ and (as discussed below) $D=4$, this C-theorem has so far been proven \emph{only} by this entropic method; there is no proof based on conventional field-theoretic quantities such as correlation functions, scattering amplitudes, or spectral densities.

In a 4d CFT in flat space, the EE takes the following form:
\begin{equation}\label{4dCFT}
S(A) = \int d^2x\sqrt{h}\left[\frac\alpha{\epsilon^2}+\frac{\ln\epsilon}{2\pi}\left(a\tilde R + c(K_{\mu\nu}K^{\mu\nu}-\frac12K^2)\right)\right] + \text{constant}\,,
\end{equation}
where the integral is over the entangling surface, $\alpha$ is a (non-meaningful) dimensionless constant, $a$ and $c$ are the Weyl anomaly coefficients (the analogues of the central charge in 4 dimensions), and $h_{\mu\nu}$, $\tilde R$, and $K_{\mu\nu}$ are the induced metric and intrinsic and extrinsic curvatures respectively of the entangling surface \cite{Solodukhin:2008dh}. As in 2d, because of the $\ln\epsilon$ term, the constant term is not meaningful on its own; only differences between different states or configurations are meaningful. If we consider the entropy of a round ball, then we can see from \eqref{4dCFT} that the coefficient of $c$ vanishes, so the universal $\ln\epsilon$ term is proportional to $a$. Again, this is an RG monotone:
\begin{equation}
a_{\rm UV}\ge a_{\rm IR}\,.
\end{equation}
This has been proven using both conventional \cite{Komargodski:2011vj} and entropic \cite{Casini:2017vbe} techniques.

The C-theorems in $D=2,3,4$ suggest that the universal term in the vacuum ball EE (constant for odd $D$, coefficient of $\ln\epsilon$ for even $D$) might be an RG monotone in any dimension. It is not known if this is true, or indeed if there exist any monotones in $D>4$.

This concludes our grand tour of EEs in quantum field theories. Although the examples and properties we have discussed in this section have just scratched the surface of the topic of entanglement in quantum field theories, they have hopefully served to give some idea of how EEs reflect and illuminate some of the most important physics in quantum field theories. We now turn to a specific class of theories: holographic ones.

\section{Entropy, entanglement, fields, and gravity}
\label{sec:EEfieldsgravity}

Two things are hopefully clear to the reader who has read the previous section: First, entanglement entropies (EEs) provide significant physical insight into quantum field theories. Second, they are difficult or impossible to calculate except in a few simple cases. This is obviously a somewhat frustrating situation.

There is, however, a surprising exception to this rule. In a \textbf{holographic} theory---a field theory admitting a dual description in terms of a higher-dimensional gravitational theory---it is possible to compute an EE just by solving a certain classical geometry problem, which is to find the area of a minimal surface with a given boundary. This statement is the celebrated \textbf{Ryu-Takayanagi (RT) formula} \eqref{RT} \cite{Ryu:2006bv,Ryu:2006ef}. RT is much more than a calculational tool, however. It is a profound statement about quantum gravity, one that we do not yet fully understand but that has already been fruitfully exploited to advance our understanding of the mysteries of holography. In fact, RT provides such a direct link between spatial entanglement and bulk geometry that it has even been suggested that entanglement is the basic ingredient in the field theory from which the bulk spacetime emerges \cite{VanRaamsdonk:2010pw}.

We will not discuss such investigations here. Instead, our focus will be on how the RT works in practice. Specifically, after stating the formula and giving it some motivation and context in subsection \ref{sec:RT formula}, we will show in subsection \ref{sec:RT examples} how the behavior of the EEs explored in the previous section is realized geometrically by minimal surfaces, and then in subsection \ref{sec:RT properties} how the general properties of EEs are satisfied. Along the way we will also discover some new behaviors of properties that are special to holographic theories.

\subsection{Holographic dualities}
\label{sec:holographic}

We start with a brief review of the basic facts we will need about holographic dualities. Such a duality relates a $D$-dimensional quantum field theory to a $D+1$-dimensional quantum gravity theory with negative cosmological constant. The gravity theory is subject to asymptotically AdS boundary conditions on the metric, with conformal boundary equal to the manifold $M$ on which the field theory lives, meaning that near the boundary the metric takes the form
\begin{equation}\label{aAdS}
ds^2 = \frac{\ell^2}{z^2}\left(ds_M^2+dz^2\right)
\end{equation}
where $ds_M^2$ is the metric on $M$, $\ell$ is the AdS radius (related to the cosmological constant by $\Lambda = -\frac12D(D-1)\ell^{-2}$), and the coordinate $z$ takes positive values, with the boundary at $z\to0$.\footnote{For simplicity, in this description we have performed a Kaluza-Klein reduction on any compact manifold in the bulk, such as the $S^5$ in AdS${}_5\times S^5$, so the modes of the fields on this manifold become fields on the AdS$_{D+1}$ spacetime.} There are also boundary conditions on the other bulk fields depending on the values of the coupling constants in the field theory; the details here will not be important to us.

In the limit of a large number of field-theory degrees of freedom (e.g.\ large $N$ for a gauge theory, large $c$ for a 2-dimensional CFT), the gravitational theory becomes classical, with that number controlling the hierarchy between the AdS radius $\ell$ and the Planck length $(\GN\hbar)^{1/(D-1)}$. Specifically, for a two-dimensional CFT,
\begin{equation}\label{ell over GN}
\frac \ell{\GN\hbar} = \frac{2c}3\,,
\end{equation}
while for a gauge theory
\begin{equation}\label{ell over GN2}
\frac{\ell^{D-1}}{\GN\hbar}\propto N^2\,.
\end{equation}
In a further limit of the field theory parameters (generally a strong-coupling limit), the low-energy sector of the gravitational theory becomes Einstein gravity coupled to some matter fields. We will be working in both of these limits. States of the system that are classical (from the bulk point of view) are described by asymptotically AdS${}_{D+1}$ solutions to the Einstein-matter equations. Depending on the bulk matter content, the boundary manifold $M$, and the boundary conditions for the matter fields, the ground state may be AdS${}_{D+1}$ (or a patch thereof), in which case the theory is conformal, or it may be more complicated, in which case the theory has a non-trivial RG flow. Roughly speaking, the region of the bulk ``close to'' the boundary corresponds to the UV of the field theory while the region ``far'' from the boundary corresponds to the IR. A UV cutoff on the field theory can be implemented by cutting off the bulk geometry at $z=\epsilon$.

The large-$N$ (or large-$c$) limit is a kind of thermodynamic limit, and the large number of field-theory degrees of freedom allows phase transitions to occur even in finite volume. For example, there is typically such a phase transition as a function of temperature from a ``confined'' phase with free energy and entropy of order 1 to a ``deconfined'' phase with free energy and entropy of order $N^2$ or $c$. In the confined phase the bulk geometry is the same as in the ground state (with the entropy accounted for by a thermal gas of gravitons and other light particles), while in the deconfined phase there is a black hole in the bulk (with the order-$N^2$ part of the entropy accounted for by its Bekenstein-Hawking entropy). For example, a CFT on $\R\times S^{D-1}$ undergoes a first-order transition, called the \textbf{Hawking-Page transition}, from a low-temperature state described by global AdS${}_{D+1}$ to a high-temperature one described by an AdS-Schwarzschild black hole \cite{Witten:1998zw}.

\subsection{Ryu-Takayanagi formula}
\label{sec:RT formula}

\subsubsection{Motivation: Bekenstein-Hawking as entanglement entropy}

RT relates EEs in holographic theories to the areas of minimal surfaces in the bulk. We will motivate it by starting with the Bekenstein-Hawking entropy, interpreting the latter as an EE, noting that the horizon is a minimal surface, and then writing down the most natural general formula consistent with this example.

Consider a holographic field theory in its deconfined phase. The bulk spacetime is a static, asymptotially AdS black hole. The entropy $S_{\rm BH}$ is given (to leading order in $\hbar$) by the \textbf{Bekenstein-Hawking entropy}, which is one-quarter the area of the (bifurcate) horizon $m_{\rm hor}$ in Planck units (from now on we suppress $\hbar$):
\begin{equation}\label{bhentropy}
S_{\rm BH} = \frac1{4\GN}\area(m_{\rm hor})\,.
\end{equation}

\begin{figure}[tbp]
\centering
\includegraphics[width=0.6\textwidth]{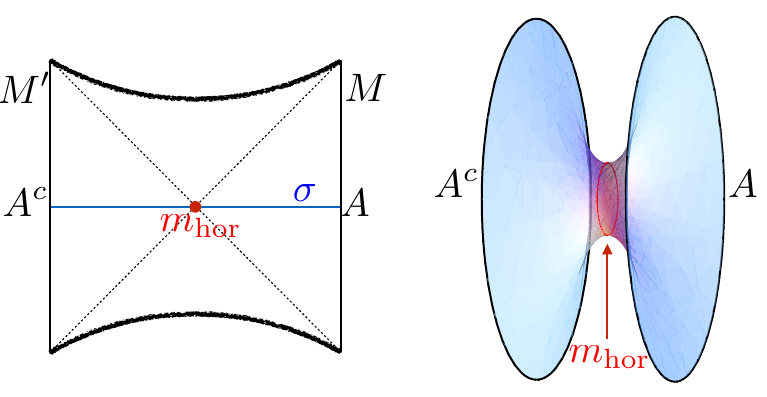}
\caption{\label{fig:twosided}
Left: Penrose diagram of two-sided static asymptotically AdS black hole. Its boundary consists of two disconnected spacetimes $M$ and $M'$. The bulk Cauchy slice $\sigma$ is time-reflection-invariant and passes through the bifurcate horizon $m_{\rm hor}$. Its conformal boundaries $A$, $A^c$ are Cauchy slices for $M$, $M'$ respectively. Right: The induced geometry on $\sigma$ is an Einstein-Rosen bridge connecting $A$ and $A^c$. The bifurcate horizon, in red, is the minimal surface on $\sigma$ between $A$ and $A^c$.
}
\end{figure}

$S_{\rm BH}$ is a thermal entropy. But, as we discussed in subsection \ref{sec:purification}, the entropy of any mixed state can be understood as an ``entanglement'' entropy by purifying the state, and for a thermal state a particularly natural purification is the thermofield double state \eqref{TFD}. The thermofield double is a pure state on two copies of the system, in this case two copies of the field theory, or, to say it differently, the field theory on two copies $M$ and $M'$ of the boundary spacetime. The thermofield double state is represented holographically by the symmetric two-sided extended black hole spacetime, which includes the original black hole representing the thermal state as one of its exterior regions; see fig.\ \ref{fig:twosided} \cite{Maldacena:2001kr}. The two exterior regions are static, but the extended spacetime is not globally static, so we cannot define canonical ``constant-time'' slices. Nonetheless it has distinguished Cauchy slices, namely those that are invariant under a time reflection symmetry of the spacetime. Such an invariant slice $\sigma$ passes through the bifurcate horizon $m_{\rm hor}$, and its restriction to either exterior region is a constant-time slice there. It intersects the full boundary $M\cup M'$ on a Cauchy slice $\Sigma$ which is the union of Cauchy slices $A$ for $M$ and $A^c$ for $M'$. By the definition of the thermofield double state, the reduced density matrix $\rho_A$ is the same as the thermal density matrix on one copy of the field theory, so $S(A)$ is the black hole entropy \eqref{bhentropy}:
\begin{equation}\label{RT1}
S(A) = S_{\rm BH} = \frac1{4\GN}\area(m_{\rm hor})\,.
\end{equation}

Now we want to generalize \eqref{RT1}, so we need some bulk surface that will stand in for $m_{\rm hor}$ in a more general situation. The bifurcate horizon is distinguished in various ways; for example, it is the locus of fixed points of a Killing vector (which is timelike in the exterior regions and spacelike in the interior regions). But we can also characterize $m_{\rm hor}$ without appealing to the continuous symmetries of the spacetime: it is a \emph{minimal-area} surface on $\sigma$, and more specifically it is the minimal surface separating $A$ from $A^c$, i.e.\ dividing $\sigma$ into two regions, one bounded by $A$ and the other by $A^c$.

\subsubsection{Statement}
\label{sec:RT statement}

We now guess that this \emph{minimal-area} property is the key feature that associates the surface $m_{\rm hor}$ with the entropy $S(A)$, and that this holds more generally than in the thermofield double example. For a holographic spacetime with a time-reflection symmetry, let $\sigma$ be a reflection-invariant bulk Cauchy slice and let $A$ be a region of its conformal boundary $\Sigma$. We assume for the moment that the full boundary (or set of boundaries) is in a pure state, as in the thermofield double example. The natural generalization of that example is to look for the minimal surface $m(A)$ in the bulk that separates $A$ and $A^c$, in other words that divides $\sigma$ into a region $r(A)$ bounded by $A$ and another one $r(A^c)$ bounded by $A^c$, and assert that the area of $m(A)$ gives the entropy of $A$:\footnote{By \emph{area}, we always mean the area calculated in the Einstein-frame metric. Only the Einstein-frame area is consistent with the Bekenstein-Hawking formula and is invariant under dimensional reduction, T-duality, and S-duality.} 
\begin{equation}\label{RT}
S(A) = \frac1{4\GN}\area(m(A))\,.
\end{equation}
This is the RT formula. The gist of it is that the reduced density matrix $\rho_A$ can be treated as if it were described by a black hole with horizon $m(A)$, \emph{even when there is no Killing horizon}. RT thus asserts that the Bekenstein-Hawking formula applies far beyond the domain where it was originally derived.

It is useful to be slightly more careful in the definition of $m(A)$, by keeping track of orientations. Assign an orientation to the bulk slice $\sigma$. The boundary region $A\subseteq\partial\sigma$ inherits an orientation from $\sigma$. We say an oriented surface $m$ is \textbf{homologous}\footnote{The notion of \emph{homologous} should not be confused with that of \emph{homotopic} (continuously deformable into). Homotopic implies homologous, but not the converse. An example highlighting the distinction will be given in subsection \ref{sec:CFT thermal}. We should also note that, technically, we are working in homology \emph{relative to} $\partial A$.} to $A$, and write $m\sim A$, if there exists a bulk region $r\subseteq\sigma$ such that
\begin{equation}
\partial r = A\cup-m\,,
\end{equation}
where $-m$ is the surface $m$ with its orientation reversed. $m(A)$ is then the minimal-area surface homologous to $A$; we call the corresponding region $r(A)$ the \textbf{homology region}.\footnote{The \textbf{entanglement wedge}, which has played a prominent role in investigations into subregion duality, is defined as the causal domain of the homology region $r(A)$ \cite{Headrick:2014cta}. The term ``entanglement wedge'' is sometimes used loosely to refer to the homology region in circumstances where time is not playing an essential role.} In this notation we can also write \eqref{RT} as
\begin{equation}\label{RT2}
S(A) = \frac1{4\GN}\min_{m\sim A}\area(m)\,.
\end{equation}

A few comments are in order. First, the requirement of time symmetry was necessary to pick out a distinguished slice, as we can see in the thermofield-double case: a general bulk Cauchy slice of the extended black hole spacetime shown in fig.\ \ref{fig:twosided} does not pass through the bifurcate horizon, and the minimal surface on it does not have area equal to $S_{\rm BH}$. In practice this is not such an onerous requirement, as one is very often interested in entropies in \emph{static} spacetimes, where any constant-time slice is automatically time-reflection invariant; all the examples we discuss in the rest of this section are like this. However, as we will mention in subsection \ref{sec:generalizations}, this restriction can be relaxed in a natural way, allowing one to compute entropies in time-dependent spacetimes.

Second, there is a clear symmetry in the RT formula between $A$ and $A^c$. The minimal surface for $A^c$ is the same as the one for $A$, or more precisely the same with the orientation reversed,
\begin{equation}\label{pure m equal}
m(A^c) = -m(A)\,,
\end{equation}
since if $m\sim A$ then $-m\sim A^c$. So RT says
\begin{equation}\label{pure S equal}
S(A^c) = S(A)\,,
\end{equation}
as expected given the assumption that the full system is in a pure state. (We will explain how to relax this assumption in subsection \ref{sec:CFT thermal}.)

\begin{figure}[tbp]
\centering
\includegraphics[width=0.5\textwidth]{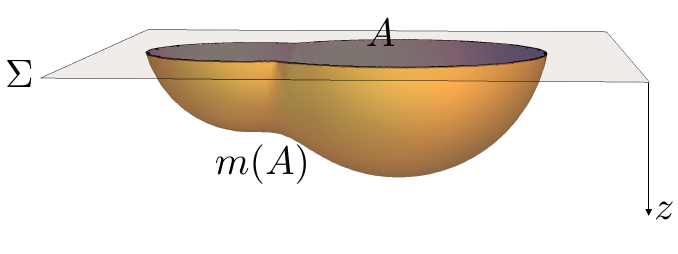}
\caption{\label{fig:3Ddressed}
Illustration of RT formula \eqref{RT}. $A$ is a region on the boundary Cauchy slice $\Sigma$. The space below $\Sigma$ is the time-reflection-invariant bulk Cauchy slice $\sigma$. $m(A)$ is the minimal surface in the bulk homologous to $A$. In particular, being homologous to $A$ implies that $m(A)$ is anchored on the entangling surface $\partial A$. It ``hangs down'' into the bulk in order to minimize its area subject to that boundary condition, given that the prefactor $1/z^2$ in the metric makes areas smaller deeper in the bulk.
}
\end{figure}

Finally, in the thermofield double example we used to motivate RT, the region $A$ was an entire boundary so there was no entangling surface $\partial A$. However, as we saw in the previous section one is often interested in studying the entropy of a region that does have an entangling surface. The homology condition then has an important implication for the minimal surface. Since
\begin{equation}\label{homology}
A\cup -m(A)= \partial r(A)\,, 
\end{equation}
and $\partial r(A)$ (being a boundary) is necessarily closed, $A$ and $m(A)$ must have the same boundary:
\begin{equation}\label{m(A) BC}
\partial m(A)=\partial A\,.
\end{equation}
So the minimal surface must reach all the way to the conformal boundary, which it must touch on the entangling surface; see fig.\ \ref{fig:3Ddressed} for an illustration. Given the $1/z^2$ factor in the metric \eqref{aAdS}, its area must diverge. Since the small-$z$ region of the bulk corresponds to the UV of the field theory, this is a UV divergence, and (as we will see in the next subsection) it accounts precisely for the expected UV divergence in $S(A)$.

The reader may feel uncomfortable about having the minimal surface reach all the way to the boundary. Recall, however, that there are horizons in asymptotically AdS spacetimes, with entropy calculated by the ordinary Bekenstein-Hawking formula, that reach the boundary. The simplest example is the AdS-Rindler wedge,
\begin{equation}
ds^2 = \frac{\ell^2}{z^2}(ds^2_M + dz^2)\,,
\end{equation}
where $M$ is $D$-dimensional Rindler wedge (the part of Minkowski spacetime with $|t|<x^1$). The boundary state is the one obtained by tracing the Minkowski vacuum on the complementary region, and the Minkowski vacuum in turn is its thermofield double, as discussed in subsection \ref{sec:higherdim} above. Its (UV-divergent) entropy is computed by applying the Bekenstein-Hawking entropy to the bifurcate horizon of the AdS-Rindler wedge, the surface $t=x^1=0$, which reaches the boundary. Thus the innovation in RT is not the fact that the minimal surface sometimes reaches the boundary, but rather that it applies in the absence of the Killing symmetries required to apply the Bekenstein-Hawking formula.

\subsubsection{Checks}

Should you believe such a bold generalization of the Bekenstein-Hawking formula? Yes, you should. We already saw in the previous subsection that it passes a few basic sanity checks. It has been checked and confirmed in many more ways, some of which we will review in these lectures:
\begin{itemize}
\item RT agrees with calculations of EEs or parts thereof (e.g.\ certain divergent terms) in cases where these can be calculated from first principles in the field theory [far too many to cite]. We will cover a few of these in subsection \ref{sec:RT examples}.
\item As we will discuss in subsection \ref{sec:RT properties}, RT obeys highly non-trivial properties of EEs, such as strong subadditivity \cite{Headrick:2007km,Headrick:2013zda}. In fact, it obeys all known applicable properties of EEs \cite{Hayden:2011ag}.
\item RT is consistent with holographic calculations of R\'enyi entropies carried out using the replica trick in the framework of Euclidean quantum gravity \cite{Headrick:2010zt,Faulkner:2013yia,Lewkowycz:2013nqa}.\footnote{In holographic theories it is much harder to calculate R\'enyi than von Neumann entropies, the opposite of the usual situation in field theories. This is related to the fact that holographic theories are in a certain sense thermodynamic, with the bulk geometry representing the macrostate. Von Neumann entropies are (as discussed in subsection \ref{sec:Renyis}) good thermodynamic quantities and are therefore captured by this bulk geometry. The R\'enyi entropies are not thermodynamic quantities, so their calculation requires re-solving the saddle-point equation, in this case the Einstein equation with different boundary conditions. See \cite{Headrick:2013zda} for further discussion on this point.}
\end{itemize}

In addition, RT is implicitly tested by its numerous applications and extensions (again, there are far too many papers to cite here):
\begin{itemize}
\item It has been applied and given reasonable results in probably hundreds of different spacetimes (in many cases informing our understanding of field-theory EEs more generally).
\item It admits sensible generalizations, briefly discussed in subsection \ref{sec:generalizations}, which have also been checked similarly to (thought not quite as extensively as) RT itself.
\item It sits at the center of a broader (but still incomplete) theoretical framework concerning the detailed relation between the bulk and boundary theories, which includes such topics as: derivation of the Einstein equation from RT; subregion duality and entanglement wedge reconstruction; relation between bulk and boundary modular Hamiltonians; holographic error-correcting codes; bit threads; tensor networks models of holography; hole-ography, differential entropy, and kinematic space; other entanglement-related quantities such as logarithmic negativity and entanglement of purification; and much more.
\end{itemize}

\subsubsection{Generalizations}
\label{sec:generalizations}

The RT formula is limited in three key ways: It requires the bulk theory to be \emph{classical} \emph{Einstein} gravity and the bulk spacetime to possess a \emph{time-reflection symmetry} under which the boundary spatial region $A$ is invariant. All three of these restrictions can be relaxed:
\begin{itemize}
\item For a general bulk spacetime, without assuming any symmetries, and a general boundary spatial region $A$, the entropy is given by the area of the minimal bulk \emph{extremal} spacelike surface homologous to $A$ \cite{Hubeny:2007xt}. (Extremal means an extremum of the area functional.) This generalization is called the Hubeny-Rangamani-Takayanagi (HRT) formula.
\item Moving away from Einstein gravity means including higher-derivative (e.g.\ stringy) corrections to the bulk gravitational action. This can be taken into account by correcting the functional of the surface to be minimized to include terms beyond the area \cite{Dong:2013qoa}. In particular, for a Lovelock-type term in the action, the correction takes a particularly simple form: it is the next-lower Lovelock term applied to the surface \cite{deBoer:2011wk,Hung:2011xb}. For example, a Gauss-Bonnet term in the gravitational action adds an Einstein-Hilbert term (integrated induced Ricci scalar) to the functional to be minimized. A generalization of RT has also been conjectured for three-dimensional Chern-Simons gravity \cite{deBoer:2013vca}.
\item Moving away from the classical limit means including $\GN$ corrections to the RT formula. At order $\GN^0$, this correction is given by treating the bulk fields (including metric perturbations) as quantum fields on a fixed background and computing the EE of the homology region $r(A)$ \cite{Faulkner:2013ana}. This is called the Faulkner-Lewkowycz-Maldacena (FLM) formula. An all-orders generalization is conjectured in \cite{Engelhardt:2014gca}. Little is known about non-perturbative quantum corrections to RT, aside from the fact that they play the important role of smoothing out the phase transitions predicted by RT, discussed in subsections \ref{sec:gapped} and \ref{sec:MI holography} below. 
\end{itemize}
Unfortunately, we will not have time in the current lectures to discuss these generalizations further.

\subsection{Examples}
\label{sec:RT examples}

In section \ref{sec:EEfields}, we either estimated $S(A)$ or computed it exactly for a variety of configurations and states of two-dimensional field theories, and saw in each case how the result reflected interesting qualitative features of the physics. To gain practice and intuition with the RT formula, in this subsection we will calculate the same quantities in the holographic setting.  We will see that the results agree with those derived earlier from first principles, providing a highly non-trivial check on the formula.\footnote{Most of the calculations in this section were first done in the original RT paper \cite{Ryu:2006ef}, although not everything was understood at that time. For example, the phase transition in the gapped case (see subsection \ref{sec:gapped interval}) was first clearly understood in \cite{Klebanov:2007ws}, while the two-interval case (subsection \ref{sec:MI holography}) was first clearly understood in \cite{Headrick:2010zt}.} In fact, it seems almost miraculous how the minimal surface manages to realize each aspect of the physics.

The bulk in each case is a static asymptotically AdS$_3$ spacetime, so the constant-time slice $\sigma$ is an asymptotically hyperbolic two-dimensional space and the ``minimal-area surface'' $m(A)$ is a minimal-length geodesic. We will also make a few comments about the higher-dimensional case as we go along.

\subsubsection{CFT vacuum: interval}
\label{sec:vacuum line}

\begin{figure}[tbp]
\centering
\includegraphics[width=0.4\textwidth]{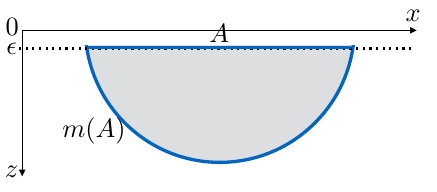}
\caption{\label{fig:vacuumdressed}
The geodesic $m(A)$ on the hyperbolic plane \eqref{hyperbolic} between the endpoints of a boundary interval $A$ is a coordinate semi-circle, truncated at the cutoff surface $z=\epsilon$. The homology region $r(A)$ that interpolates between $A$ and $m(A)$ is shown in gray.
}
\end{figure}

The metric dual to the vacuum of a two-dimensional CFT on the line is (the Poincar\'e patch of) AdS$_3$, whose constant-time slice is the hyperbolic plane, with metric
\begin{equation}\label{hyperbolic}
ds^2 = \frac{\ell^2}{z^2}(dx^2+dz^2)\,.
\end{equation}
The geodesics in this space are easily found to be coordinate semicircles centered on the boundary,
\begin{equation}\label{semicircle}
(x-x_0)^2+z^2 = r^2\,.
\end{equation}
If we put the endpoints of the interval at $x=\pm L/2$, then, in order to have the endpoints meet the cutoff surface $z=\epsilon$ at the endpoints, we should choose
\begin{equation}
x_0 = 0\,,\qquad r^2 = \left(\frac L2\right)^2+\epsilon^2\,.
\end{equation}
(See fig.\ \ref{fig:vacuumdressed}.) Note that this geodesic is (rather trivially) homologous to the interval $A$, with the region $r(A)$ being the half-disk bounded by $A$ and $m(A)$. Its length is easily computed to be $2\ell\ln(L/\epsilon)$ (dropping terms of order $\epsilon$). ({\bf Exercise:} Check these statements.) Using \eqref{ell over GN}, RT says
\begin{equation}
S(A) = \frac c3\ln\frac L\epsilon\,,
\end{equation}
in agreement with \eqref{HLW}.

The interesting thing here is how the UV divergence arises. The blow-up of the metric near the boundary forces the geodesic to hit it orthogonally. Given that the length element is $ds\sim \ell dz/z$, the total length is logarithmically divergent, with the correct coefficient. In fact, this generalizes to higher dimensions. Due to the $1/z^2$ factor in the near-boundary metric \eqref{aAdS}, the minimal surface $m(A)$ approaches $\partial A$ vertically as shown in fig.\ \ref{fig:3Ddressed}. Using \eqref{aAdS}, this gives us the leading divergent part of the area:
\begin{equation}
\int_{m(A)}\sqrt h \approx \ell^{D-1}\int_{\partial A}\sqrt{h_{\partial A}}\int_\epsilon dz\, z^{1-D} = \frac{\ell^{D-1}\sigma}{(D-2)\epsilon^{D-2}}\,,
\end{equation}
where $h$ is the determinant of the induced metric on $m(A)$, $h_{\partial A}$ is the determinant of the metric on $\partial A$ induced from $ds^2_M$ (the boundary metric), and $\sigma$ is the total area of the entangling surface (with respect to $ds^2_M$). By RT, using \eqref{ell over GN2}, and dropping constants, we find that the leading divergence in $S(A)$ is $N^2\sigma/\epsilon^{D-2}$, in agreement with \eqref{area law}. In other words, in RT the area-law UV divergence comes from the part of the minimal surface near the boundary; this part of the minimal surface represents the short-distance entanglement of nearby degrees of freedom on either side of the entangling surface. Furthermore, the number of degrees of freedom entangled in this way is $N^2$, the number of \emph{elementary} (``gluon'') fields.

Here we are discussing only the leading (area-law) divergence, but the subleading terms (such as in \eqref{4dCFT}) can also be matched, as well as divergences associated with various types of corners.

We can also reproduce the result \eqref{circle EE} for the entropy of an interval of length $L$ in the vacuum of the theory on a circle $\Sigma$ of radius $R$. This state is described by global AdS$_3$, a spatial slice of which is the hyperbolic plane, with metric 
\begin{equation}\label{global AdS}
ds^2 = \frac{\ell^2}{z^2}\left(dx^2+\frac{dz^2}{f(z)}\right),\qquad
f(z) = 1+\frac{z^2}{R^2}
\end{equation}
where $x\sim x+2\pi R$. {\bf Exercise:} Find the geodesics in the metric \eqref{global AdS}, compute the entropy of an interval, and show that they agree with \eqref{circle EE}. As noted in subsection \ref{sec:RT statement}, the fact that $S(A^c)=S(A)$ is realized geometrically by the fact that $m(A^c)=-m(A)$.

There is one other case that, while somewhat trivial, is worth mentioning, which is when $A$ is the entire circle, $A=\Sigma$. (Since there is no entangling surface and therefore no UV divergence, this case cannot be considered the $L\to2\pi R$ limit of the interval of length $L$.) Since the state is pure, we should find $S(A)=0$. Indeed, the boundary in this case is null-homologous, $A\sim\emptyset$, with the homology region being the whole bulk, $r=\sigma$. Since the empty set has area 0, the smallest any surface can have, it must be the minimal surface:
\begin{equation}\label{empty}
m(A) = \emptyset\,,\qquad r(A) = \sigma\,.
\end{equation}

\subsubsection{CFT thermal state: interval on circle}
\label{sec:CFT thermal}

A holographic CFT on a circle $\Sigma$ of radius $R$ undergoes a Hawking-Page transition at a temperature $T_{\rm HP} = 1/(2\pi R)$. Above this temperature, the bulk spacetime is the BTZ black hole:
\begin{equation}\label{BTZ}
ds^2 = \frac{\ell^2}{z^2}\left(-f(z)dt^2+dx^2+\frac{dz^2}{f(z)}\right)\,,\qquad
f(z) = 1-\frac{z^2}{z_h^2}\,,
\end{equation}
where $x\sim x+2\pi R$ and the horizon position $z=z_h$ is related to the temperature by
\begin{equation}
T = \frac1{2\pi z_h}\,.
\end{equation}
The thermal state $\rho_\Sigma$ is, of course, mixed. Since the RT formula, as stated in subsection \ref{sec:RT statement}, assumes the full system is pure, to apply it to some region $A\subseteq\Sigma$, we should first purify $\rho_\Sigma$. Furthermore we should do it in such a way that the full system is describable holographically, i.e.\ as a solution to the Einstein-matter equations (with a time-reflection symmetry). The spacetime representing any such purification must have a horizon with an external region with metric \eqref{BTZ}. Furthermore, in order to correctly reproduce the Bekenstein-Hawking entropy, the horizon must be the minimal surface homologous to the boundary $\Sigma$,\footnote{If we consider, for example, a solution containing the BTZ external region but with no other boundary, then $\Sigma$ is null-homologous and therefore has vanishing entropy. Thus such a solution is \emph{not} a purification of the thermal state $\rho_\Sigma$, but rather a pure state on $\Sigma$, although it may be indistinguishable for many observables from the thermal state.}
\begin{equation}
r(\Sigma) = \sigma\,,\qquad m(\Sigma) = m_{\rm hor}
\end{equation}
(where $\sigma$ is the constant-time slice of the BTZ metric bounded by $\Sigma$ and the horizon). One option is the thermofield double state, dual to the two-sided BTZ black hole. However, there are many other options, and on the other side of the horizon, any number of different things can happen; for example, there are multi-boundary wormhole solutions which have one horizon for each boundary and the metric \eqref{BTZ} in each exterior region.

This plethora of possible holographic purifications could lead to disaster for the RT formula, if for a region $A\subseteq\Sigma$ the minimal surface $m(A)$ depended on the choice of purification, since of course in reality $\rho_A$ and therefore $S(A)$ do not depend on that choice. Luckily, RT elegantly evades this paradox, by a property of homology regions called \textbf{nesting} \cite{Headrick:2013zda}, which says that if $A\subseteq\Sigma$ then
\begin{equation}\label{nesting1}
r(A) \subseteq r(\Sigma) = \sigma\,.
\end{equation}
According to \eqref{nesting1}, for the purposes of computing $S(A)$, the choice of purification is irrelevant; we can simply work within the external slice $\sigma$, without worrying about what happens behind the horizon.

More generally, the nesting property says that for any (disjoint) regions $A,B$,
\begin{equation}\label{nesting2}
r(A)\subseteq r(AB)\,.
\end{equation}
Hence the geometry of the homology region of any region containing $A$ is sufficient data for the purposes of computing $S(A)$; the rest of the spacetime is irrelevant. \eqref{nesting2} is thus apparently the geometrical reflection of the fact that $\rho_{AB}$ determines $\rho_A$; the state of any larger system that $AB$ happens to be part of is irrelevant once $\rho_{AB}$ is fixed. This analogy led to the notion of \emph{subregion duality}, that the bulk region $r(A)$ is the holographic dual of $\rho_A$ \cite{Czech:2012bh,Headrick:2013zda,Headrick:2014cta}.\footnote{Including the time direction, subregion duality says that the causal domain of $r(A)$, the entanglement wedge, is the dual of $\rho_A$, which as we saw in subsection \ref{sec:subsystem relativistic} is associated not just with $A$ but with its causal domain $D(A)$. Subregion duality is closely related to the idea of \emph{entanglement wedge reconstruction}, which says that the bulk local field operators in the entanglement wedge can be written in terms of field operators of the boundary field theory restricted to $D(A)$.} This idea is already familiar in the thermal case, where it is a standard fact (already used several times in these notes) that the thermal state (above the Hawking-Page transition) is represented holographically by the black hole (exterior) spacetime. Subregion duality is as much a generalization of this fact as RT is a generalization of Bekenstein-Hawking.

The version of the RT formula we gave in subsection \ref{sec:RT statement} required the full system to be in a pure state. We now see that this assumption was unnecessary. We merely need to find the minimal surface $m(A)$ homologous to $A$ within $\sigma$, where $\sigma$ is the homology region for whatever region $\Sigma$ containing $A$ that we are considering to be the ``full system'' (which could itself be a subregion of a larger manifold). Equations \eqref{RT} and \eqref{RT2} are unchanged.

\begin{figure}[tbp]
\centering
\includegraphics[width=\textwidth]{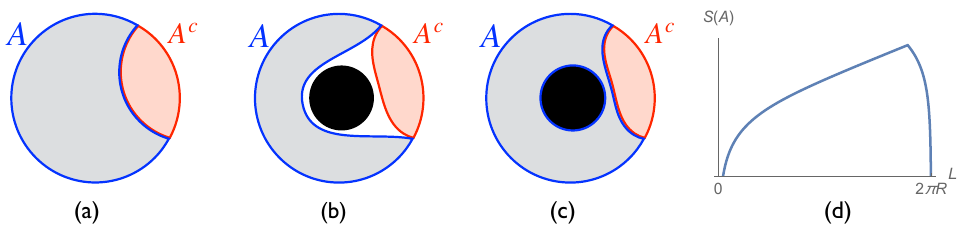}
\caption{\label{fig:thermalcircle}
RT surfaces $m(A)$, $m(A^c)$ and homology regions $r(A)$, $r(A^c)$ for an interval $A$ and the complementary interval $A^c$ on a circle. In the vacuum or other pure state, shown in (a), $m(A)$ and $m(A^c)$ coincide (more precisely $m(A^c) = -m(A)$, although orientations are not shown on the diagram). In a black hole geometry, there are two possible configurations for $m(A)$, neither of which equals $m(A^c)$: a single geodesic which is homotopic to $A$ (b), or the union of the horizon and $m(A^c)$ (c). $S(A)$, plotted in (d), is given by the smaller of the areas of the surfaces shown in (b) and (c) ((c) when $A$ covers almost the entire circle, otherwise (b)).
}
\end{figure}

However, there is an important change when the full state is not pure, or more specifically when $m(\Sigma)$ is not empty. Eq.\ \eqref{pure m equal} (with $A^c:=\Sigma\setminus A$) is no longer valid, so \eqref{pure S equal} need not hold; see fig.\ \ref{fig:thermalcircle} for examples. But this is exactly as it should be: in a mixed state, $S(A^c)$ need not equal $S(A)$. The basic physics is again perfectly captured by the RT formula.

The examples shown in fig.\ \ref{fig:thermalcircle} illustrate several other important features of minimal surfaces. First, there are two topologically distinct locally minimal surface that are both homologous to $A$. One of them, (b), has a single connected component; its length is
\begin{equation}\label{thermal circle1}
2\ell\ln\frac{\sinh(\pi TL)}{\pi T\epsilon}\,,
\end{equation}
while the other, (c), has two components, one of which coincides with the horizon and the other with $m(A^c)$; the total length is
\begin{equation}\label{thermal circle2}
(2\pi)^2\ell RT + 2\ell\ln\frac{\sinh(\pi T(2\pi R-L))}{\pi T\epsilon}\,.
\end{equation}
({\bf Exercise:} Find the geodesics and reproduce \eqref{thermal circle1}, \eqref{thermal circle2}.) Whichever surface has smaller total area is $m(A)$, and this depends on $L$, with the first being smaller for smaller $L$ and the second for larger $L$:
\begin{equation}\label{thermal circle}
S(A) = \frac c3\min\left\{\ln\frac{\sinh(\pi TL)}{\pi T\epsilon},2\pi^2 RT + \ln\frac{\sinh(\pi T(2\pi R-L))}{\pi T\epsilon}\right\}
\end{equation}
(plotted in fig.\ \ref{fig:thermalcircle}(d)). Thus the minimal surface will switch as $A$ is continuously deformed, leading to a sort of ``phase transition'' in $S(A)$. We will see such jumps again, and discuss their meaning, in subsections \ref{sec:gapped interval} and \ref{sec:MI holography}. Fig.\ \ref{fig:thermalcircle}(c) illustrates the fact that the minimal surface, although homologous to $A$, need not be \emph{homotopic}, i.e.\ continuously deformable, to it. Note that the entropy here also saturates the Araki-Lieb inequality \eqref{arakilieb}: since $m(A)$ consists of a piece that wraps the horizon, with area $S(\Sigma)$, and a piece that coincides with $m(A^c)$, we have
\begin{equation}
S(A)-S(A^c) = S(AA^c)\,.
\end{equation}

\subsubsection{CFT thermal state: interval on line}

\begin{figure}[tbp]
\centering
\includegraphics[width=0.8\textwidth]{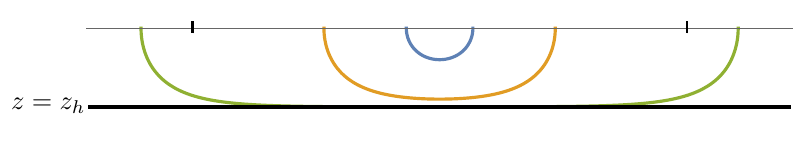}
\caption{\label{fig:thermal}
Geodesics on a spatial slice of the planar BTZ black hole geometry \eqref{BTZ} for various boundary intervals. The separation between the ticks is the thermal length $1/T$. For intervals much shorter than that, the geodesic is close to being a semi-circle as in the vacuum case of fig.\ \ref{fig:vacuumdressed}; the geodesic stays close to the boundary, where the metric is approximately the same as \eqref{hyperbolic}. On the other hand, for intervals longer than $1/T$, the geodesic skirts the horizon over a long distance, resulting in the linear ``volume-law'' behavior of the entropy, with slope equal to the thermal entropy density. (In the figure, the curve appears, due to its finite thickness, to coincide with the horizon; in fact it comes very close---in this case, to $z=0.999z_h$---but does not touch the horizon.)
}
\end{figure}

As noted in subsection \ref{sec:CFT thermal circle}, there is no universal result for the entropy of an interval in a CFT in a thermal state on a circle, so the above result for $S(A)$ is a new prediction of RT, rather than a check of it. However, on the line the result becomes universal, and is given by \eqref{thermal line}. We can easily check this by taking the $R\to\infty$ limit of \eqref{thermal circle}; in this limit the first surface (whose length is actually $R$-independent) is always shorter, and indeed we reproduce \eqref{thermal line}. The geometry is simply \eqref{BTZ} without the periodic identification of $x$, and the geodesics are shown in fig.\ \ref{fig:thermal}. {\bf Exercise:} Find the geodesics and reproduce \eqref{thermal line}.

As noted in subsection \ref{sec:CFT thermal circle}, for intervals much shorter than the thermal length $1/T$, the entropy is close to that in the vacuum. Here this is due to the fact that the corresponding geodesics stay close to the boundary, where the metric is very close to the vacuum metric \eqref{hyperbolic}. 

Another feature noted in subsection \ref{sec:CFT thermal circle} is the linear, ``volume-law'' part of the entropy for $L\gg1/T$, with slope equal to the thermal entropy density. Again, this property of the entropy is manifested in the behavior of the geodesic. As can be seen in fig.\ \ref{fig:thermal}, for large $L$, the geodesic basically has three parts: a part that goes from the boundary almost to the horizon and which in the $L\to\infty$ limit is independent of $L$, a part that is very close to (but not intersecting) the horizon, and a part that goes from almost the horizon back to the boundary and again is independent of $L$. The first and last parts are UV divergent and contribute the $\ln\epsilon$ part of the entropy, while the part that hugs the horizon contributes an amount equal to $L$ times the thermal entropy density (which is the Bekenstein-Hawking entropy per unit of boundary distance). This is a general phenomenon: The minimal surface for a large region in a deconfined state hugs the horizon, reproducing the expected volume-law thermal part of the EE.

\subsubsection{Gapped theory: half-line}
\label{sec:gapped}

In the last two examples (CFT at finite temperature and on a circle), conformal invariance was broken either by the state or by the geometry of the space on which the field theory lived. Conformal invariance can also be broken by the theory itself. A holographic theory that flows from a UV fixed point is represented by a gravitational theory with boundary conditions such that the ground state (on Minkowski space, say) is not AdS$_{D+1}$ but rather some asymptotically AdS$_{D+1}$ spacetime, on which typically some scalar field depends on the holographic coordinate $z$. If the theory has a non-trivial IR fixed point, then the bulk has a domain wall parallel to the boundary which interpolates between the UV AdS$_{D+1}$ region near the boundary and another one (with different cosmological constant) in the interior. A simpler case, however, is a wall on which space simply ends. This represents an RG flow to a trivial theory with no degrees of freedom in the deep IR, in other words a gapped or massive theory (also known in the gauge-theory context as a \emph{confining} theory). There exist many holographic constructions of gapped theories, of which some are honest solutions to known quantum gravity theories and others are constructed in an ad hoc fashion, often in order to model some real-world phenomenon.

Since our purpose here is merely to illustrate the qualitative behavior of EEs in gapped theories, we will consider the simplest possible ad hoc theory. This is a two-dimensional theory on Minkowski space, whose vacuum is represented by the AdS$_3$ metric, but where the $z$ coordinate extends only from $\epsilon$ to some finite value $\xi$, at which point the bulk space simply ends:
\begin{equation}\label{confining}
ds^2 = \frac{\ell^2}{z^2}\left(-dt^2+dx^2+dz^2\right),
\qquad \epsilon<z\le\xi\,.
\end{equation}
(In honest models of gapped theories, the radial direction ends in a more graceful manner, for example with a compact internal manifold smoothly capping off.) Since as a general rule the region farther from the boundary represents the IR of the field theory, it is intuitive that if the space ends then the theory would have no degrees of freedom in the IR. More quantitatively, by putting appropriate boundary conditions on the bulk fields at $z=\xi$, one can show that correlation functions of boundary fields die off at distances larger than $\xi$ like $e^{-\Delta x/\xi}$ or faster, so $\xi$ is the correlation length.

We emphasized in the previous subsection that a horizon in the bulk influenced the homology condition, so that $A$ and $A^c$ would not have the same minimal surface, reflecting the fact that the full system is not in a pure state. Given the presence of the confining wall, the same argument would seem to lead to the same conclusion, yet the system is in the vacuum, a pure state. The resolution to this puzzle is that, unlike a horizon, a confining wall $X$ (an internal wall on which space ends) does \emph{not} count as a ``boundary'' for the purpose of imposing the homology constraint. In other words, we only require that \eqref{homology} is satisfied modulo $X$. (Technically, we work in homology \emph{relative to} $X$.) In the case where the wall is due to an internal compact manifold capping off, this is very natural, since from the higher-dimensional viewpoint there is indeed no boundary at $X$. More generally, the physical reason for this rule is that a wall, unlike a horizon, does not carry entropy. For example, if we apply RT to the whole time slice $\Sigma$, we find that it is null-homologous, $\Sigma\sim\emptyset$, with $r(\Sigma)=\sigma$, the whole bulk time slice, giving the correct answer $S(\Sigma)=0$, as in \eqref{empty}.

\begin{figure}[tbp]
\centering
\includegraphics[width=0.5\textwidth]{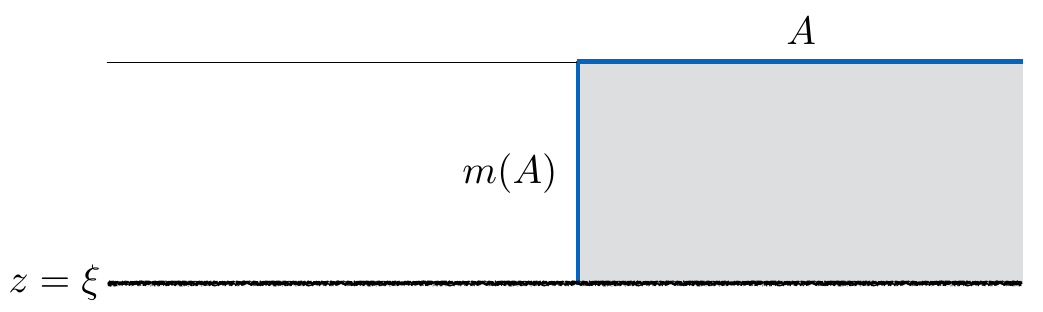}
\caption{\label{fig:gappedhalfline}
The minimal surface $m(A)$ for the half-line $A=\{x\ge0\}$ is the straight line in the bulk at $x=0$, extending from the boundary to the wall.
}
\end{figure}

A more interesting case is the half-line, $A=\{x\ge0\}$, studied in a general field theory in subsection \ref{sec:Rindler}. To apply RT, we look for a geodesic in the geometry \eqref{confining} that begins at $(x,z)=(0,\epsilon)$  and does not end anywhere else on the boundary, so it must end on the wall. As discussed in subsection \ref{sec:vacuum line}, such geodesics are coordinate semicircles (see \eqref{semicircle}). The shortest one is just the straight line (a ``semicircle'' with infinite radius),
\begin{equation}
m(A) = \{x=0, \epsilon<z\le\xi\}
\end{equation}
(see fig.\ \ref{fig:gappedhalfline}). Its length is
\begin{equation}
\int_\epsilon^\xi dz\frac\ell z = \ell\ln\frac\xi\epsilon\,.
\end{equation}
Using \eqref{ell over GN}, we find
\begin{equation}
S(A) = \frac c6\ln\frac\xi\epsilon\,,
\end{equation}
in agreement with the general result \eqref{S(A) estimate}. (The fact that we got the value on the right-hand side of \eqref{S(A) estimate} \emph{exactly} is  due to the artificial simplicity of our spacetime and should not be taken as significant.) The homology region is the half-space
\begin{equation}
r(A) = \{x\ge0,\epsilon< z\le\xi\}\,.
\end{equation}

\subsubsection{Gapped theory: interval}
\label{sec:gapped interval}

\begin{figure}[tbp]
\centering
\includegraphics[width=0.9\textwidth]{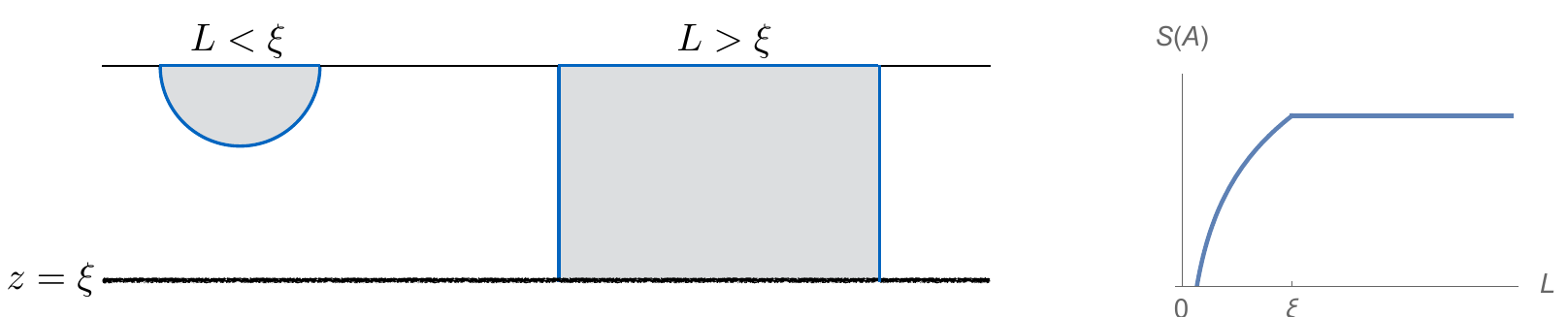}
\caption{\label{fig:deconfinement}
Left: In a bulk with a confining wall at $z=\xi$, the minimal surface $m(A)$ for an interval $A$ of length $L$ takes two different topologies depending on $L$: for $L<\xi$ it is a semicircle connecting the endpoints of the interval, while for $L>\xi$ it is two line segments connecting each endpoint to the wall. Right: As a result, the entropy $S(A)$ shows a first-order transition (continuous with discontinuous first derivative) at $L=\xi$ from a logarithmic growth to a constant; see \eqref{gapped holo}.
}
\end{figure}

We now move on to the case of a finite interval, $A=\{-L/2\le x\le L/2\}$. For a general massive theory, we concluded in subsection \ref{sec:one interval general} that $S(A)$ was given for short and long intervals by \eqref{gapped short} and \eqref{gapped long} respectively. Here we encounter again, as we saw in subsection \ref{sec:CFT thermal}, the phenomenon of competing minimal surfaces. One \emph{locally} minimal surface homologous to $A$ is the union of vertical segments connecting each endpoint to the wall:
\begin{equation}
m_1 = \left\{x=-\frac L2, \epsilon<z\le\xi\right\} \cup \left\{x=\frac L2, \epsilon<z\le\xi\right\}\,.
\end{equation}
The corresponding homology region is the rectangle
\begin{equation}
r_1 = \left\{-\frac L2\le x\le \frac L2, \epsilon<z\le\xi\right\} \,.
\end{equation}
$m_1$ has length
\begin{equation}
l_1 = 2\ell\ln\frac\xi\epsilon\,.
\end{equation}
On the other hand, for sufficiently short intervals, there is another homologous geodesic, namely the semicircle connecting the endpoints:
\begin{equation}
m_2 =\left \{x^2+z^2 = \left(\frac L2\right)^2+\epsilon^2\right\}.
\end{equation}
$m_2$ is only locally minimal for $L<2\xi$. For $L\ge 2\xi$, this semicircle collides with the wall; the arcs in the physical region $0<z\le\xi$ are not minimal-length since they can be shortened by straightening them out, returning us to $m_1$. For $L<2\xi$ it has length
\begin{equation}
l_2 = 2\ell\ln\frac L\epsilon\,.
\end{equation}
We see that $m_2$ is shorter for $L<\xi$ while $m_1$ is shorter for $L>\xi$, giving
\begin{equation}\label{gapped holo}
S(A) = \frac c3\times\begin{cases}\ln\frac L\epsilon\,,\quad &L<\xi \\ \ln\frac\xi\epsilon\,,\quad & L>\xi\end{cases}\,.
\end{equation}
So as we vary $L$, the entropy is continuous but has a discontinuous first derivative at $L=\xi$; we can think of this as a phase transition, as discussed below. Basically, the short- and long-interval estimates \eqref{gapped short} and \eqref{gapped long} become \emph{exact} in this holographic set-up. The behavior is shown in fig.\ \ref{fig:deconfinement}.

While the details of the calculation depend on the bulk geometry, the existence of a phase transition does not: in any gapped two-dimensional holographic theory, $S(A)$ undergoes a phase transition as we increase $L$ past a critical value $L_c$ which is of the order of the correlation length. For $L<L_c$, $m(A)$ connects the endpoints of $A$, while for $L>\xi$ it consists of segments extending directly from each endpoint to the wall. In higher dimensions, analogous behavior is seen: roughly speaking, when $A$ is small compared to the correlation length, $m(A)$ has the same topology as $A$, hanging down in to the bulk but not touching the wall; when $A$ is large, $m(A)$ has the topology of $\partial A$ times an interval, and directly connects $\partial A$ to the wall.

We can address this behavior of the EE using our intuition from statistical mechanics by writing $\rho_A$ as the thermal state at unit temperature with respect to the modular Hamiltonian $H_A$, $\rho_A=e^{-H_A}/Z$. As we vary $L$, the parameters in this Hamiltonian change, changing the entropy. (Note that, unlike for a half-line or an interval in a CFT, $H_A$ is not the integral of a local operator, but rather some complicated non-local operator which we don't have an explicit expression for. This does not affect the discussion that follows.) Based on the behavior of $S(A)$, we conclude that the system must undergo a first-order phase transition at $L=L_c$. This is surprising, since the system has finite spatial extent and therefore (in the presence of the UV cutoff $\epsilon$) would seem to have a finite number of degrees of freedom. The resolution to this puzzle is that we are working in the \emph{classical} limit $\GN\hbar\to0$ of the bulk, which represents the limit $c\to\infty$ in the field theory. This is indeed a kind of thermodynamic limit---not the infinite-volume thermodynamic limit typically encountered for example in lattice condensed-matter systems, but a limit with an infinite \emph{density} of degrees of freedom. A more familiar example is the Hawking-Page transition. Such transitions cannot occur at finite $c$, so they must be smoothed out by $1/c$ corrections, which translate holographically to quantum corrections. More precisely, only \emph{non-perturbative} corrections can smooth out a phase transition. In the case of the Hawking-Page transition, these corrections can be computed using the framework of Euclidean quantum gravity. Unfortunately, we do not yet know how to compute non-perturbative quantum corrections to holographic EEs.

We can get another point of view on the phase transition in EE by comparing it to the Page curve, fig.\ \ref{fig:Page}. Recall that this curve represents the EE $S(A)$ in a \emph{random} pure state of a large bipartite system, as a function of the fractional size of subsystem $A$. This entropy is as large as it can be subject to the two constraints \eqref{SAbound1}, \eqref{SAbound2}, and the phase transition occurs when the active constraint switches. Holographic states are certainly not random states in the full field-theory Hilbert space---their EEs are far smaller than that of a random state. Nonetheless, the similarity of figs.\ \ref{fig:thermalcircle}(d) and \ref{fig:deconfinement} to the Page curve suggests that perhaps holographic states are essentially generic states subject to some constraints. The simplest possibility consistent with RT is that every bulk surface $m$ homologous to $A$ imposes a constraint on $S(A)$ proportional to its area,
\begin{equation}\label{surfconstraints}
S(A) \le \frac1{4G_{\rm N}}\area(m)\,,
\end{equation}
and $S(A)$ is as large as possible subject to all the constraints \eqref{surfconstraints}; a phase transition is simply the point at which the active constraint switches from one surface to another. Unfortunately, we don't yet know if this picture is correct, and if so what the nature of the constraints is.

\subsubsection{CFT: Two intervals}
\label{sec:MI holography}

\begin{figure}[tbp]
\centering
\includegraphics[width=\textwidth]{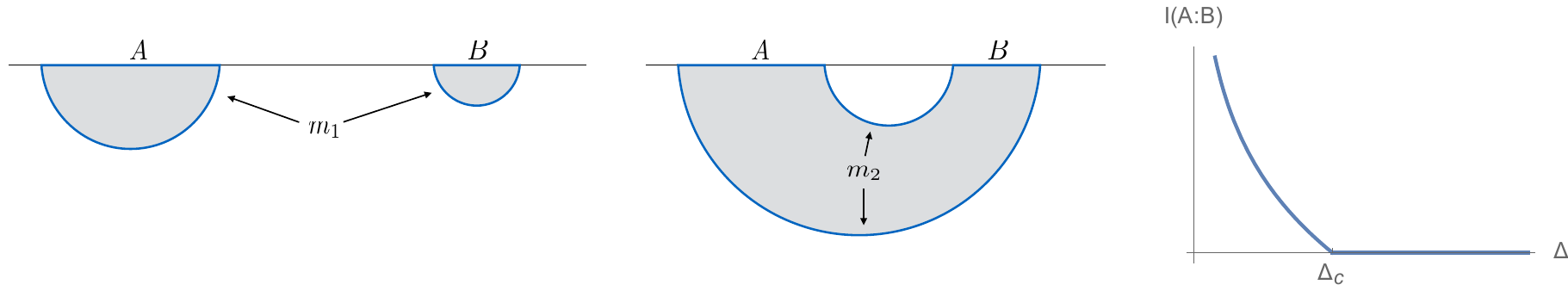}
\caption{\label{fig:twointervals}
Left and middle: Candidate minimal surfaces for two intervals in a CFT vacuum. $m_1$ is the union of the minimal surfaces for $A$ and $B$ respectively, while $m_2$ connects $A$ and $B$. Right: Mutual information \eqref{holoMI} calculated from these minimal surfaces as a function of the separation $\Delta$ between the intervals.
}
\end{figure}

We conclude this subsection with the simplest example of a disconnected region, namely the union of two disjoint intervals $A,B$ on a line in a CFT vacuum. This was discussed in the context of a general field theory in subsection \ref{sec:two intervals}. In computing $S(AB)$, a locally minimal surface consists of geodesics connecting the four endpoints of $AB$. According to the boundary condition \eqref{m(A) BC}, and keeping track of orientations, each geodesic must begin on the left endpoint of an interval and end on the right endpoint of an interval. There are two possible configurations, depending on whether each geodesic connects the endpoints of the same or different intervals (see fig.\ \ref{fig:twointervals}):
\begin{enumerate}
\item One geodesic connects the two endpoints of $A$ and the other the two endpoints of $B$; this is the union of the respective minimal surfaces for $A$ and $B$, $m_1=m(A)\cup m(B)$, and it is homologous to $AB$ via the union of the two homology regions, $r_1=r(A)\cup r(B)$. The ``area'' (or length) of $m_1$ is the sum of the lengths of $m(A)$ and $m(B)$:
\begin{equation}
\area(m_1)=\frac\ell2\ln\frac{L_A}\epsilon+\frac\ell2\ln\frac{L_B}\epsilon\,.
\end{equation}
\item The geodesics connect the left endpoint of $A$ to the right endpoint of $B$ and vice versa. Here the homology region $r_2$ connects $A$ to $B$. We have
\begin{equation}
\area(m_2)=\frac\ell2\ln\frac\Delta\epsilon+\frac\ell2\ln\frac{L_A+L_B+\Delta}\epsilon\,.
\end{equation}
\end{enumerate}
So, as in the single interval in the gapped theory, we see a phase transition: $m_1$ is the minimal surface if $\Delta>\Delta_c$ while $m_2$ is the minimal surface if $\Delta<\Delta_c$, where the critical separation is
\begin{equation}
\Delta_c:=\sqrt{\left(\frac{L_A+L_B}2\right)^2+L_AL_B}-\frac{L_A+L_B}2\,.
\end{equation}
From this we easily calculate the mutual information,
\begin{equation}\label{holoMI}
I(A:B)=\begin{cases}
\frac c3\ln\frac{L_AL_B}{\Delta(L_A+L_B+\Delta)}\,,\quad&\Delta\le\Delta_c \\
0\,,\quad&\Delta\ge\Delta_c
\end{cases}\,,
\end{equation}
plotted in fig.\ \ref{fig:twointervals}. As in the gapped theory, this phase transition must be smoothed out at finite $c$ by non-perturbative quantum corrections to the RT formula.

In subsection \ref{sec:two intervals}, we made several predictions for properties of $I(A:B)$ that should hold in any CFT. We see that the holographic mutual information \eqref{holoMI} obeys all but one of them, namely number 2: it vanishes for finite separations. But this is impossible, since vanishing mutual information implies $\rho_{AB}=\rho_A\otimes\rho_B$, implying that all correlators factorize, $\ev{\mathcal{O}_A\mathcal{O}_B}=\ev{\mathcal{O}_A}\ev{\mathcal{O}_B}$, which is obviously not true. The resolution to this apparent contradiction lies in remembering that the RT formula only tells us the \emph{leading}, order $c$, entropy in a $1/c$ expansion. So all we know from this calculation is that the order-$c$ part of the mutual information vanishes for $\Delta\ge\Delta_c$. The order-1 term arises from a one-loop correction to the RT formula. (It can also be computed directly in the CFT using large-$c$ techniques.) The result depends on the matter content of the bulk theory, but it is non-zero in any theory, as we would expect.

\subsection{General properties}
\label{sec:RT properties}

We have already noted several general properties of EEs that the RT formula obeys. These include the leading area divergence (subsection \ref{sec:vacuum line}), and the fact that in a pure state $S(A^c)=S(A)$ (subsection \ref{sec:RT statement}) whereas in a mixed state this need not be the case (subsection \ref{sec:CFT thermal}). We also saw in one particular case---two intervals in the vacuum of a CFT on the line (subsection \ref{sec:MI holography})---that the subadditivity property
\begin{equation}\label{subadditivity2}
S(AB)\le S(A)+S(B)\,,
\end{equation}
is satisfied.

In fact it is straightforward to see that RT always satisfies \eqref{subadditivity2}. As always, $A$ and $B$ are disjoint regions. First we note that $r(A)$ and $r(B)$ don't overlap. This follows from the nesting property: since $B\subseteq A^c$, $r(B)\subseteq r(A^c)=r(A)^c$. (The last equality assumes the full system is pure. If it isn't, purify it first; this doesn't affect $r(A)$ or $r(B)$.) Let their union be $r_1$:
\begin{equation}
r_1 := r(A)\cup r(B)\,.
\end{equation}
$r_1$ intersects the conformal boundary on $AB$, so it's a candidate homology region. Let its other boundary (in the bulk) be $m_1$:
\begin{equation}
\partial r_1 = AB\cup-m_1\,.
\end{equation}
Then, since $m_1\sim AB$,
\begin{equation}
S(AB)\le\frac1{4G_{\rm N}}\area(m_1)\,.
\end{equation}
Now, just because $r(A)$ and $r(B)$ don't overlap doesn't mean they don't touch. For example, if the full system is pure and $B=A^c$, then $r(A)$ and $r(B)$ meet along $m(B)=-m(A)$ (fig.\ \ref{fig:thermalcircle}(a)). Let $m'$ be their common boundary; thus $m'\subseteq m(A)$, $-m'\subseteq m(B)$. The surface $m_1$ is the union of the remaining parts of $m(A)$ and $m(B)$:
\begin{equation}
m_1 = (m(A)\setminus m')\cup(m(B)\setminus -m')\,.
\end{equation}
So
\begin{equation}
\area(m_1) \le\area(m(A))+\area(m(B))\,,
\end{equation}
establishing \eqref{subadditivity2}.

\begin{figure}[tbp]
\centering
\includegraphics[width=\textwidth]{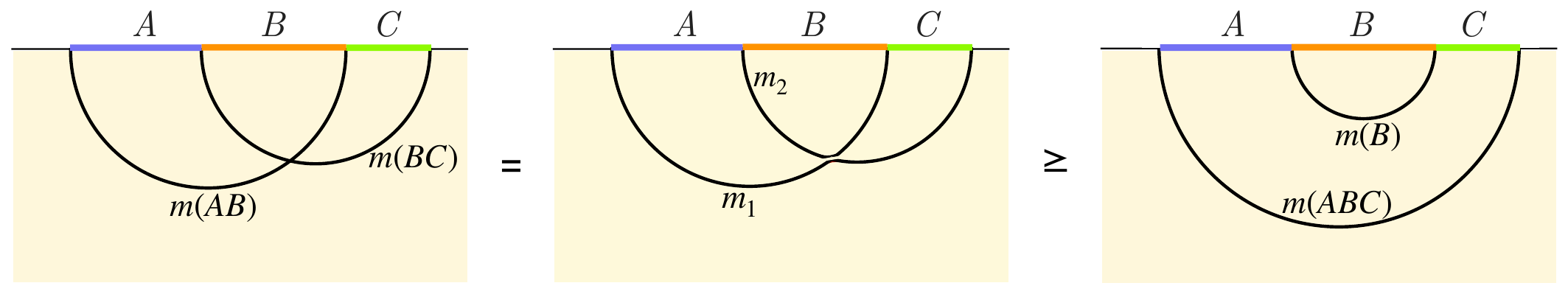}
\caption{\label{fig:SSA}
Illustration of proof that RT obeys the strong subadditivity inequality \eqref{SSA3}. The two surfaces $m(AB)$, $m(BC)$ are cut and pasted to give surfaces $m_1$, $m_2$ with the same total area. $m_1$ and $m_2$ are homologous to $ABC$ and $B$, via the regions $r(AB)\cup r(BC)$ and $r(AB)\cap r(BC)$, so their areas bound $S(ABC)$ and $S(B)$, respectively.
}
\end{figure}

Similar but more elaborate arguments establish two other important inequalities. The first is strong subadditivity \cite{Headrick:2007km,Headrick:2013zda}:
\begin{equation}\label{SSA3}
S(AB)+S(BC)\ge S(B)+S(ABC)\,.
\end{equation}
To avoid cluttering the proof, we will neglect the possibility of regions that touch; the reader should check that the presence of touching regions only strengthens the inequality. We define bulk regions $r_{1,2}$ as follows:
\begin{equation}
r_1 := r(AB)\cup r(BC)\,,\qquad
r_2:=r(AB)\cap r(BC)\,.
\end{equation}
These define bulk surfaces $m_{1,2}$, that are homologous to $AB\cup BC=ABC$ and $AB\cap BC=B$ respectively:
\begin{equation}
m_1\sim ABC\,,\qquad m_2\sim B\,.
\end{equation}
Therefore we have
\begin{equation}\label{SSAproof1}
S(ABC)\le\frac1{4G_{\rm N}}\area(m_1)\,,\qquad
S(B)\le\frac1{4G_{\rm N}}\area(m_2)\,.
\end{equation}
On the other hand, $m_1$ and $m_2$ are obtained by swapping parts of $m(AB)$ and $m(BC)$, so they have the same total area:
\begin{equation}\label{SSAproof2}
\area(m_1)+\area(m_2) = \area(m(AB))+\area(m(BC))\,;
\end{equation}
this can be shown by decomposing $m(AB)$ into the part inside $r(BC)$ and the part outside it, and similarly for $m(BC)$, and rearranging the parts to obtain $m_1$ and $m_2$. Together, \eqref{SSAproof1} and \eqref{SSAproof2} establish \eqref{SSA3}. See fig.\ \ref{fig:SSA} for an illustration of the proof. Once again, we see that RT, in an elegant and almost effortless manner, encodes geometrically a fundamental and highly non-trivial aspect of EEs. A surprising fact about the proof is how little physics is involved: we did not invoke any equations of motion, asymptotically AdS boundary conditions, or any other aspect of holography aside from the RT formula itself.

The second inequality is called \textbf{MMI}\footnote{MMI stands for \emph{monogamy of mutual information}. The meaning of the long name is somewhat obscure and unimportant for our purposes.} \cite{Hayden:2011ag}:
\begin{equation}\label{MMI}
S(AB)+S(BC)+S(AC)\ge S(A)+S(B)+S(C)+S(ABC)\,.
\end{equation}
The proof of MMI uses the same techniques as for strong subadditivity; we leave it as an exercise to the reader. While MMI looks like a natural generalization of subadditivity and strong subadditivity, there is a crucial difference: unlike those inequalities (and the other properties of RT discussed so far), MMI is \emph{not} a general property of EEs. In fact, it is quite easy to find a quantum state that violates \eqref{MMI}:
\begin{equation}
\rho_{ABC} = \frac12\left(\ket{000}\bra{000}+\ket{111}\bra{111}\right).
\end{equation}
Nor does MMI hold in general quantum field theories \cite{Casini:2008wt}. Instead, it is a special property of \emph{holographic} theories. To give some intuition for the inequality, it is helpful to write it in terms of mutual informations:
\begin{equation}
I(A:BC)\ge I(A:B)+I(A:C)\,.
\end{equation}
Thus MMI is saying that the mutual information is \emph{superadditive}. A simple example illustrates how this works. Suppose that $A,B,C$ are all intervals in a 2d CFT, with $B$ and $C$ adjacent and $A$ somewhat separated. It can easily be arranged that $I(A:B)=I(A:C)=0$, yet $I(A:BC)>0$. It is not entirely clear what is behind the MMI property, in other words why holographic states obey this inequality. While MMI is not a general property of entropies, it does imply an infinite set of constrained inequalities that \emph{are} general properties \cite{Cadney2011}. This is actually the complete set of known general properties of the von Neumann entropy; thus, rather remarkably, RT obeys \emph{all} known general properties of entropy.

MMI is just the first in an infinite set of special inequalities obeyed by RT, which (along with subadditivity and strong subadditivity) define the so-called \textbf{holographic entropy cone} \cite{Bao:2015bfa}.

\acknowledgments{
The author's work was supported by the National Science Foundation through Career Award No.\ PHY-1053842; by the Simons Foundation through \emph{It from Qubit: Simons Collaboration on Quantum Fields, Gravity, and Information} and through a Simons Fellowship in Theoretical Physics; and by the Department of Energy through grant DE-SC0009987. I would also like to thank the MIT Center for Theoretical Physics, where much of these notes were written, for hospitality throughout the year 2017. I would like to thank the organizers of the various schools where the lectures upon which these notes are based were delivered, for inviting me to speak, as well as the students at those schools, whose many interesting questions and comments improved the lectures and my own understanding of the subject.
}

\bibliographystyle{JHEP}
\bibliography{refs}

\end{document}